**Prediction bias in biological ageing markers**

**Authors:** Yiqun Lin[1,2,3,4], Ariel Yuhan Ong[1,2], Matthew Yu Heng Wong[5], Yilan Wu[1,6], Wenyi Hu[7,8], Shubhank Shobhit Verma[1], Broder Poschkamp[1,9], Lie Ju[1,2], Akshay Narayan[1], Sophie A. Martin[3,4], Maitrei Kohli[3,4], Fares Antaki[11,12,13], Alexander C. Heatley[1], Josef Huemer[1,2,14], Meng Wang[15,16], Ke Zou[15,17], Zheyuan Wang[18,19], Eden Ruffell[1,2], Dominic Williamson[1,2], Justin Engelmann[1], Hyunmin Kim[1,2], Takahiro Ninomiya[1,20], Rahul A. Jonas[1], Yansong Liu[4], Zijie Cheng[10], Hanyuan Zhang[3,21], Hejie Cui[22], Jiayang Xu[23], Hualiang Wang[23], Zuozhu Liu[24,25], Bin Pu[26], Chubo Liu[26], Kenli Li[26], Xinpeng Ding[27], Guokai Zhang[28], Weiru Wang[1,2,3,4], Jiaxiang Wang[1,29], Huazhu Fu[30], Zongyuan Ge[31,32], Xujia Liu[33], Mohammad Eslami[34,35], Milen Raytchev[34,35], Tobias Elze[34,35], Michael G. Morley[34,35], Neil P. Oxtoby[3,4], Daniel C. Alexander[3,4], Anthony P. Khawaja[1,2], Ching-Yu Cheng[15,16], Lisa Zhuoting Zhu[7,8], Yih Chung Tham[15,16,17], James H. Cole[3,4], Carol Y. Cheung[33], Pearse A. Keane[1,2], Siegfried K. Wagner[1,2], Yukun Zhou[1,2,3,4+]

**Affiliations:**

[1] Institute of Ophthalmology, University College London, London, UK

[2] NIHR Biomedical Research Centre, Moorfields Eye Hospital NHS Foundation Trust, London, UK

[3] UCL Hawkes Institute, University College London, London, UK

[4] Department of Computer Science, University College London, London, UK

[5] University of Cambridge, Cambridge, UK

[6] Beijing Visual Science and Translational Eye Research Institute, Eye Center of Beijing Tsinghua Changgung Hospital, School of Clinical Medicine, Tsinghua Medicine, Tsinghua University, Beijing, China

[7] Centre for Eye Research Australia, Melbourne, VIC, Australia

[8] Department of Surgery (Ophthalmology), The University of Melbourne, Melbourne, VIC, Australia

[9] Institute for Community Medicine, University Medicine Greifswald, Greifswald, Germany

[10] University College London, London, UK

[11] Cole Eye Institute, Cleveland Clinic, Cleveland, OH, USA

[12] Department of Ophthalmology, Centre Hospitalier de l'Université de Montréal, Montréal, QC, Canada

[13] The CHUM School of Artificial Intelligence in Healthcare, Centre Hospitalier de l'Université de Montréal, Montreal, QC, Canada

[14] Department of Ophthalmology and Optometry, Kepler University Hospital, Linz, Austria

[15] Department of Ophthalmology, Yong Loo Lin School of Medicine, National University of Singapore, Singapore

[16] Singapore Eye Research Institute, Singapore National Eye Centre, Singapore

[17] Centre for Innovation and Precision Eye Health, National University of Singapore, Singapore

[18] Department of Computer Science and Engineering, Shanghai Jiao Tong University, Shanghai, China

[19] MOE Key Laboratory of Artificial Intelligence, School of Electronic, Information, and Electrical Engineering, Shanghai Jiao Tong University, Shanghai, China

[20] Graduate School of Medicine, Tohoku University, Sendai, Japan

[21] Department of Medical Physics and Biomedical Engineering, University College London, London, UK

[22] Stanford University, Stanford, CA, USA

[23] Department of Electronic and Computer Engineering, The Hong Kong University of Science and Technology, Hong Kong SAR, China

[24] ZJU-Angelalign R&D Center for Intelligent Healthcare, ZJU-UIUC Institute, Zhejiang University, Haining, China

[25] Zhejiang Key Laboratory of Medical Imaging Artificial Intelligence, Zhejiang University, Hangzhou, China

[26] College of Computer Science and Electronic Engineering, Hunan University, Changsha, China

[27] Hangzhou Institute of Technology, Xidian University, Hangzhou, China

[28] University of Shanghai for Science and Technology, Shanghai, China

[29] Department of Neurosurgery, Xiangya Hospital, Central South University, Changsha, China

[30] Institute of Advanced Intelligence and Computing, Agency for Science, Technology and Research (A*STAR), Singapore

[31] Department of Data Science and AI, Monash University, Melbourne, VIC, Australia

[32] AIM for Health Lab, Monash University, Melbourne, VIC, Australia

[33] Department of Ophthalmology and Visual Sciences, Faculty of Medicine, The Chinese University of Hong Kong, Hong Kong SAR, China

[34] Department of Ophthalmology, Harvard Medical School, Boston, MA, USA

[35] Schepens Eye Research Institute of Massachusetts Eye and Ear, Boston, MA, USA

[+] **Corresponding author:** Yukun Zhou (yukun.zhou.19@ucl.ac.uk)

**Abstract**

Biological ageing markers have attracted growing interest, with models estimating age from organ imaging or blood biomarkers. An estimated age above chronological age or the age-specific population expectation is assumed to reflect accelerated ageing and poorer health. Previous research has supported this assumption through positive associations between disease and age gaps or acceleration. However, in this study, we identified a widespread health-dependent prediction bias in ageing markers that affects their key interpretation and application. Specifically, we investigated five ageing markers derived from retinal images, brain MRI, chest radiographs, abdominal CT, and blood tests, and evaluated them using association analyses. We observed the well-recognised phenomenon of regression to the mean (RTM) in the four organ-image based markers, whereby estimated ages were shifted towards the mean age of the training cohort. More importantly, we revealed that the strength of RTM varied with health status, with stronger RTM in unhealthy than in healthy individuals. This differential RTM introduced a health-dependent prediction bias that persisted after calibration and systematically altered associations across age subgroups, suggesting that whole cohort associations may not reflect those observed within individual age subgroups. Additionally, we showed that the tested ageing markers, including PhenoAge derived from blood biomarkers, had limited ability to distinguish health status at the individual level. These findings call for careful interpretation of biological ageing markers and their use in clinical studies, and highlight the need for further development and validation before these ageing markers can reliably inform individual health assessments.

## Introduction

Ageing is a complex biological process characterised by the accumulation of cellular damage and physiological decline [1,2]. While chronological age reflects the time since birth, it does not fully capture an individual's true health status, which is influenced by a combination of factors such as socioeconomic determinants, genetic background, and comorbidity burden [3,4]. In contrast, biological ageing markers, such as brain age [5–9], face age [10], and epigenetic clocks [11,12] better reflect physiological status predictive of health outcomes [7,13–16]. Such ageing markers are now derived from several parts of the body and from blood tests, and are increasingly used to interpret health status in clinical association studies. Their overall associations with health outcomes have been reported widely, but their prediction bias, and the mechanism behind it, has rarely been studied.

Ageing leads to widespread changes in tissue structure, metabolism, and physiological function. Retinal manifestations include reduced thickness of the retinal nerve fibre layer and decreased calibre of arterioles [17,18], while changes in the brain include cortical thinning, ventricular enlargement, and white matter damage [19,20]. Cardiovascular and respiratory changes include aortic stiffening and calcification, alongside reduced lung elasticity [21,22]. Body composition also shifts, with declining muscle mass and increasing fat accumulation observed around abdominal organs [23,24]. Alongside these structural changes, blood biomarkers reflect increased inflammation, impaired glucose regulation, and declining renal filtration [25–27]. These manifestations provide the basis for biological ageing markers, often termed "ageing clocks", in which a model learns age-related patterns in organ imaging, physiological measurements, and blood biomarkers to estimate biological age at the systemic or organ level [28,29]. Differences between model-estimated age and chronological age, commonly termed "age gaps", provide candidate markers of individual variation in ageing and have been associated with subsequent disease and mortality [15,30–34] (Fig. 1a). Recent extensions to multiple organs highlight growing interest in characterising the heterogeneity of ageing and identifying imaging and molecular signatures associated with declining organ health [16,28,29,35–37].

However, many ageing markers carry a prediction bias. The estimates of a regression-based ageing model are pulled towards the mean of its training distribution, a phenomenon known as regression to the mean (RTM) [36,38–43]. RTM can cause a model to overestimate the age of younger individuals and underestimate the age of older individuals, so that estimated ages are more concentrated around the mean age, as illustrated in Fig. 1b. Because RTM shifts estimated ages towards the mean, the resulting age gap may partly reflect the estimation shift rather than health. RTM has been recognised in previous studies, and several corrections have been proposed to address it [39–42,44,45]. However, whether this shift varies across unhealthy and healthy individuals, which we term differential RTM, has not been explored (Fig. 1b). Differential RTM shifts the age gap difference between healthy and unhealthy groups, so the association between the age gap and health status changes with age. This age dependence is particularly important given that previous studies commonly report an association for an entire cohort and therefore obscure these differences and fail to capture accurate associations for fine-grained subgroups.

In this study, we aimed to investigate the prediction bias in biological ageing markers and examined its implications for interpretation and clinical application. We analysed five ageing markers, Retinal Age from colour fundus photographs, Chest Age from chest radiographs, Abdominal Age from abdominal computed tomography, Brain Age from brain magnetic resonance imaging, and PhenoAge from a blood test, in five cohorts, ranging from 808 to 41,512 participants with 1,454 to 65,360 records. Consistent with previous studies [28,31,32,46–49], a larger age gap was associated with poorer health status in each cohort. However, we observed differential RTM in every ageing marker that estimates chronological age: the estimated ages were pulled towards the training-cohort mean more strongly in unhealthy than in healthy participants (Fig. 1b). This prediction bias made the association between the age gap and health status inconsistent across age subgroups in many of the ageing markers, where the positive association was weakened, and even reversed to negative, in the oldest subgroups (Fig. 1c, d). We further showed that this prediction bias could not be removed by calibrating the age gap or by rebalancing the age distribution of the training cohort. In addition, none of the five ageing markers could accurately reflect the health status of an individual. These findings highlight a prediction bias in ageing markers, and suggest that their association with health status should be reported within age subgroups. More broadly, ageing markers require more accurate interpretation and further development before they can be used to inform health at the individual level.

## Results

### Ageing markers and study cohorts

We studied five ageing markers, four derived from images of the eye, the chest, the abdomen and the brain, and one from a blood test (Fig. 1a). Retinal Age was estimated from colour fundus photograph (CFP) images (AlzEye [50]), Chest Age from chest radiograph images (ChestX-ray14 [51]), Abdominal Age from abdominal computed tomography (CT) volumes (Merlin [52]) and Brain Age from brain magnetic resonance imaging (MRI) volumes (OASIS-3 [53]), each by a deep learning model fitted to estimate chronological age. For each of these four ageing markers, the age gap was defined as estimated age minus chronological age [32,46–48]. PhenoAge [49] was computed from chronological age and nine blood biomarkers in UK Biobank [54]. Its age gap, PhenoAgeAccel, was defined as PhenoAge minus the mean PhenoAge of same age peers. The models of the four organ-derived ageing markers were trained on healthy participants only and evaluated on a test subset containing both healthy and unhealthy participants. PhenoAge involves no training in this study. We used the published equation, whose coefficients were fitted to predict mortality in NHANES III [55], and evaluated it on the whole cohort without refitting. Healthy and unhealthy status were defined per cohort from the disease records available in each dataset (see Methods).

The test subset comprised 65,360 images from 12,141 patients (mean age 68.38±13.34 years) for Retinal Age, 21,733 images from 4,623 patients (49.77±15.00 years) for Chest Age, 4,984 volumes (55.17±18.44 years) for Abdominal Age and 1,454 volumes from 808 patients (73.11±8.22 years) for Brain Age, and the first blood tests of 41,512 patients (57.72±7.99 years) for PhenoAge. The full age distributions, for each subset and for the healthy and unhealthy groups separately, are shown in Supplementary Fig. 7-16.

### Age estimation performance and disease associations

Age estimation performance varied across the four organ-derived ageing markers (Fig. 2a-d). Squared Pearson correlation between estimated and chronological age was 0.745 for Retinal Age, 0.796 for Chest Age, 0.918 for Abdominal Age and 0.643 for Brain Age, with corresponding MAE of 5.31, 5.15, 4.04 and 3.81 years and RMSE of 6.80, 6.80, 5.32 and 4.92 years.

PhenoAge is not an estimate of chronological age and therefore has no corresponding error metrics. We assessed PhenoAge against all-cause mortality, the outcome it was fitted to predict (Fig. 2e). The Kaplan-Meier estimate of 10-year survival was 92.25% (95% CI: [91.48%, 92.96%]) in the lowest tertile of PhenoAgeAccel, 90.15% (95% CI: [89.28%, 90.95%]) in the middle tertile and 83.55% (95% CI: [82.56%, 84.48%]) in the highest, corresponding to 542, 702 and 1,310 deaths within 10 years among 13,838, 13,837 and 13,837 patients respectively. The absolute difference in 10-year survival between the extreme tertiles was 8.70 percentage points, and the three curves were separated ($p<0.001$, log-rank).

All five ageing markers reproduced the positive association between a larger age gap (or PhenoAgeAccel for PhenoAge) and unhealthy status when estimated on the whole cohort (Fig. 2f). After adjustment for chronological age, each 1-year increase in the age gap was associated with higher odds of unhealthy status, expressed as odds ratios (OR), for Retinal Age (OR = 1.049, 95% CI: [1.032, 1.067]), Chest Age (OR = 1.011, 95% CI: [1.003, 1.019]), Abdominal Age (OR = 1.032, 95% CI: [1.019, 1.045]), Brain Age (OR = 1.238, 95% CI: [1.182, 1.297]) and PhenoAge (OR = 1.076, 95% CI: [1.072, 1.080]); $p<0.05$ for all associations. Associations with different disease categories are shown in Fig. 2g-k. Because the disease definitions available in each cohort are different, these disease categories are reported per ageing marker and are not compared across them.

**Differential RTM can lead to inconsistent associations**

All four organ-derived ageing markers showed differential RTM, a prediction bias in which the age gap decreases faster with chronological age in the unhealthy than in the healthy group (Fig. 3a-d). The unhealthy group had the steeper negative slope in every ageing marker (Fig. 3e-h): the healthy and unhealthy slopes were −0.184 (95% CI: [−0.217, −0.151]) and −0.314 (95% CI: [−0.326, −0.303]) years per year for Retinal Age, −0.202 (95% CI: [−0.223, −0.180]) and −0.257 (95% CI: [−0.283, −0.231]) for Chest Age, −0.073 (95% CI: [−0.085, −0.060]) and −0.115 (95% CI: [−0.125, −0.105]) for Abdominal Age, and −0.372 (95% CI: [−0.416, −0.328]) and −0.527 (95% CI: [−0.583, −0.471]) for Brain Age. The slope difference between the unhealthy and healthy groups was −0.130 (95% CI: [−0.165, −0.096]) for Retinal Age, −0.056 (95% CI: [−0.073, −0.039]) for Chest Age, −0.042 (95% CI: [−0.058, −0.026]) for Abdominal Age and −0.155 (95% CI: [−0.222, −0.087]) for Brain Age, and was significant in all four ageing markers (all $p<0.001$). Differential RTM was also present within each Retinal Age disease category and grew with the number of disease categories (Supplementary Fig. 17 and 18; Supplementary Tab. 10).

We then examined the association between the age gap and health status within age subgroups (Fig. 3i-l). We calculated the OR in each age subgroup, and used $p_{trend}$ to assess whether the OR increases or decreases with chronological age and $p_{heterogeneity}$ to assess whether the subgroup ORs are all equal (see Methods). For Retinal Age, Chest Age and Abdominal Age, the OR decreased significantly with age (all $p_{trend}<0.001$) and was inconsistent across age subgroups (all $p_{heterogeneity}<0.05$). For Retinal Age, the association was positive below age 70 and not significant from age 70 onwards. The OR decreased from 1.114 (95% CI: [1.085, 1.145], $p<0.001$) below age 50 to 0.955 (95% CI: [0.890, 1.025], $p=0.200$) at age 80 and above. For Chest Age, the association was positive below age 40, not significant from age 40 to 70, and negative at age 70 and above. The OR decreased from 1.051 (95% CI: [1.024, 1.079], $p<0.001$) below age 30 to 0.975 (95% CI: [0.953, 0.998], $p=0.035$) at age 70 and above. For Abdominal Age, the association was positive below age 50 and in the 60-70 subgroup, and not significant in the other age subgroups. The OR decreased from 1.080 (95% CI: [1.048, 1.112], $p<0.001$) below age 40 to 1.005 (95% CI: [0.963, 1.049], $p=0.804$) at age 80 and above. For Brain Age, the association was positive in every age subgroup (all $p<0.001$), and there was no significant trend ($p_{trend}=0.504$) or heterogeneity ($p_{heterogeneity}=0.439$). The OR was 1.190 (95% CI: [1.076, 1.316]) below age 65 and 1.181 (95% CI: [1.078, 1.294]) at age 80 and above. These results indicate that differential RTM can lead to inconsistent associations across age subgroups, but does not necessarily do so.

### Age gap variance and group difference jointly determine association strength

The OR is determined by two components: the difference in the mean age gap between the unhealthy and healthy groups, and the age gap variance within the two groups (Fig. 4). The OR increases with the difference and decreases with the variance (under a normal approximation with equal variance in the two groups; see Methods). In all four ageing markers, the OR implied by the difference and the variance closely matched the observed OR in every age subgroup (Fig. 4i-l). In Retinal Age, Chest Age and Abdominal Age, the difference decreased with age and became negative in one or more of the older age subgroups (Fig. 4a-c), while the variance showed no consistent trend (Retinal Age and Chest Age) or increased with age (Abdominal Age) (Fig. 4e-g). The OR therefore decreased with age in these three ageing markers, and the association was no longer significantly positive in the older age subgroups (e.g., age 80 and above in Retinal Age and Abdominal Age, and age 70 and above in Chest Age; Fig. 4i-k). In Brain Age, the difference and the variance decreased from the youngest to the oldest age subgroup in almost the same proportion (Fig. 4d, h), so the OR did not decrease with age and the association remained significantly positive in every age subgroup (Fig. 4l).

### Differential RTM persists after calibration and balanced training

We examined whether differential RTM could be removed by calibrating the age gap to chronological age, using Retinal Age as the example (Fig. 5a). In Fig. 5b, c, we applied linear and quadratic age calibrators fitted on the healthy patients of the downstream training subset. Both calibrations largely flattened the trend of the age gap with chronological age in the healthy group: the healthy slope was 0.017 (95% CI: [−0.015, 0.050]) after linear calibration and 0.018 (95% CI: [−0.015, 0.051]) after quadratic calibration. However, a significant negative slope persisted in the unhealthy group under both calibrations: −0.113 (95% CI: [−0.124, −0.101]) after linear calibration and −0.101 (95% CI: [−0.113, −0.090]) after quadratic calibration. Regardless of the calibration method, the slope difference between the two groups remained significant ($p<0.001$). These results indicate that age calibration can only remove the average trend of the age gap with chronological age in the healthy group, but cannot remove differential RTM between the healthy and unhealthy groups.

We examined whether differential RTM could be removed by balancing the age distribution of the training data. In Fig. 5d, e, we constructed a training subset with an even age distribution through inverse frequency-weighted sampling, and used it to retrain the Retinal Age model. In Fig. 5f, we evaluated the retrained model on the same test subset. Differential RTM remained: the healthy and unhealthy slopes were −0.190 (95% CI: [−0.217, −0.162]) and −0.344 (95% CI: [−0.355, −0.333]), respectively. The slope difference of −0.154 (95% CI: [−0.184, −0.124]) remained significant ($p<0.001$), larger in magnitude than the −0.130 obtained without balancing. The same holds for Chest Age, Abdominal Age and Brain Age: the slope difference remained significant after calibration and balanced training (all $p<0.001$; Supplementary Fig. 19-21 and Supplementary Tab. 7). Differential RTM therefore persisted after balancing the training age distribution.

### Organ-derived ageing markers poorly reflect individual health

We then examined whether the age gap can distinguish healthy from unhealthy individuals in the four organ-derived ageing markers (Fig. 6). Within each age subgroup, we paired every healthy image or volume with every unhealthy one and calculated the percentage of pairs in which the healthy one had the larger age gap (see Methods). In every ageing marker and every age subgroup, this was the case in at least a quarter of the pairs (25% to 59%; Fig. 6a-d). In Retinal Age, the percentage increased from 35% below age 50 to 59% at age 80 and above, so that in the oldest subgroup, the healthy one had the larger age gap in more than half of the pairs. The percentage increased from 42% to 55% in Chest Age and from 45% to 53% in Abdominal Age, and ranged from 25% to 41% in Brain Age. Fig. 6e-g shows pairs of patients of the same chronological age in which the healthy patient had an age gap 8.8 to 16.4 years larger than the unhealthy patient. These results indicate that the organ-derived ageing markers poorly reflect the health status of an individual.

**PhenoAge shows consistent associations but poorly reflects individual health**

We applied the same analysis to PhenoAge. In Fig. 7a, PhenoAge underestimated 10-year all-cause mortality risk by a similar proportion in the healthy and unhealthy groups, with observed risk about 1.5 times the predicted risk in both. Unlike the organ-derived ageing markers, PhenoAge showed no differential RTM (Fig. 7b, c). PhenoAgeAccel changed little with chronological age in either group (healthy slope −0.032, 95% CI: [−0.039, −0.025]; unhealthy slope −0.015, 95% CI: [−0.027, −0.003]), and the slope difference was small and in the opposite direction (+0.017, 95% CI: [0.003, 0.031], p=0.018).

Neither the difference nor the variance of PhenoAgeAccel decreased with age (Fig. 7d, e), and the OR implied by them closely matched the observed OR in every age subgroup (Fig. 7f). The association remained significantly positive in every age subgroup (all $p<0.001$), with ORs between 1.069 (95% CI: [1.059, 1.079]) and 1.085 (95% CI: [1.075, 1.096]) and no significant trend with age ($p_{trend}=0.291$), although the subgroup ORs were not all equal ($p_{heterogeneity}=0.036$).

However, in every age subgroup, the healthy patient had the larger PhenoAgeAccel in 39% to 41% of healthy-unhealthy pairs (Fig. 7g). Fig. 7h shows pairs of patients of the same age in which the healthy patient had the larger PhenoAgeAccel, +8.2 versus −5.8 years below age 50 and +7.2 versus −6.8 years at age 65 and above. These results indicate that PhenoAge shows a consistent association across age subgroups, but poorly reflects the health status of an individual.

## Discussion

In this study, we investigated five ageing markers derived from the eye, the chest, the abdomen, the brain and blood tests. In those that estimate chronological age, we found differential RTM: the age gap regressed towards the mean more strongly in the unhealthy than in the healthy group. This prediction bias is a fundamental characteristic of these ageing markers. In most of them, it leads to an inconsistent association between the age gap and health status, which is positive at younger ages, weakens as chronological age increases, and can reverse in the oldest age subgroup. In addition, we found that all five ageing markers capture differences in health status at the population level, but cannot accurately reflect the health status of an individual. This study highlights that the association with health status should be reported within fine-grained age subgroups, and that ageing markers require more accurate interpretation and further development before they can be used to inform individual health.

In all five ageing markers, better health corresponded to a smaller age gap (or PhenoAgeAccel for PhenoAge; Fig. 2f-k). Our observations are consistent with previous studies, which link a larger age gap to adverse health outcomes. This has been reported both for ageing markers derived from organs [15,31,32,47,56] and for blood-derived measures [33,49,57]. These findings suggest that all five ageing markers capture disease-related changes. In the organ-derived ageing markers, these appear as morphological and structural changes such as vascular alterations [28,29,37]. In PhenoAge, which is computed from blood biomarkers, they appear as physiological dysregulation [49,57].

We observed differential RTM in the ageing markers that estimate chronological age. The age gap decreased as chronological age increased in both healthy and unhealthy groups, and more strongly in the unhealthy group (Fig. 3e-h). Disease-related lesions and pathological manifestations in unhealthy participants likely present more difficult cases to a model fitted on relatively healthy participants. Therefore, the estimated ages of unhealthy participants are pulled more strongly towards the mean age of the training data. Previous studies have observed RTM [39–42,44,45], but rarely differential RTM. Some studies found inconsistent associations between the age gap and health status across age subgroups, but explained this by the accumulation of ageing effects over time and did not examine the cause [34,56,58,59]. Others noticed that the age gap decreases faster in the unhealthy than in the healthy group, and attributed this to disease pathology, RTM, or the age distribution of the training data, but did not examine its effect on the association across age subgroups [60–62]. Our study found that differential RTM is a fundamental characteristic of ageing markers that estimate chronological age, and that this prediction bias can lead to inconsistent associations between the age gap and health status across age subgroups. Our findings therefore encourage studies that use an ageing marker to report the association within each age subgroup, rather than a single association for the whole cohort.

Including an interaction term between chronological age and the age gap may provide a practical way to account for differential RTM. It helps capture age-dependent shifts in the association, and gives the OR at any chronological age (Supplementary Fig. 22a). The same applies when an ageing marker is used for prediction. Previous studies [31,63] have typically added an ageing marker to chronological age as one additional input, and have reported that it predicts mortality and morbidity better than chronological age alone. However, in our disease classification for Retinal Age in AlzEye, this approach improved overall performance slightly but degraded it in the oldest age subgroup (Supplementary Fig. 22b-f). Adding the interaction term recovered most of this loss and gave the best overall performance (Supplementary Fig. 22b-i). Our findings therefore suggest that including this interaction term can improve both the interpretation and the application of ageing markers in the presence of this prediction bias.

Differential RTM does not necessarily produce an inconsistent association. It makes the age gap difference between the healthy and unhealthy groups decrease with chronological age, but this decrease alone does not lead to a weaker association. The OR is determined by two components: the difference in the age gap between the two groups, and the age gap variance within the two groups. When the difference decreases with chronological age while the variance stays the same or increases, the OR decreases, as in Retinal Age, Chest Age and Abdominal Age (Fig. 4a-c, e-g, i-k). When the difference and the variance both decrease with chronological age, the OR can stay unchanged, as in Brain Age (Fig. 4d, h, l). To understand why an association changes with age, the change in age gap variance therefore has to be examined together with the change in the difference. Reporting the association separately within age subgroups remains the most straightforward and reliable way to interpret an ageing marker.

Differential RTM was not removed by calibration, or by balancing the training age distribution. Previous studies correct RTM by fitting the correction on a healthy reference group and applying it to every participant [39,44,45]. Because the same correction is applied to both groups, the difference between them is left unchanged, and differential RTM is not removed. To test this, we applied a similar approach: the calibrators were fitted on healthy participants and applied to every participant, subtracting the same value from the healthy and unhealthy groups at any given chronological age. Differential RTM remained after calibration (Fig. 5b, c). Calibrating each group separately would remove differential RTM, because the age gap would then no longer decrease with chronological age in either group. However, on one hand, calibrating each group separately would also set the difference between the two groups to zero, and remove the health-related signal that the age gap is expected to carry; on the other hand, it would require the health status of each participant as an input, which is what the age gap is meant to indicate. Differential RTM is also not a consequence of the age distribution of the training data: it remained after the model was retrained on a balanced age distribution (Fig. 5d-f). Taken together, differential RTM is a fundamental characteristic of ageing markers that estimate chronological age, and not an artefact that a better correction method will remove.

PhenoAge showed no differential RTM, unlike the organ-derived ageing markers (Fig. 7b, c). PhenoAge does not estimate age from its inputs: chronological age enters its equation directly, with the same weight for every participant, and the nine blood biomarkers adjust it [49]. Its estimates are therefore not pulled towards the mean age of a training cohort. Without differential RTM, the difference in PhenoAgeAccel between the healthy and unhealthy groups did not decrease with age (Fig. 7d), and the association between PhenoAgeAccel and unhealthy status remained positive in every age subgroup, with no trend with age (Fig. 7f). This agrees with a previous report that PhenoAge predicted mortality within every age group, from young to older adults [57].

None of these ageing markers can currently be used alone to predict an individual's health status. For example, in the youngest subgroup of Retinal Age, the mean age gap of the healthy group was well below that of the unhealthy group, yet the healthy patient had the larger age gap in 35% of healthy-unhealthy pairs. In the oldest subgroup the proportion increased to 59% (Fig. 6a). The other three organ-derived ageing markers showed a similar pattern (Fig. 6b-d). This also extends to PhenoAge (Fig. 7g), whose association is consistent across age subgroups. Overall, in every ageing marker and every age subgroup, the healthy participant had the larger age gap in at least a quarter of healthy-unhealthy pairs, so the age gap alone is not enough to accurately reflect the health status of an individual. Other measurements, such as chronological age, sex and established risk factors, therefore need to be included alongside the ageing marker.

This study also has several limitations. First, the “healthy” participants included here cannot be considered completely healthy; rather, they were defined as relatively healthy based on the absence of recorded disease categories (see Methods). Definitions of health vary substantially across studies and often depend on available disease records and diagnostic criteria. Second, the five ageing markers were evaluated on various cohorts with different age ranges. The magnitudes of the associations are therefore not comparable between ageing markers. Third, we used logistic regression to analyse the cross-sectional association between the age gap and disease as a representative example of clinical analysis. Future work could incorporate additional modelling frameworks, such as Cox regression for longitudinal risk prediction, to provide a broader evaluation of clinical utility. Fourth, except for the colour fundus photographs, we did not apply additional image-quality filtering to the released imaging datasets. Residual acquisition artefacts or image degradation may therefore introduce noise into age estimates and, if unevenly distributed across age or health status, may introduce bias. Finally, each ageing marker was evaluated on a single cohort, and whether the patterns reported here reproduce in independent cohorts of the same modality was not addressed.

In conclusion, we found a prediction bias (i.e., differential RTM) in ageing markers. This prediction bias has rarely been studied, cannot be removed by existing corrections, and in many of these ageing markers makes the association with health status inconsistent across age subgroups. This study highlights this prediction bias and emphasises that the association should be reported within age subgroups to better interpret ageing markers. It also highlights the need to further develop ageing markers by incorporating a broader range of individual-level data to more accurately reflect each person’s health status.

## Methods

### Model description

For the ageing markers derived from the eye, the chest, the abdomen and the brain, age estimation was formulated as a regression problem on chronological age, as in previous studies [15,47,48]. All four were trained separately, one per marker.

Retinal Age and Chest Age take a single 2D image (Supplementary Fig. 1a). We employed DINOv3-Large (vit_large_patch16_dinov3.lvd1689m) [64,65] as the feature extractor, a standard Vision Transformer (ViT) [66] initialised with weights pretrained on the LVD-1689M dataset, operating on 224 × 224 inputs with a patch size of 16. A regression head consisting of a linear layer reducing the dimension to 32, a ReLU activation, a dropout layer with dropout probability 0.5, and a final linear transformation produced a single scalar output, the estimated age. Colour fundus photographs (CFPs) have three channels; chest radiographs have a single greyscale channel, which was copied into the three input channels.

Abdominal Age and Brain Age take a 3D volume rather than a single image, and use the same feature extractor and the same regression head (Supplementary Fig. 1b). The volume was first reduced to a set of 2D slices, a 2.5D scheme. Each slice was encoded independently by the same backbone weights, and the resulting per-slice feature vectors were averaged into one vector before the head, an aggregation stage that the 2D models do not have. For Brain Age, each T1-weighted volume was resampled and cropped to the brain, and 32 axial slices were sampled; each slice was scaled to [0, 1] and copied into the three input channels. For Abdominal Age, each CT volume was resampled to 1.5 mm isotropic spacing and cropped to the body, and 32 axial slices were sampled; the three input channels hold three intensity windows of the same slice (soft tissue, low density and high density). Slices that contained almost no imaged anatomy were skipped. Slices were normalised with the ImageNet mean and standard deviation.

PhenoAge [49] is calculated from chronological age and nine blood biomarkers with the published equation, using the coefficients from the 2019 correction notice [57]:

$$xb = -19.907 - 0.0336 \times \text{albumin} + 0.0095 \times \text{creatinine} + 0.1953 \times \text{glucose} + 0.0954 \times \ln(\text{CRP}) - 0.0120 \times \text{lymphocyte percentage} + 0.0268 \times \text{MCV} + 0.3306 \times \text{RDW} + 0.00188 \times \text{ALP} + 0.0554 \times \text{WBC} + 0.0804 \times \text{age}$$

$$M = 1 - \exp(-1.51714 \times \exp(xb) / 0.0076927)$$

$$\text{PhenoAge} = 141.50225 + \ln(-0.00553 \times \ln(1 - M)) / 0.090165$$

where albumin is in g/L, creatinine in μmol/L, glucose in mmol/L, C-reactive protein (CRP) in mg/dL, lymphocyte percentage in %, mean cell volume (MCV) in fL, red cell distribution width (RDW) in %, alkaline phosphatase (ALP) in U/L, white blood cell count (WBC) in $10^3$ cells/μL and age in years, and M is the predicted 10-year mortality risk. UK Biobank reports CRP in mg/L, so it was divided by 10. The UK Biobank fields, units and values of the nine biomarkers are given in Supplementary Tab. 16.

PhenoAge [49] is defined on the age scale but is fitted to predict mortality, not chronological age. Its coefficients were fitted on NHANES III [55] and were applied without refitting. Its age gap,

PhenoAgeAccel, is the residual of PhenoAge on chronological age, that is, the difference between a participant's PhenoAge and the mean PhenoAge of participants of the same chronological age.

### Model training

All backbone and head parameters were jointly fine-tuned. Optimisation used AdamW with weight decay 0.05 and layer-wise learning-rate decay of 0.65, with a base learning rate of $5 \times 10^{-3}$ decayed to a minimum of $1 \times 10^{-6}$. After each epoch, performance was evaluated on that cohort's age model validation subset, and the checkpoint with the lowest mean absolute error (MAE) was retained. The 2D and 2.5D models differ in schedule and augmentation. Retinal Age and Chest Age were trained with a batch size of 32 for 50 epochs, with 10 warmup epochs followed by cosine annealing, and used the augmentation recipe of RETFound [67]: random resized crop to 224 × 224, random horizontal flipping, AutoAugment, random erasing and ImageNet normalisation. Abdominal Age and Brain Age were trained with a batch size of 16 for up to 100 epochs, with 5 warmup epochs and early stopping when validation MAE failed to improve for 15 consecutive epochs, and used geometric augmentation only, since the photometric and erasing transforms are not appropriate for computed tomography or magnetic resonance imaging.

### Datasets

Five cohorts were used, one per marker.

AlzEye [50] supplied CFPs for Retinal Age. It is a retrospective cohort that links the ophthalmic imaging of 353,157 patients aged 40 years or older, who attended Moorfields Eye Hospital NHS Foundation Trust (London, UK) between January 2008 and April 2018, to their national hospital admission records. All imaging was acquired at Moorfields, predominantly using Topcon systems (≈99%, with the remainder using Carl Zeiss Meditec); the per-device breakdown is given in Supplementary Tab. 1. The extract used here contained 2,583,012 CFPs from 233,488 patients.

ChestX-ray14 [51] supplied frontal chest radiographs for Chest Age. It is a public hospital-based cohort of 112,120 frontal radiographs from 30,805 patients, collected at the NIH Clinical Center (Bethesda, Maryland, USA) between 1992 and 2015. Each radiograph carries labels for fourteen thoracic findings, and a radiograph with none of them is labelled No Finding.

Merlin [52] supplied abdominal computed tomography (CT) for Abdominal Age. It is a hospital-based cohort of consecutive abdominal and pelvis CT examinations performed at Stanford between December 2012 and October 2018, across inpatient, emergency, outpatient and observation settings, assembled through the Stanford Medicine Research Data Repository (STARR). The dataset used here contains 25,489 abdominal and pelvis CT volumes from 18,317 patients, each paired with its radiology report, from which thirty findings are labelled per volume. Each examination is represented by one series, the one with the most axial slices, and the dataset authors verified its contrast phase with a previously validated classifier.

OASIS-3 [53] supplied T1-weighted brain magnetic resonance imaging (MRI) for Brain Age. It is a longitudinal cohort of normal ageing and cognitive decline assembled through the Washington University Knight Alzheimer Disease Research Center (St Louis, Missouri, USA), with participants spanning approximately 42 to 95 years of age. The data were downloaded from the NITRC image repository in July 2026, when the OASIS-3 project listed 1,379 patients; the extract used here provided 2,633 T1-weighted MRI volumes from 1,338 patients. All volumes were acquired on Siemens research scanners, most at 3 T. MRI volumes are linked to Clinical Dementia Rating (CDR) assessments and diagnostic visits, allowing health status to be assigned at the time of imaging. The CDR is scored 0 for no impairment, 0.5 for very mild or questionable impairment, and 1, 2 or 3 for mild, moderate and severe dementia, respectively.

UK Biobank [54] supplied the nine blood biomarkers and mortality follow-up used for PhenoAge. It is a large prospective population cohort of 503,317 participants recruited at ages 40-69 years between 2006 and 2010 across 22 assessment centres in England, Scotland and Wales. The extract used here covers the patients who underwent the ophthalmic assessment introduced in 2009, and contained 68,672 blood tests from 66,560 patients, taken at the baseline visit in 2009-2010 and at repeat assessments in 2012-2013. The biomarkers were measured centrally by UK Biobank under standard laboratory quality control.

**Dataset preprocessing**

Image quality was graded only for AlzEye, whose archive of routine fundus photographs includes many ungradable images. ChestX-ray14, Merlin and OASIS-3 were used as released, and no image or volume was excluded for image quality: each radiograph and CT volume was acquired in clinical practice and reported by a radiologist, Merlin retains the most complete series of each examination, and OASIS-3 volumes were acquired under research protocols (see Datasets).

For AlzEye [50], CFP images were graded as Good, Usable or Reject using AutoMorph [68]. Of 2,583,012 images from 233,488 patients, 1,059,876 images classified as Reject were excluded, leaving 1,523,136 Good or Usable images from 195,642 patients. A further 1,195,372 images without a confirmed disease record were excluded, leaving 327,764 images from 60,865 patients (Supplementary Fig. 2).

For ChestX-ray14 [51], radiograph images were restricted to patients aged 18 to 100. Of the 112,120 released images, 5,241 were excluded for age under 18 and 16 for age above 100. This left 106,863 images from 29,363 patients (Supplementary Fig. 3).

For Merlin [52], CT volumes were restricted to patients aged 18 to 100 with a confirmed disease record. Of the 25,489 volumes in the release, 77 had no age, 366 were from patients outside 18 to 100 years and 214 had no confirmed disease record. This left 24,832 volumes, of which 7,736 carried no positive finding and 17,096 carried at least one (Supplementary Fig. 4).

For OASIS-3 [53], of the 2,633 volumes from 1,338 patients available, 61 without a confirmed disease record were excluded, leaving 2,572 volumes from 1,322 patients (Supplementary Fig. 5).

For UK Biobank [54], we retained the first blood test of each patient and required complete measurements for all nine biomarkers used to calculate PhenoAge, a confirmed disease record and follow-up after the blood test. Of 68,672 blood tests from 66,560 patients, 2,112 repeat blood tests were excluded, followed by 9,772 patients with missing biomarkers, 11,127 without a confirmed disease record and 4,149 without follow-up, leaving 41,512 patients for analysis (Supplementary Fig. 6).

### Disease labelling

Each image, volume or blood test was labelled unhealthy if at least one disease record was positive and healthy if all were negative; those without a confirmed disease record were excluded.

For AlzEye [50], disease labels were obtained through linked hospital records using International Classification of Diseases, 10th Revision (ICD-10) codes [69], organised into four categories: cardiovascular, chronic, ocular and neurodegenerative diseases. Cardiovascular diseases included atrial fibrillation, heart failure, myocardial infarction, intracerebral haemorrhage, ischaemic stroke, aortic stenosis, aortic regurgitation, mitral stenosis and mitral regurgitation. Chronic diseases included diabetes mellitus and hypertension. Ocular diseases included glaucoma, diabetic retinopathy and age-related macular degeneration. Neurodegenerative diseases included dementia and Parkinson's disease. The full ICD-10 definitions are in Supplementary Tab. 15. Neurodegenerative disease was used in defining the overall unhealthy group but not analysed as a subgroup, because of limited sample size. Cardiovascular, chronic and ocular disease are also the categories used in the per-disease analysis for this marker.

For ChestX-ray14 [51], the released per-image finding labels were used. The fourteen findings are infiltration, effusion, atelectasis, nodule, mass, pneumothorax, consolidation, pleural thickening, cardiomegaly, emphysema, oedema, fibrosis, pneumonia and hernia. A radiograph was labelled unhealthy if any of them was positive, and healthy if none was, which corresponds to the released No Finding label. Each finding is also a category in the per-disease analysis, in which a radiograph is a case for that category when that finding is positive. Only the four with the most cases are reported.

For Merlin [52], per-study findings were used, and a volume was labelled unhealthy if any was positive. The findings are submucosal oedema, renal hypodensities, aortic valve calcification, coronary calcification, thrombosis, metastatic disease, pancreatic atrophy, renal cyst, osteopenia, surgically absent gallbladder, atelectasis, abdominal aortic aneurysm, anasarca, hiatal hernia, lymphadenopathy, prostatomegaly, biliary ductal dilation, cardiomegaly, splenomegaly, hepatomegaly, atherosclerosis, ascites, pleural effusion, hepatic steatosis, appendicitis, gallstones, hydronephrosis, bowel obstruction, free air and fracture. As for ChestX-ray14 [51], these findings are also the categories used in the per-disease analysis, taken individually, and only the four with the most cases are reported.

For OASIS-3 [53], health status was taken from the clinical visit nearest in time to each MRI volume, within 365 days. A volume was labelled healthy when that visit recorded CDR 0, normal cognition, and no flag for mild cognitive impairment or dementia. It was labelled unhealthy when the visit recorded CDR above 0, or a flag for mild cognitive impairment, dementia, or impairment not amounting to mild cognitive impairment. This cohort has no disease categories, so the per-disease analysis uses CDR severity strata instead: very mild or questionable impairment (CDR 0.5), and mild or greater dementia (CDR of 1 or above).

For UK Biobank [54], disease information was obtained using ICD-10 [69] and self-report codes (Data-Field 20002), covering atrial fibrillation, heart failure, myocardial infarction, intracerebral haemorrhage, ischaemic stroke, aortic stenosis, aortic regurgitation, mitral stenosis, mitral regurgitation, diabetes mellitus, hypertension, dementia, Parkinson's disease, glaucoma, cataract and diabetic retinopathy. The full definitions are in Supplementary Tab. 15. The categories used in the per-disease analysis are cardiovascular and chronic. Cardiovascular comprises atrial fibrillation, heart failure, myocardial infarction, intracerebral haemorrhage, ischaemic stroke, aortic stenosis, aortic regurgitation, mitral stenosis and mitral regurgitation. Chronic comprises diabetes mellitus and hypertension.

### Dataset splits

AlzEye [50], ChestX-ray14 [51], Merlin [52] and OASIS-3 [53] each have three subsets: an age model training subset and an age model validation subset, both containing healthy patients only, and a test subset on which every reported association is estimated. AlzEye and ChestX-ray14 have an additional downstream training subset, used to fit the age calibrator; in AlzEye, this subset was also used to fit the disease classification models. AlzEye, ChestX-ray14 and OASIS-3 were divided at the patient level, so that no patient appears in more than one subset. Merlin did not release any patient identifiers, so its age model training, age model validation and test subsets were taken from the official recommended training, validation and test partitions, respectively, keeping all CT volumes of a patient in the same partition [52]. These partitions could not be divided further without the risk of a patient appearing in two subsets, so Merlin has no downstream training subset. OASIS-3 is small, so no downstream training subset was split off. For both Merlin and OASIS-3, the age model validation subset was used to fit the age calibrator. UK Biobank [54] was not divided, because PhenoAge involves no training; all 41,512 patients with the nine biomarkers, a confirmed disease record and follow-up form its test subset (Supplementary Fig. 6).

AlzEye [50]. The age model training subset holds 9,844 healthy patients (53,680 images) and the validation subset 2,462 healthy patients (14,038 images). The downstream training subset holds 36,418 patients (194,686 images), of which 154,819 images were labelled unhealthy, and the test subset 12,141 patients (65,360 images), of which 51,603 images were labelled unhealthy; both were drawn from the remaining healthy patients together with all unhealthy patients. The split is shown in Supplementary Fig. 2.

ChestX-ray14 [51]. The age model training subset holds 8,698 healthy patients (12,622 images) and the validation subset 2,175 (3,174 images). The downstream training subset holds 13,867 patients (69,334 images) and the test subset 4,623 patients (21,733 images), of which 11,828 images were labelled unhealthy. The split is shown in Supplementary Fig. 3.

Merlin [52]. The age model training subset holds 4,633 healthy volumes and the validation subset 1,597. The 13,618 unhealthy volumes of the official training and validation partitions were not used. The test subset holds 4,984 volumes, of which 3,478 were labelled unhealthy. The split is shown in Supplementary Fig. 4.

OASIS-3 [53]. The age model training subset holds 411 healthy patients (889 volumes) and the validation subset 103 healthy patients (229 volumes). The test subset holds 808 patients (1,454 volumes), of which 521 volumes were labelled unhealthy. The split is shown in Supplementary Fig. 5.

### Evaluation of age estimation

We evaluated the age estimation models for the eye, the chest, the abdomen and the brain with the mean absolute error (MAE), the root mean squared error (RMSE) and the squared Pearson correlation coefficient (Pearson's $r^2$) between estimated and chronological age. PhenoAge [49] is not an estimate of chronological age, so these metrics do not apply to it. It was instead evaluated against the outcome it was fitted to predict (all-cause mortality). Patients were divided into tertiles of PhenoAgeAccel, and survival up to 10 years was compared between the tertiles with Kaplan-Meier curves and the log-rank test. In the healthy and unhealthy groups separately, patients were ordered by predicted 10-year mortality risk and grouped into overlapping windows, and the observed risk in each window, the Kaplan-Meier estimate at 10 years, was compared with the predicted risk.

**Age subgroups**

Age subgroups were defined before the association analysis, in 10-year bands for AlzEye, ChestX-ray14 and Merlin and in 5-year bands for OASIS-3 and UK Biobank, whose age ranges are narrower, with open-ended youngest and oldest bands. An end band was merged into its neighbour when it held too few observations to estimate its odds ratio precisely. As a result, the youngest and oldest subgroups do not cover the same ages across ageing markers.

**Association analyses**

*Overall association.* The unit of analysis is the image, volume or blood test; a patient can contribute more than one, which is accounted for in the confidence intervals and p-values. For each ageing marker we performed a logistic regression with the disease label (healthy vs. unhealthy) as the dependent variable and the age gap as the primary independent variable of interest, adjusted for chronological age as a covariate. The strength of the association was quantified using odds ratios (OR) per +1 year increase in the age gap with corresponding 95% CIs, and statistical significance was determined by the p-value calculated using the Wald test.

*Association within age subgroups.* Within each age subgroup we fitted a logistic regression on that subgroup's data, with chronological age as a covariate, giving an OR per +1 year increase in the age gap for that subgroup. Two tests are reported. $p_{trend}$ tests whether the log OR changes linearly with chronological age, using the Wald test of the interaction term (chronological age × age gap) in a logistic regression of the disease label on chronological age, the age gap and their interaction, fitted over all ages. $p_{heterogeneity}$ tests whether the ORs of all age subgroups are equal, using a joint Wald test. The regression used for $p_{trend}$ also gives the OR per +1 year increase in the age gap as a function of chronological age, which shows how the association varies with age.

*OR decomposition.* Within each age subgroup, the OR was decomposed into two quantities, $\Delta$ and $\sigma^2$. $\Delta$ is the difference in mean age gap between the unhealthy and healthy groups after adjusting for chronological age. It was estimated with a linear regression of the age gap on chronological age and group (0 for healthy, 1 for unhealthy), fitted to data in the subgroup, and $\Delta$ is the coefficient of group (i.e., the difference in age gap between the two groups at the same chronological age). $\sigma^2$ is the variance of the age gap at the same chronological age. For the healthy and the unhealthy group separately, we fitted a linear regression of the age gap on chronological age and calculated the residuals, that is, each age gap minus the age gap predicted by its group's regression line at that chronological age. $\sigma^2$ is the variance of the residuals of the two groups combined into one set. Residuals are used rather than the age gaps themselves because the age gap still changes with chronological age within an age subgroup, and this change would otherwise be counted in $\sigma^2$. If the age gap in each group is normally distributed with the same variance, the log OR per 1-year increase in the age gap equals $\Delta/\sigma^2$. The OR implied by $\Delta$ and $\sigma^2$ is therefore $\exp(\Delta/\sigma^2)$: a larger difference $\Delta$ moves the OR further from 1, and a larger variance $\sigma^2$ moves it closer to 1.

**Linear regression of the age gap on chronological age**

For each marker, we fitted a single ordinary least squares regression on the combined data of healthy and unhealthy groups,

$$Age\ Gap = \beta_0 + \beta_1 \cdot Age + \beta_2 \cdot Group + \beta_3 \cdot Age \times Group + \varepsilon,$$

with "Group" coded 0 for healthy and 1 for unhealthy, and Age mean-centred (original age minus mean age) so that $\beta_0$ and $\beta_2$ correspond to the mean age. The healthy ($\beta_1$) and unhealthy ($\beta_1 + \beta_3$) slopes were reported with 95% confidence intervals. The Wald p-value on the interaction coefficient $\beta_3$ was reported as the test for slope difference between healthy and unhealthy groups.

**Pairwise ordering of healthy and unhealthy patients**

To assess how well the age gap distinguishes the health status of individual patients, within each age subgroup we paired every healthy image, volume or blood test with every unhealthy one, and calculated the percentage of pairs in which the unhealthy one had the larger age gap (or PhenoAgeAccel for PhenoAge), and the percentage in which the healthy one did.

**Confidence intervals and statistical tests**

All 95% confidence intervals and p-values for slopes and ORs were calculated with a cluster-robust (CR1) covariance, with the patient as the cluster, because a patient can contribute more than one image or volume and these are not independent. The p-values are Wald tests based on the same covariance. Merlin does not release patient identifiers, so for Abdominal Age each volume was treated as its own cluster, which treats volumes from the same patient as independent. In UK Biobank each patient contributes one blood test.

For quantities that have no such covariance, 95% confidence intervals were obtained with a patient-level bootstrap. Patients (volumes for Merlin) were sampled with replacement, all images or volumes of each sampled patient were kept, and the quantity was recalculated on the resampled data; this was repeated 1,000 times, and the 2.5th and 97.5th percentiles of the recalculated values give the interval. The bootstrap was used for MAE, RMSE and $r^2$ in age estimation, for AUC, F1 score, sensitivity and specificity in disease classification, and for $\Delta$ and $\sigma^2$ in the OR decomposition. The p-values for the differences in AUC, F1 score, sensitivity and specificity between the two classification models were calculated from each difference divided by its bootstrap standard error.

**Age calibration of the age gap**

Two age calibrators were fitted on healthy patients outside the test subset (see Dataset splits) by regressing the age gap on chronological age: a linear calibrator

$$Calibrator_{Linear}(Age) = \alpha_l + \beta_l \cdot Age,$$

and a quadratic calibrator

$$Calibrator_{quadratic}(Age) = \alpha_q + \beta_1 \cdot Age + \beta_2 \cdot Age^2.$$

The calibrated age gap was then defined as

$$Age\ Gap_{cal} = Age\ Gap - Calibrator(Age).$$

Within-group slopes, the interaction p-value and the subgroup odds ratios were re-computed on the test subset with the calibrated age gap in place of the raw age gap.

**Balanced training**

To assess whether the differential RTM was driven by the non-uniform age distribution of the age model training subset, we retrained each age model from the same pretrained weights on a uniform-age training distribution. Each training sample was drawn with a weight inversely proportional to the number of training samples in its 5-year age bin, so that the sampled training data had a nearly uniform age distribution. All other training hyperparameters were unchanged. The balanced model was evaluated on the same test subset and within-group slopes and subgroup odds ratios were re-computed.

**Disease classification**

For disease classification, we considered three logistic regression models:

- $logit(disease) \sim Age$
- $logit(disease) \sim Age + Age\ Gap$
- $logit(disease) \sim Age + Age\ Gap + Age \times Age\ Gap$

All three models were fitted on the AlzEye downstream training subset, separately for overall unhealthy and for each of the three disease categories versus healthy. Discrimination on the AlzEye test subset was reported as the area under the receiver operating characteristic curve (AUC). For threshold-based metrics (sensitivity, specificity, and F1 score), the operating threshold was selected by the Youden criterion on the AlzEye downstream training subset.

**Ethics approval**

Ethical approval for this study was obtained from the London-Central Research Ethics Committee (18/LO/1163, approved 1 August 2018), Confidential Advisory Group for Section 251 support (18/CAG/0111, approved 13 September 2018). The National Health Service Health Research Authority gave final approval on 13 September 2018. Moorfields Eye Hospital (MEH) NHS Foundation Trust validated the de-identifications for MEH data. Only de-identified retrospective data were used for research. The requirement for informed consent using AlzEye data was waived following section 251 supported by the UK CAG.

ChestX-ray14, Merlin and OASIS-3 are publicly released, de-identified datasets and were used under their release terms. ChestX-ray14 is distributed by the NIH Clinical Center. For Merlin, the use of retrospective clinical data was approved by the Institutional Review Board at Stanford University with a waiver of informed consent, the data were de-identified before release, and the dataset is distributed by Stanford AIMI under a non-commercial research data use agreement. OASIS-3 participants were consented under procedures approved by the Institutional Review Board of Washington University School of Medicine, and the data were obtained under the OASIS data use agreement.

**Data availability**

The MEH data consists of routinely collected healthcare data. Owing to their sensitive nature, the dataset is subject to controlled access by means of a structured application process. The AlzEye dataset is subject to the contractual restrictions of the data sharing agreements between National Health Service England, Moorfields Eye Hospital and University College London due to data privacy and ethical requirements. Researchers interested in accessing MEH data should contact the chief investigator P.A.K. at p.keane@ucl.ac.uk and submit a formal application at https://www.insight.hdrhub.org/insight-data. The data Trust Advisory Board will then review all the requests and grants. A formal data transfer agreement will be required upon approval, and the cost will be generated. Generally, all these requests for access to the data will be responded to within 3 months.

This research has been conducted using the UK Biobank Resource under application number 36741. The UK Biobank data are available for approved researchers through the UK Biobank Access Management System (https://www.ukbiobank.ac.uk).

Publicly available datasets used in this study include ChestX-ray14, accessible at https://nihcc.app.box.com/v/ChestXray-NIHCC; the Merlin Abdominal CT Dataset, accessible at https://stanfordaimi.azurewebsites.net/datasets/60b9c7ff-877b-48ce-96c3-0194c8205c40 after completion of its data use agreement; and OASIS-3, accessible at https://www.oasis-brains.org after completion of the OASIS data use agreement, which requires the following acknowledgement. Data were provided in part by OASIS-3: Longitudinal Multimodal Neuroimaging: Principal Investigators: T. Benzinger, D. Marcus, J. Morris; NIH P30 AG066444, P50 AG00561, P30 NS09857781, P01 AG026276, P01 AG003991, R01 AG043434, UL1 TR000448, R01 EB009352. AV-45 doses were provided by Avid Radiopharmaceuticals, a wholly owned subsidiary of Eli Lilly.

### Code availability

The code for the development of this study is available at our GitHub repository at https://github.com/HORIZONHealthcare/AG-RTM. We implemented the network design, model training, and model evaluation using PyTorch (https://pytorch.org) v2.2.0. The DINOv3 model weights are available at https://github.com/facebookresearch/dinov3. Images were processed with the automated retinal image analysis tool AutoMorph v.1.0 (https://github.com/rmaphoh/AutoMorph). Primary Python (v3.10.18) packages used in this study include Seaborn v0.13.2, NumPy v1.26.4, Matplotlib v3.10.7, Pandas v2.3.2, SciPy v1.15.3, and Scikit-Image v0.25.2.

### Acknowledgement

Y.Z. is supported by Wellcome Trust Early-career Award (318987/Z/24/Z) and Moorfields Eye Charity Equipment Grant (EQR-25B-102). Z.Z. is supported by the National Health and Medical Research Council (NHMRC) Investigator Grant (Emerging Leadership 1, APP2010072), NHMRC Investigator Grant (Emerging Leadership 2, APP2041559) and Medical Research Future Fund (MRFF) Early to Mid-Career Researchers Grant (APP2027411). Y.C.T. is supported by the National Medical Research Council of Singapore (NMRC/MOH/ HCSAINV21nov-0001). C.Y.C. is supported by the Research Grants Council Hong Kong (General Research Fund, ref.14101324). D.C.A. is supported by the Engineering and Physical Sciences Research Council (EP/M020533/1, EP/R014019/1 and EP/V034537/1). P.A.K. is supported by a UK Research & Innovation Future Leaders Fellowship (MR/T019050/1), the Moorfields Eye Charity with The Rubin Foundation Charitable Trust (GR001753), and an Alcon Research Institute Senior Investigator Award. We acknowledge the computational resources supported by the UCL Computer Science Cluster and UCL Advanced Research Computing UAI platform. The research was supported by the National Institute for Health and Care Research (NIHR) Biomedical Research Centre based at Moorfields Eye Hospital NHS Foundation Trust and UCL Institute of Ophthalmology. The views expressed are those of the author(s) and not necessarily those of the NHS, the NIHR or the Department of Health and Social Care.

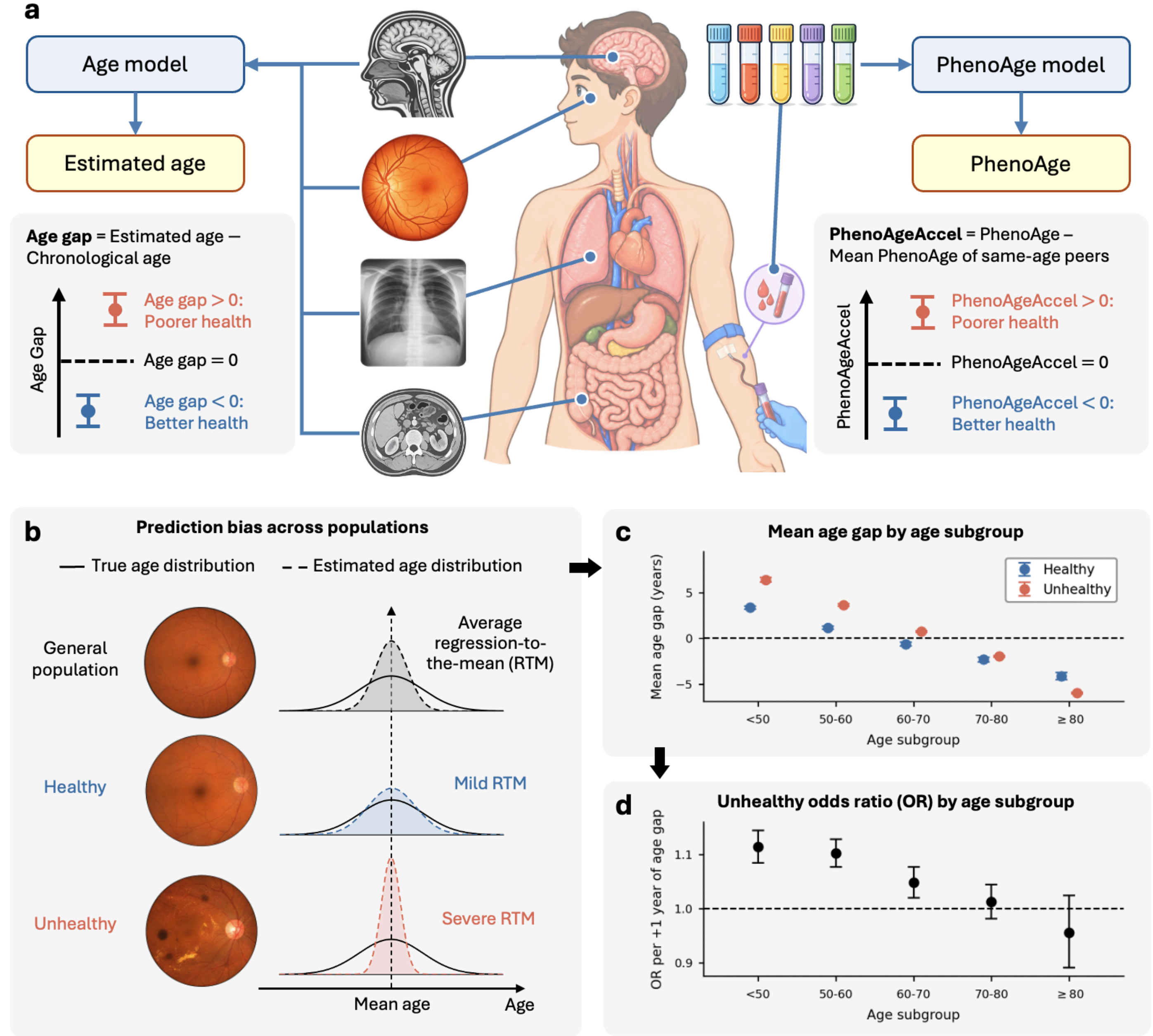

Fig. 1. **Study overview and prediction bias in ageing markers. a.** Ageing markers and their age gaps. Brain Age, Retinal Age, Chest Age and Abdominal Age are estimated from brain magnetic resonance imaging, colour fundus photographs, chest radiographs and abdominal computed tomography, respectively, and PhenoAge is computed from nine routine blood biomarkers and chronological age. **b.** Prediction bias across populations. The general population exhibits average regression to the mean (RTM), the healthy population shows mild RTM and the unhealthy population shows severe RTM, with estimated ages more strongly concentrated around the mean age. **c.** Mean age gap by age subgroup, for Retinal Age on the AlzEye [1] test subset. Differential RTM leads to distinct age-varying mean age gap patterns in the healthy and unhealthy groups: the age gap decreases with age in both groups but declines more steeply in the unhealthy group, reducing or reversing the difference at older ages. **d.** Odds ratio (OR) for unhealthy status by age subgroup, for the same data. These age-varying age gap patterns induce an age-dependent association between the age gap and unhealthy status: the OR per 1-year increase in the age gap attenuates with age and may reverse in the oldest age subgroup. Full results are reported in Supplementary Tab. 4.

**Ageing model performance and disease associations**

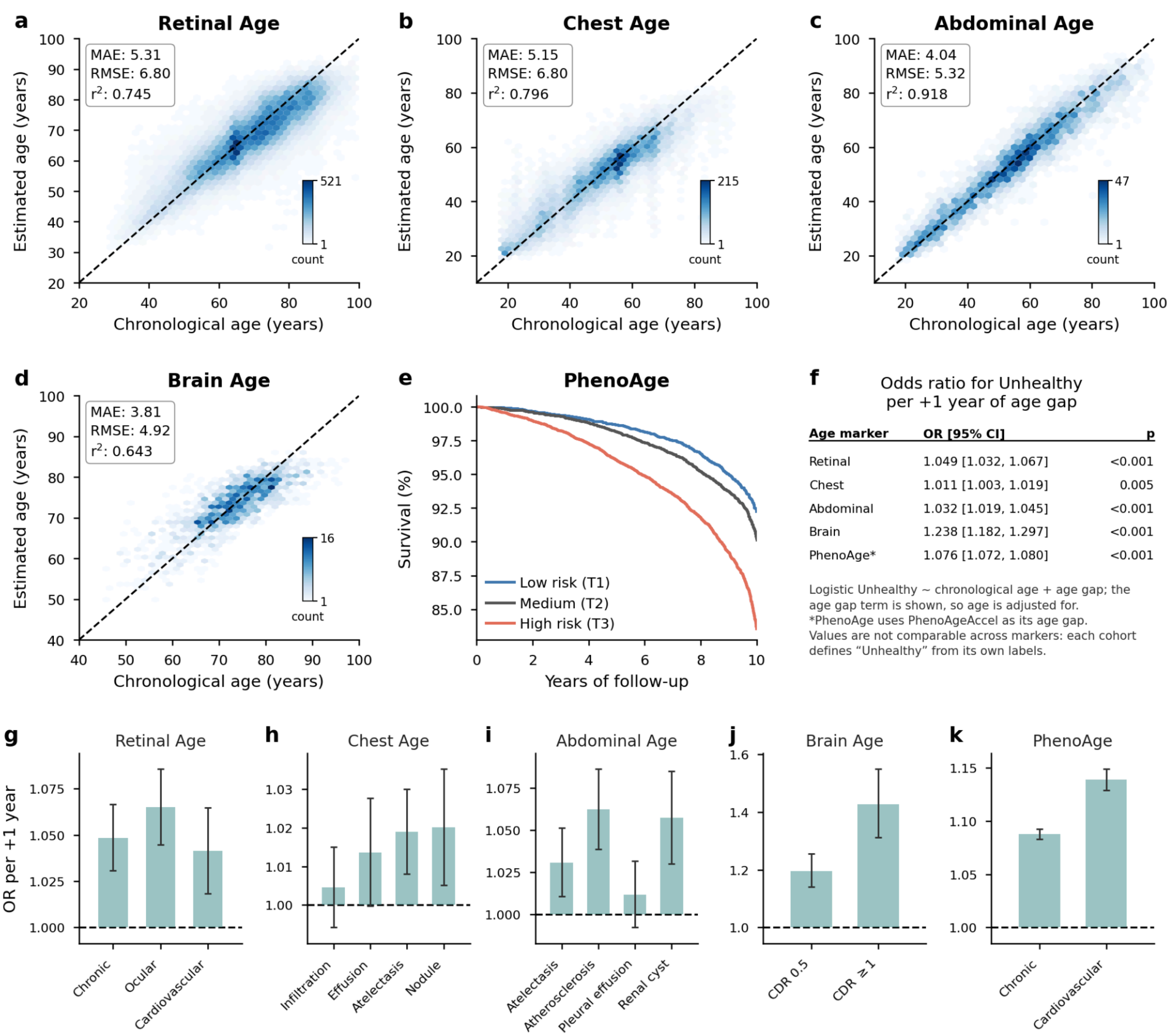

| Age marker | OR [95% CI] | p |
|---|---|---|
| Retinal | 1.049 [1.032, 1.067] | <0.001 |
| Chest | 1.011 [1.003, 1.019] | 0.005 |
| Abdominal | 1.032 [1.019, 1.045] | <0.001 |
| Brain | 1.238 [1.182, 1.297] | <0.001 |
| PhenoAge* | 1.076 [1.072, 1.080] | <0.001 |

Fig. 2. **Age estimation performance and disease associations for each ageing marker. a-d.** Hexbin density of estimated versus chronological age on the test subsets, for Retinal Age (AlzEye [1], **a**), Chest Age (ChestX-ray14 [2], **b**), Abdominal Age (Merlin [3], **c**) and Brain Age (OASIS-3 [4], **d**). The dashed line denotes the identity line (estimated = chronological). Annotated metrics: Mean Absolute Error (MAE), Root Mean Square Error (RMSE) and the squared Pearson correlation coefficient (Pearson's $r^2$) between estimated and chronological age. Lower MAE and RMSE indicate smaller estimation error. Higher Pearson's $r^2$ indicates a stronger linear correlation between estimated and chronological age. **e.** PhenoAge risk stratification. Kaplan-Meier survival curves up to 10 years after the blood test for UK Biobank [5] patients divided into tertiles of PhenoAgeAccel. **f.** Association between the age gap (or PhenoAgeAccel for PhenoAge) and unhealthy status. ORs per 1-year increase in the age gap, with 95% confidence intervals and Wald p-values, are from logistic regression with the age gap as the independent variable, unhealthy status as the outcome and chronological age as a covariate. **g-k.** ORs for the disease categories recorded in the cohort of each ageing marker, for Retinal Age (**g**), Chest Age (**h**), Abdominal Age (**i**), Brain Age (**j**) and PhenoAge (**k**), estimated as in **f**. Error bars denote 95% confidence intervals. Full results are reported in Supplementary Tab. 2 and 3.

**Differential RTM can lead to inconsistent associations**

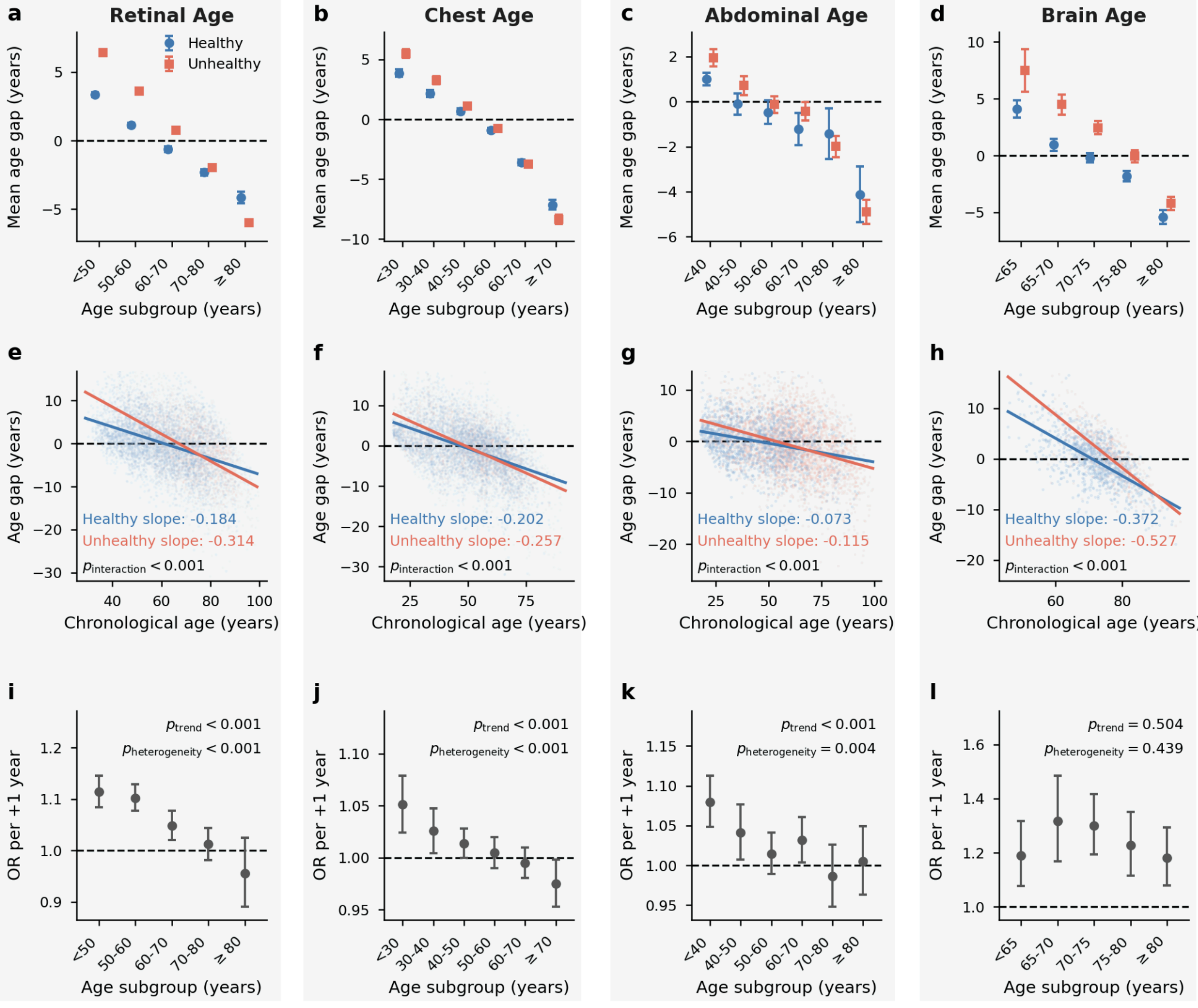

Fig. 3. **Differential RTM can lead to inconsistent associations in the four organ-derived ageing markers. a-d.** Mean age gap by age subgroup in the healthy and unhealthy groups, for Retinal Age (AlzEye [1], **a**), Chest Age (ChestX-ray14 [2], **b**), Abdominal Age (Merlin [3], **c**) and Brain Age (OASIS-3 [4], **d**) on the test subsets. **e-h.** Age gap versus chronological age with ordinary least squares fits for the healthy and unhealthy groups. The slope of each group (years of age gap per year of chronological age) and the p-value for the difference between the two slopes (the Age × Group interaction) are annotated. **i-l.** OR for unhealthy status per 1-year increase in the age gap within each age subgroup, from logistic regression with chronological age as a covariate. Annotated are $p_{trend}$, which tests for a linear trend of the log OR with chronological age, and $p_{heterogeneity}$, which tests whether the ORs of all age subgroups are equal. Error bars denote 95% confidence intervals; all tests and OR confidence intervals are cluster-robust at the patient level. Full results are reported in Supplementary Tab. 4 and 5.

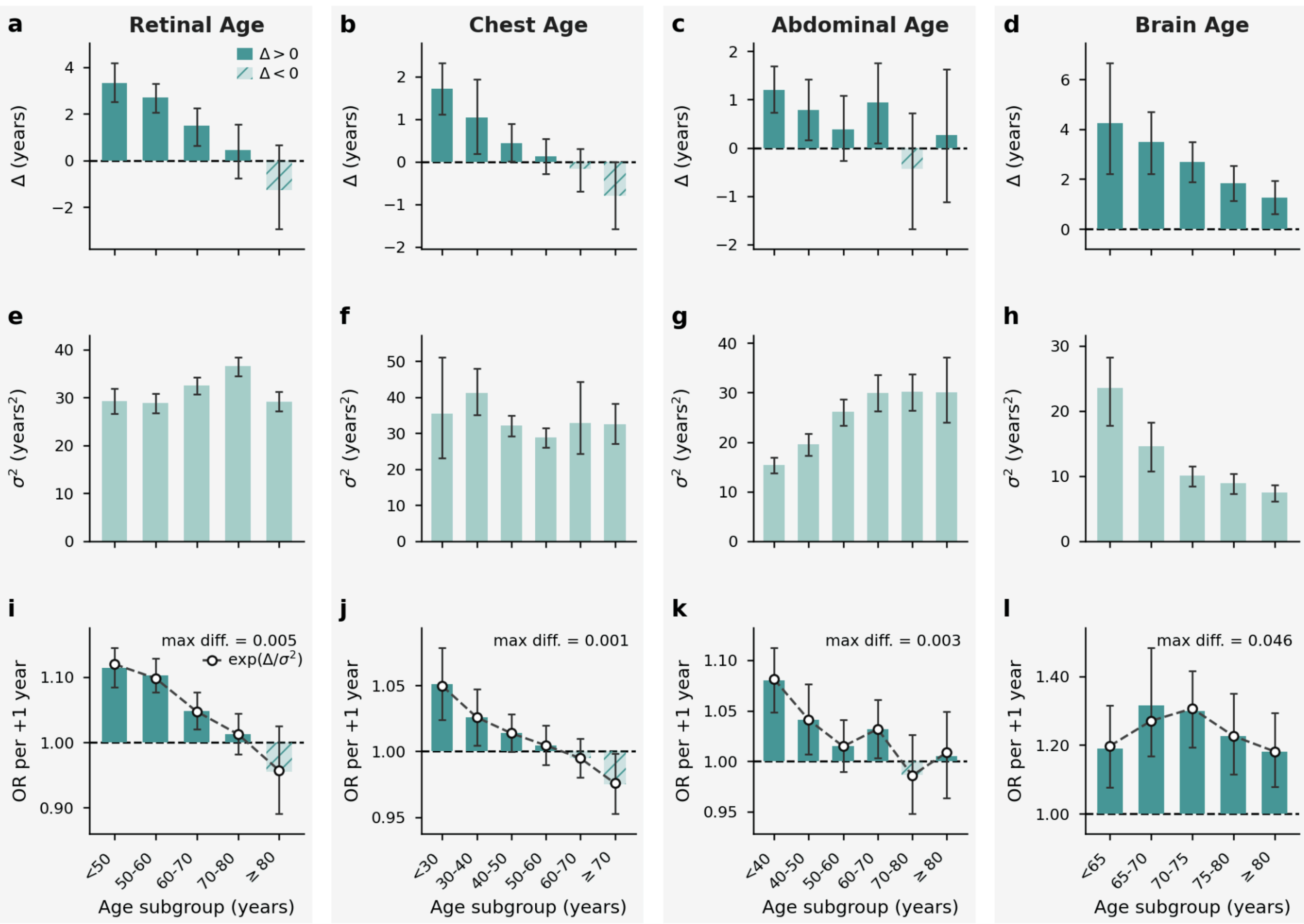

Fig. 4. **Age gap variance and group difference jointly determine association strength. a-d.** Difference in mean age gap between the unhealthy and healthy groups within each age subgroup (Δ), for Retinal Age (AlzEye [1], **a**), Chest Age (ChestX-ray14 [2], **b**), Abdominal Age (Merlin [3], **c**) and Brain Age (OASIS-3 [4], **d**) on the test subsets. **e-h.** Variance of the age gap at the same chronological age within each age subgroup, with the healthy and unhealthy groups combined ($\sigma^2$; see Methods). **i-l.** OR for unhealthy status per 1-year increase in the age gap within each age subgroup, estimated as in Fig. 3. Open circles on the dashed line show the OR implied by the difference and the variance, $\exp(\Delta/\sigma^2)$ (Methods), and max diff. is the largest absolute difference between the OR and $\exp(\Delta/\sigma^2)$ in the panel. Error bars denote 95% confidence intervals, from a patient-level bootstrap in **a-h** and cluster-robust at the patient level in **i-l**. Full results are reported in Supplementary Tab. 6.

Fig. 5. **Differential RTM persists after calibration and balanced training, shown for Retinal Age (AlzEye [1]). a-c.** Ordinary least squares fits of the age gap on chronological age for the healthy and unhealthy groups on the test subset, before calibration (**a**), after linear calibration (**b**) and after quadratic calibration (**c**). Shading marks the difference between the two groups. The slope of each group (years of age gap per year of chronological age) and the p-value for the difference between the two slopes are annotated. **d, e.** Chronological age distribution of the age model training data before (**d**) and after (**e**) resampling to a uniform age distribution. **f.** Age gap versus chronological age on the same test subset for the model retrained on the age-balanced training data, with ordinary least squares fits for the healthy and unhealthy groups, annotated as in Fig. 3. Full results are reported in Supplementary Tab. 7 and 8.

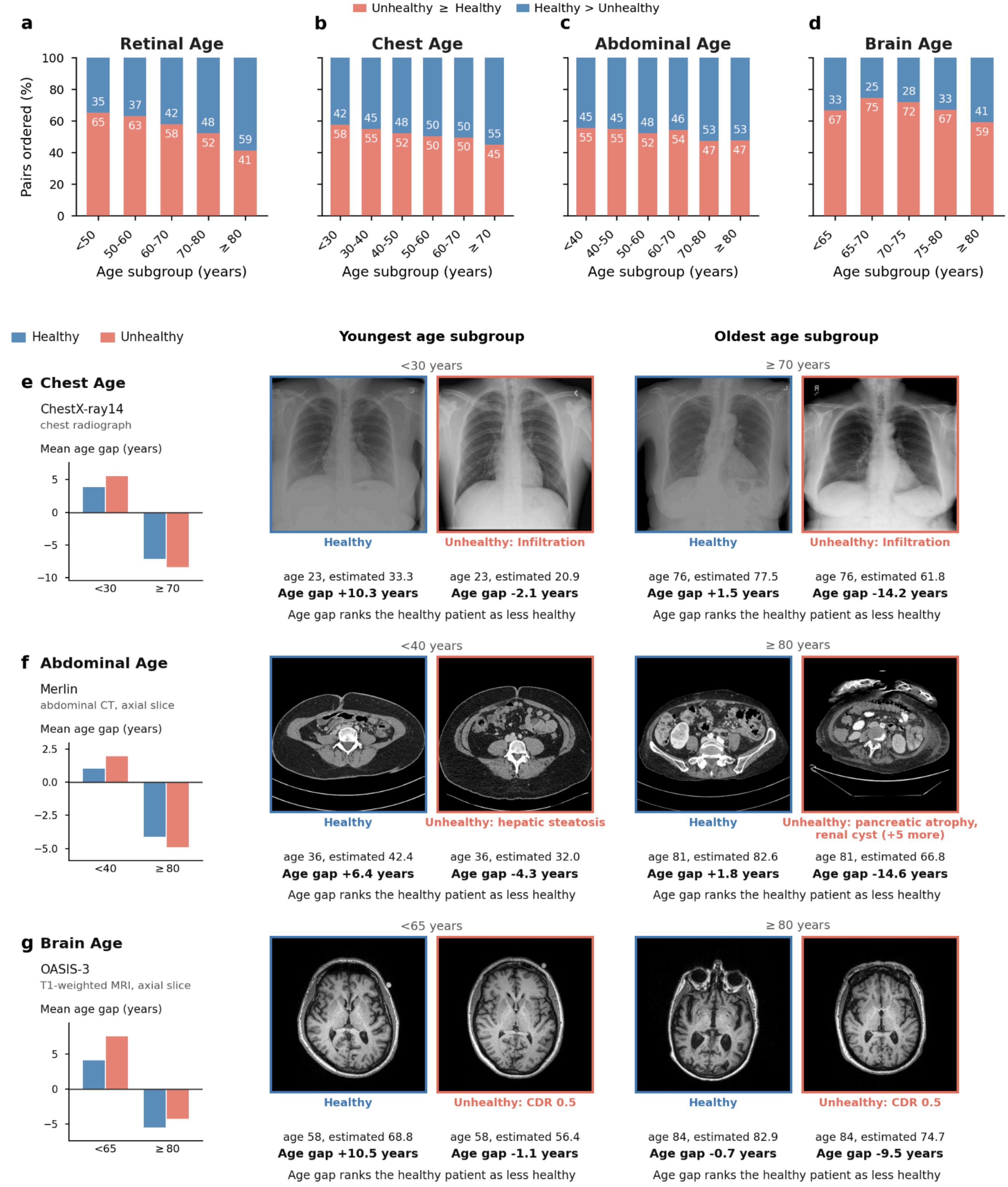

Fig. 6. **Organ-derived ageing markers poorly reflect individual health. a-d.** Percentage of healthy-unhealthy pairs within each age subgroup in which the unhealthy patient (lower block) or the healthy patient (upper block) has the larger age gap, for Retinal Age (AlzEye [1], **a**), Chest Age (ChestX-ray14 [2], **b**), Abdominal Age (Merlin [3], **c**) and Brain Age (OASIS-3 [4], **d**) on the test subsets. Age subgroups are as in Fig. 3. **e-g.** Example pairs of patients of the same chronological age, from the youngest and the oldest age subgroups, in which the healthy patient has the larger age gap, for Chest Age (**e**), Abdominal Age (**f**) and Brain Age (**g**). Under each image are the health status, with the recorded finding or Clinical Dementia Rating for the unhealthy patient, the chronological age, the estimated age and the age gap. The bars on the left give the mean age gap of the healthy and unhealthy groups in the two subgroups. CT and MRI volumes are shown as a middle axial slice. Retinal Age examples are not shown because AlzEye is a restricted-access clinical dataset. Full results are reported in Supplementary Tab. 14.

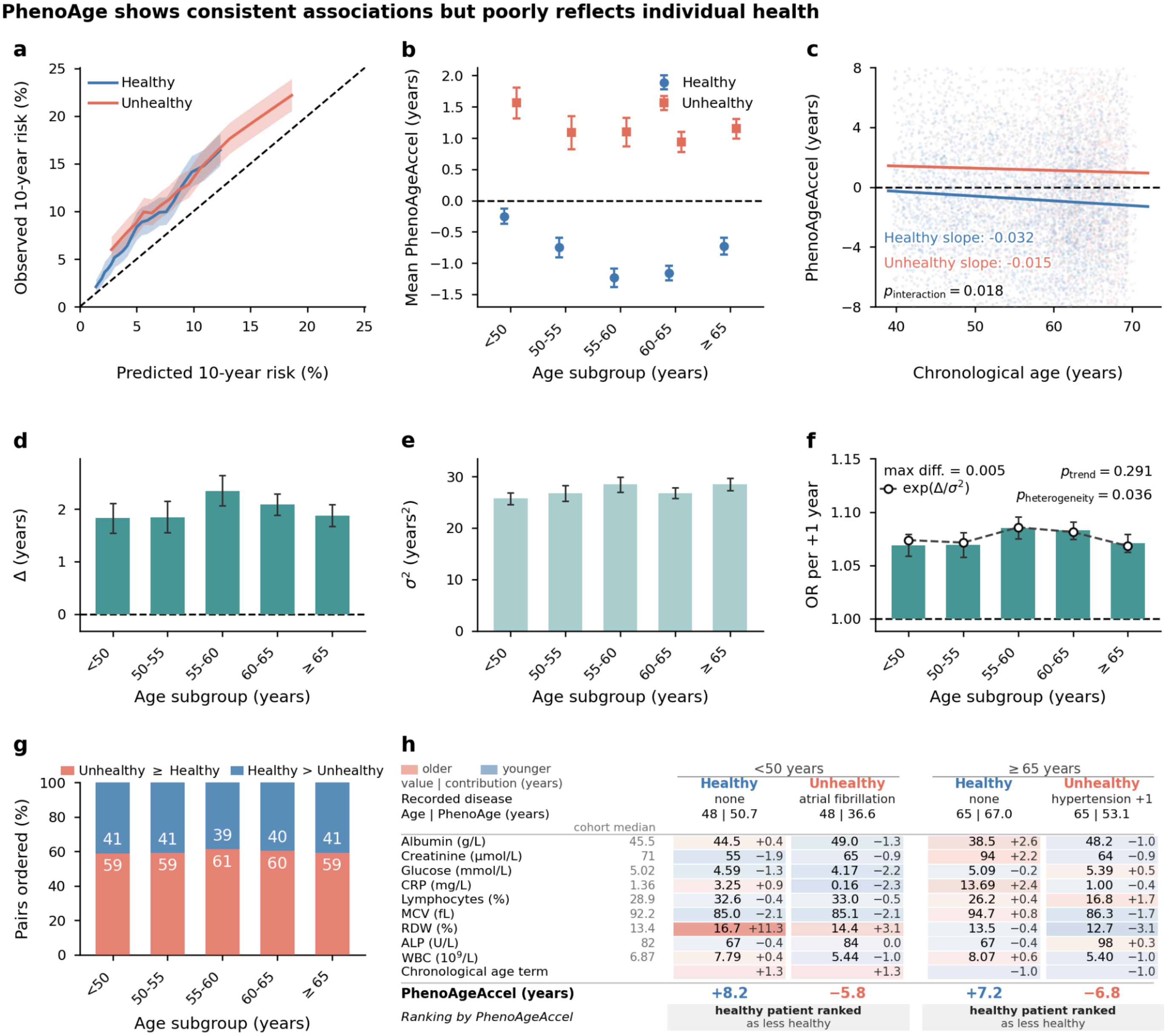

Fig. 7. **PhenoAge shows consistent associations but poorly reflects individual health in UK Biobank. a.** Observed versus predicted 10-year mortality risk in the healthy and unhealthy groups, with 95% confidence bands; observed risk is the Kaplan-Meier estimate at 10 years. The dashed line denotes observed = predicted. **b.** Mean PhenoAgeAccel by age subgroup in the healthy and unhealthy groups. **c.** PhenoAgeAccel versus chronological age with ordinary least squares fits for the healthy and unhealthy groups. The slope of each group (years of PhenoAgeAccel per year) and the p-value for the difference between the two slopes are annotated. **d-f.** Difference in mean PhenoAgeAccel between the unhealthy and healthy groups (**d**), variance of PhenoAgeAccel (**e**) and OR for unhealthy status per 1-year increase in PhenoAgeAccel (**f**) within each age subgroup, defined and estimated as in Fig. 4; $p_{trend}$ and $p_{heterogeneity}$ are as in Fig. 3. **g.** Percentage of healthy-unhealthy patient pairs within each age subgroup in which the unhealthy or the healthy patient has the larger PhenoAgeAccel. **h.** Example pairs of patients of the same chronological age, from the youngest and the oldest age subgroups, in which the healthy patient has the larger PhenoAgeAccel. Biomarker abbreviations are defined in Methods. Error bars denote 95% confidence intervals. Full results are reported in Supplementary Tab. 5 and 9.

## Supplementary Figures

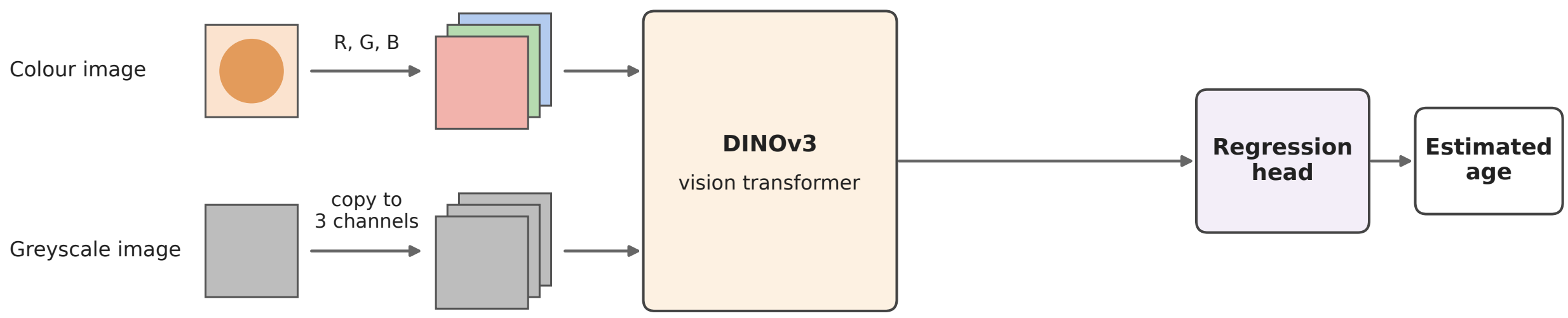

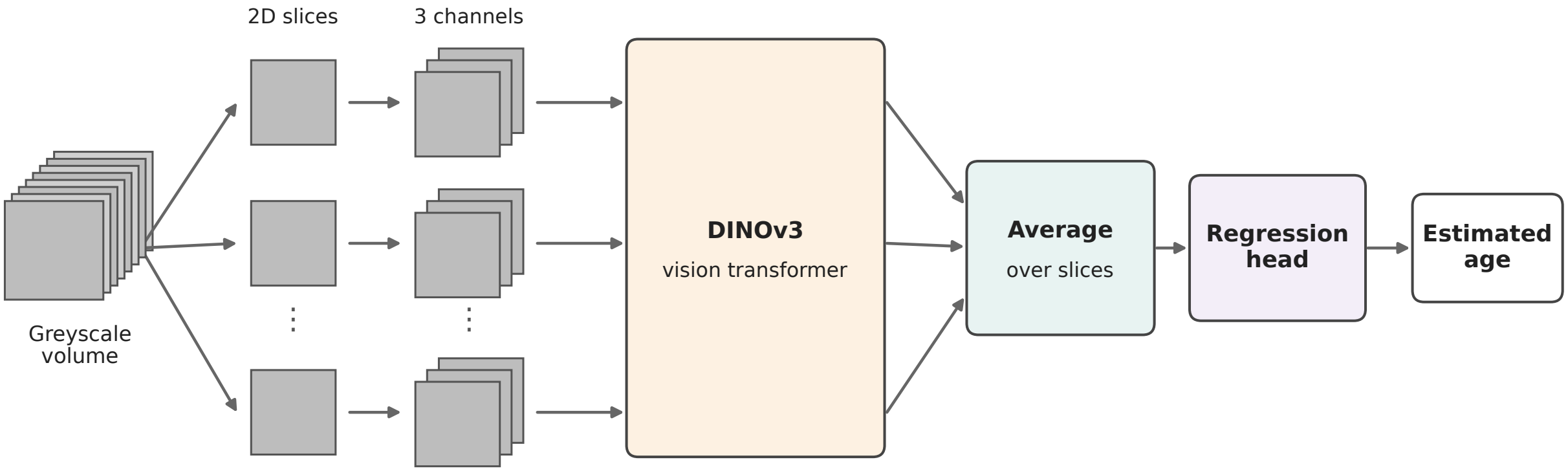

**Supplementary Fig. 1. Architecture of the organ-derived ageing models.** Each ageing model fine-tunes a DINOv3 vision transformer from its pretrained weights, followed by a regression head that outputs the estimated age. **a.** 2D images, for Retinal Age and Chest Age. A colour image enters with its three colour channels, and a greyscale image is copied into three identical channels. **b.** 3D volumes, for Abdominal Age and Brain Age. The greyscale volume is cut into 2D slices, each slice is turned into a three-channel image (copied for MRI, three intensity windows for CT) and encoded by the same DINOv3 model, and the slice features are averaged before the regression head. Preprocessing and training settings are given in Methods.

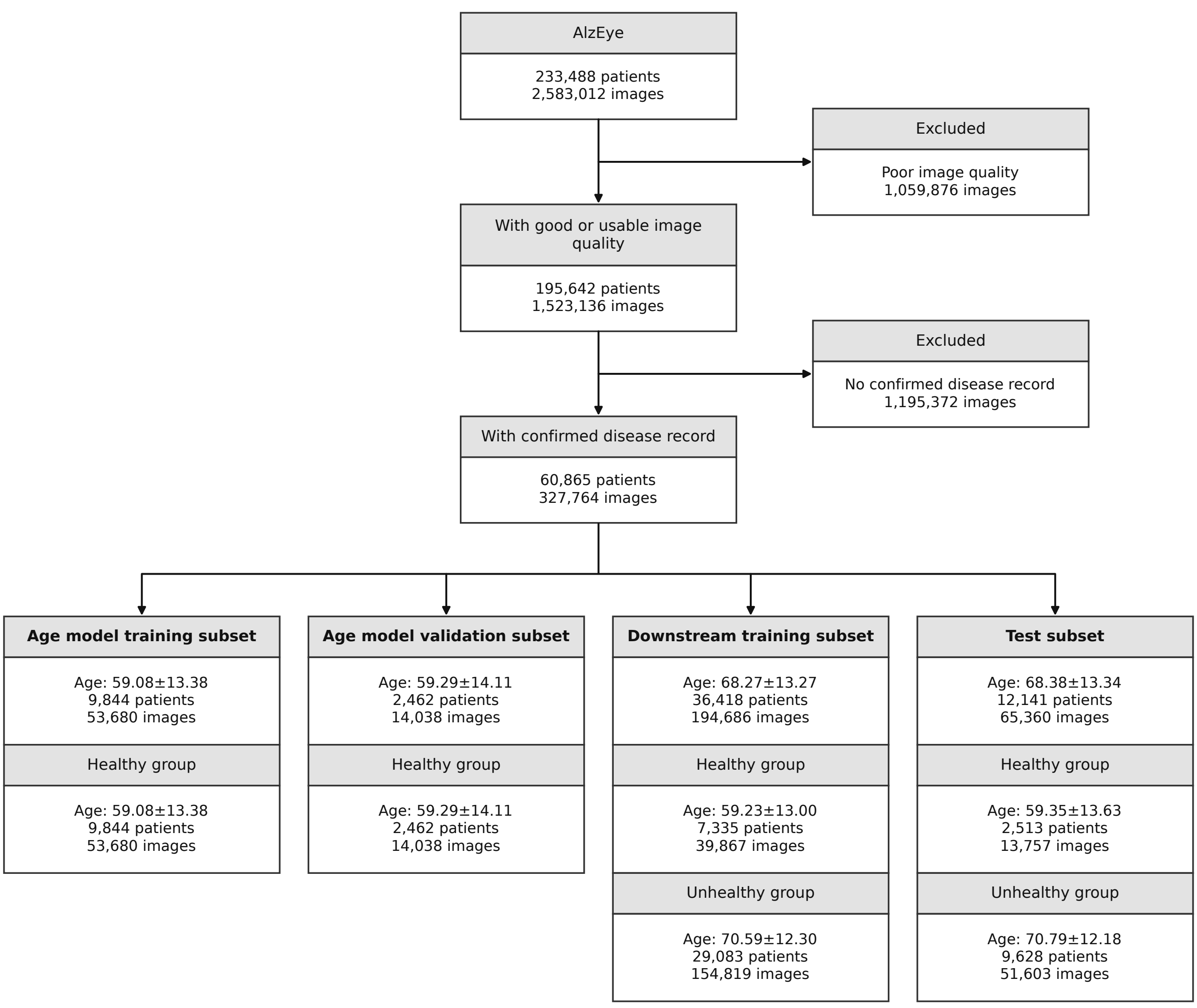

**Supplementary Fig. 2. Data curation and split for AlzEye.** Images were restricted to those of good or usable quality and then to those with a confirmed disease record, and divided into an age model training subset and an age model validation subset, both healthy only, a downstream training subset and a test subset. Age is mean±SD in years.

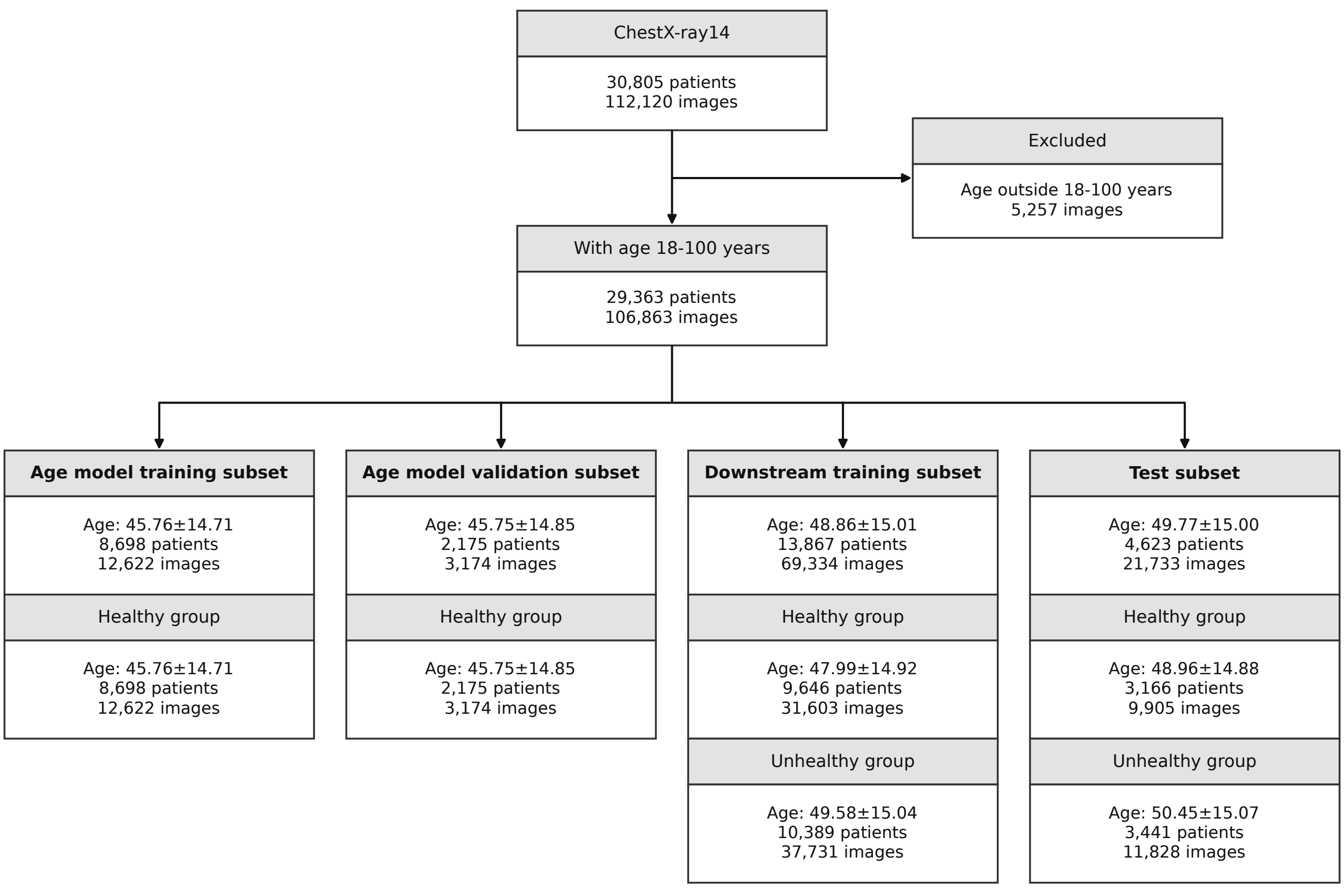

**Supplementary Fig. 3. Data curation and split for ChestX-ray14.** Images were restricted to ages 18 to 100 years and divided into the same four subsets as for AlzEye. A patient with both healthy and unhealthy images is counted in both groups, so the patients of the healthy and unhealthy groups can add up to more than the patients of the subset. Age is mean±SD in years.

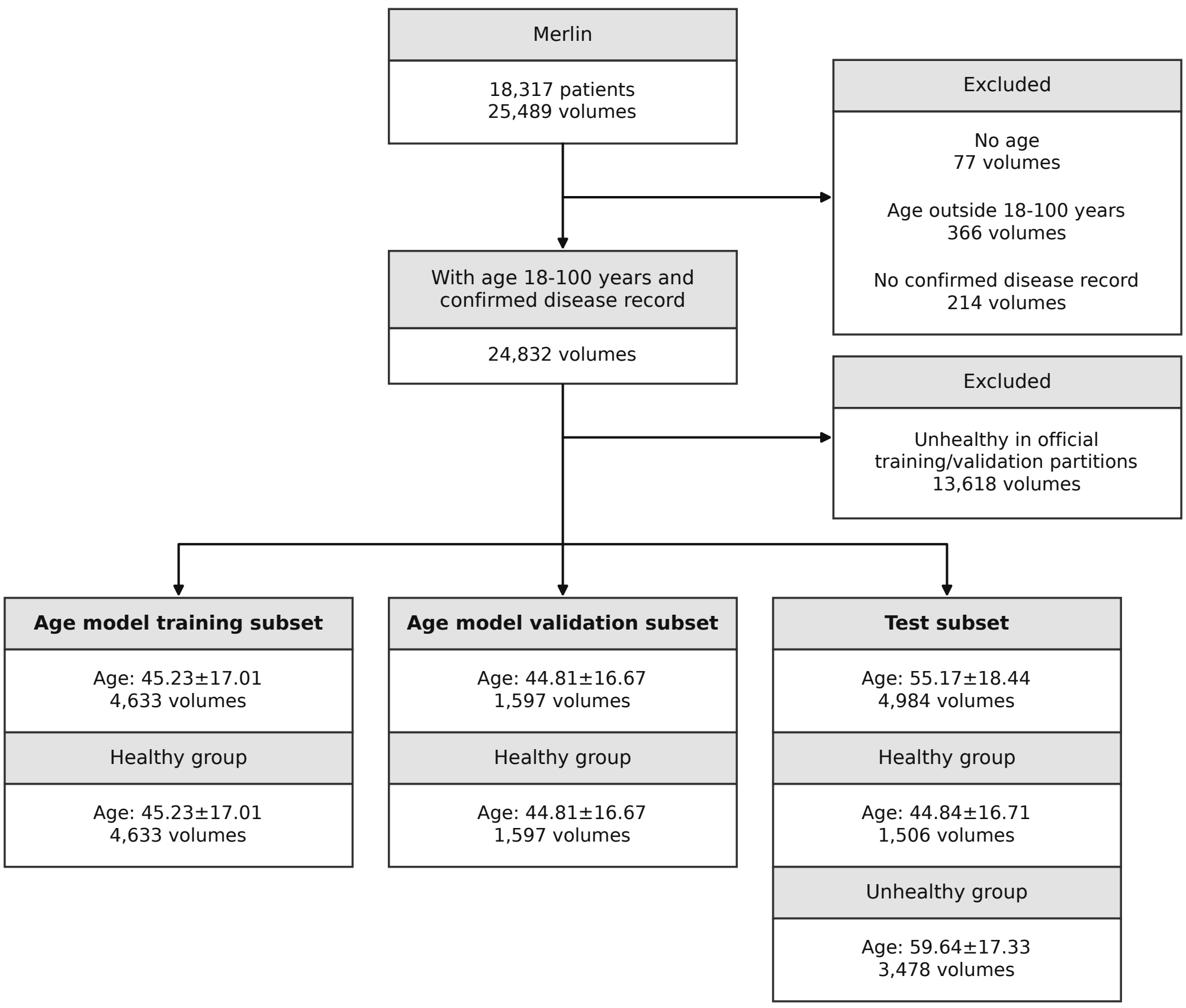

**Supplementary Fig. 4. Data curation and split for Merlin.** Volumes were restricted to those with an age of 18 to 100 years and a confirmed disease record. The subsets follow the official Merlin partitions: the age model training and validation subsets are the healthy volumes of the official training and validation partitions, and the test subset is the official test partition. Merlin releases no patient identifiers. Age is mean±SD in years.

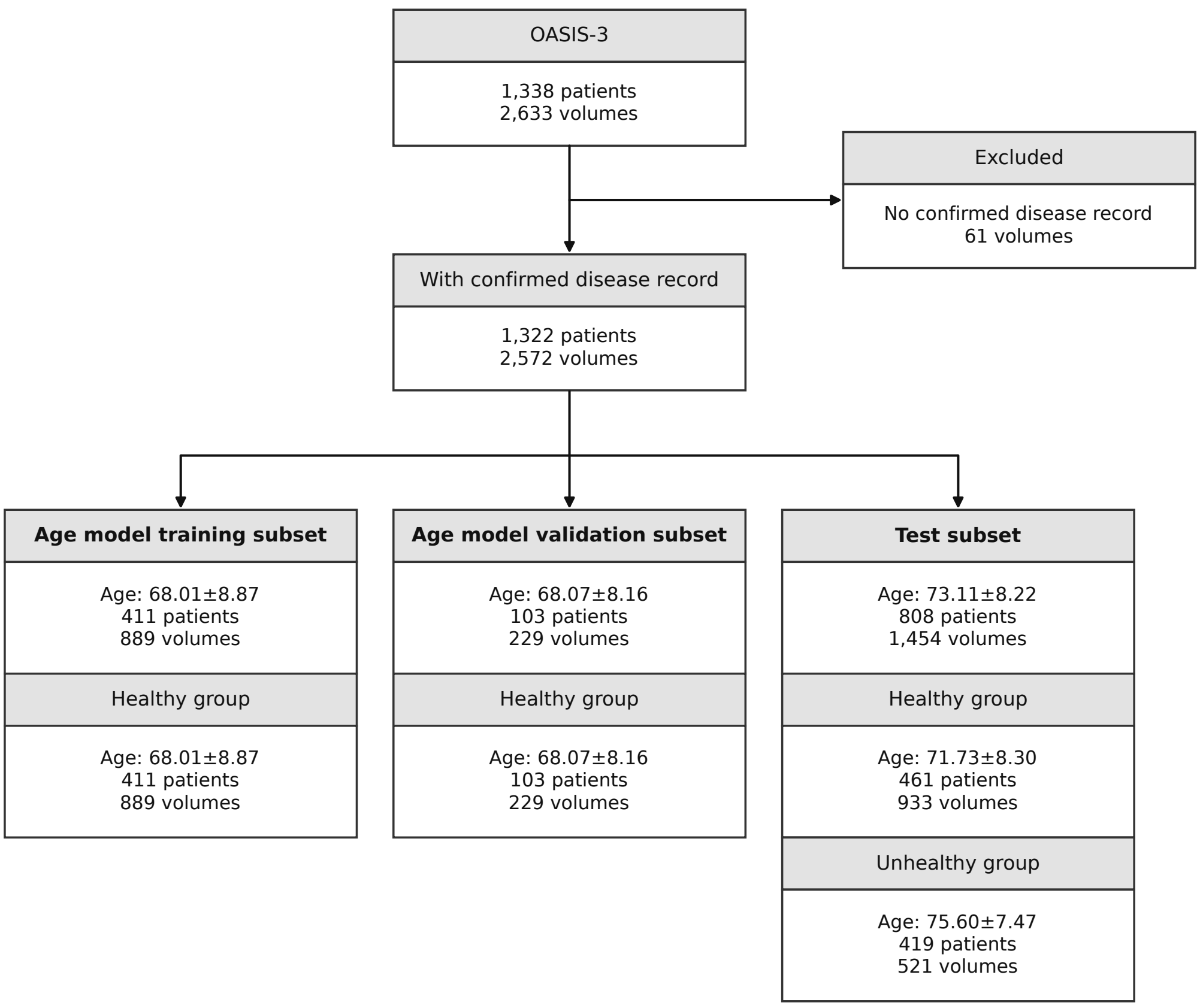

**Supplementary Fig. 5. Data curation and split for OASIS-3.** Volumes without a confirmed disease record were excluded, and the rest were divided into an age model training subset and an age model validation subset, both healthy only, and a test subset. Health status is assigned per volume, so a patient with both healthy and unhealthy volumes is counted in both groups of the test subset. Age is mean±SD in years.

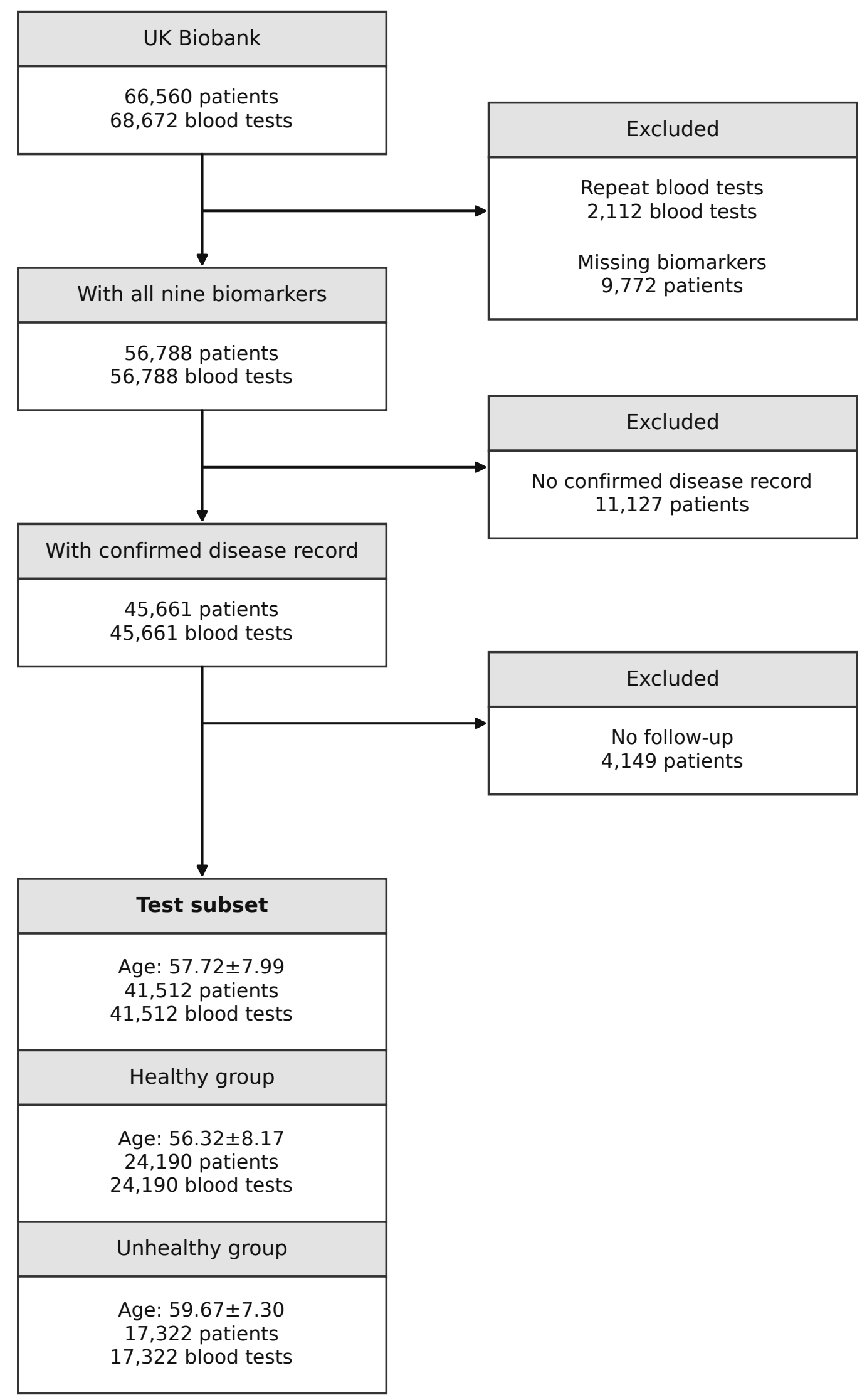

**Supplementary Fig. 6. Data curation for UK Biobank.** The first blood test of each patient was used. Patients without follow-up after the blood test were excluded because the mortality analyses (Fig. 2e and Fig. 7a) need follow-up time. UK Biobank is used for PhenoAge, which involves no training, so there are no training or validation subsets, and all patients left after curation form the test subset. Age is mean±SD in years.

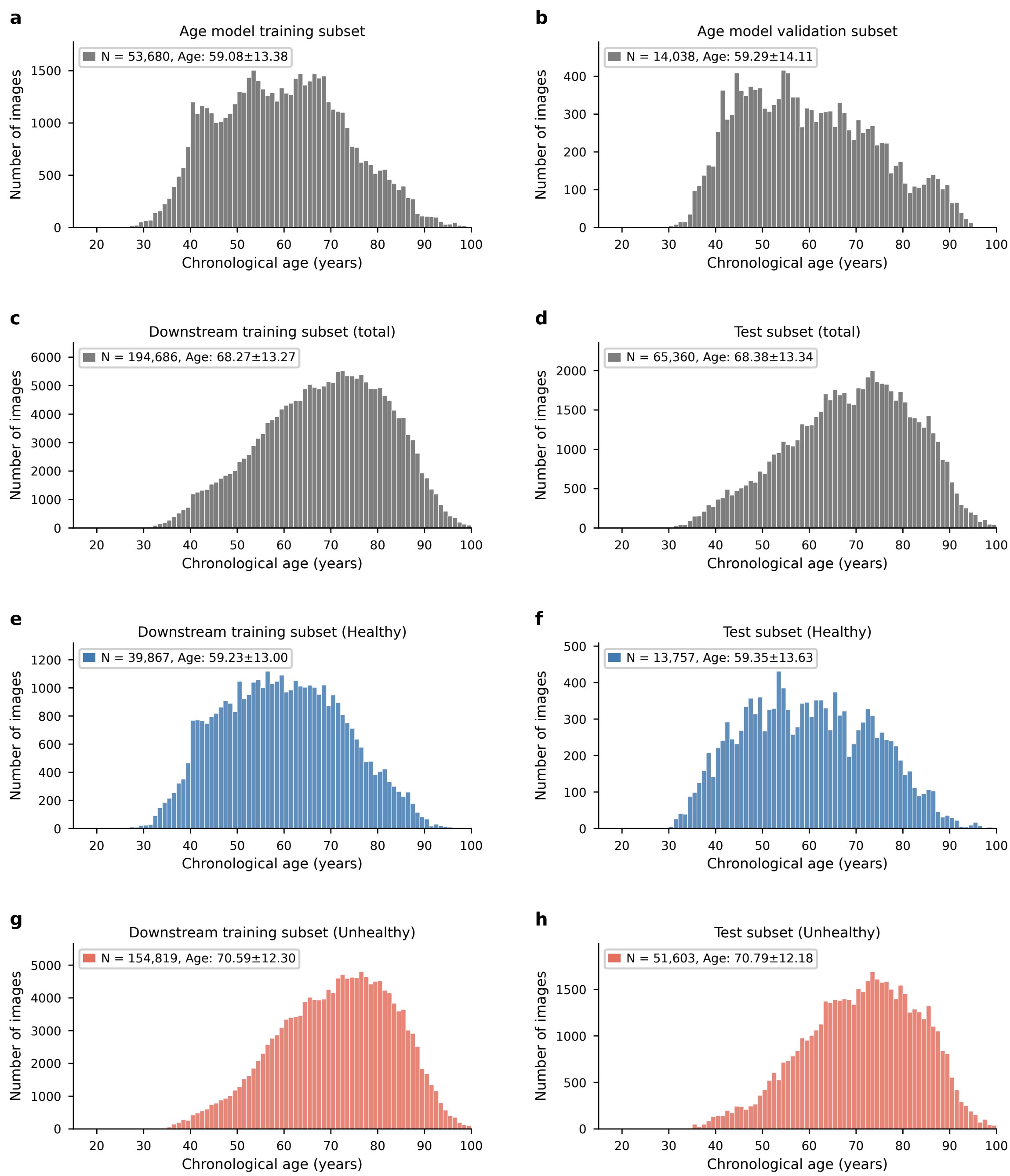

**Supplementary Fig. 7. Chronological age distributions for the AlzEye subsets.** Histograms in 1-year bins of age at imaging for the age model training subset (**a**) and validation subset (**b**), and for the downstream training subset and the test subset in total (**c, d**) and separately for the healthy (**e, f**) and unhealthy (**g, h**) groups. N and mean±SD in years are annotated.

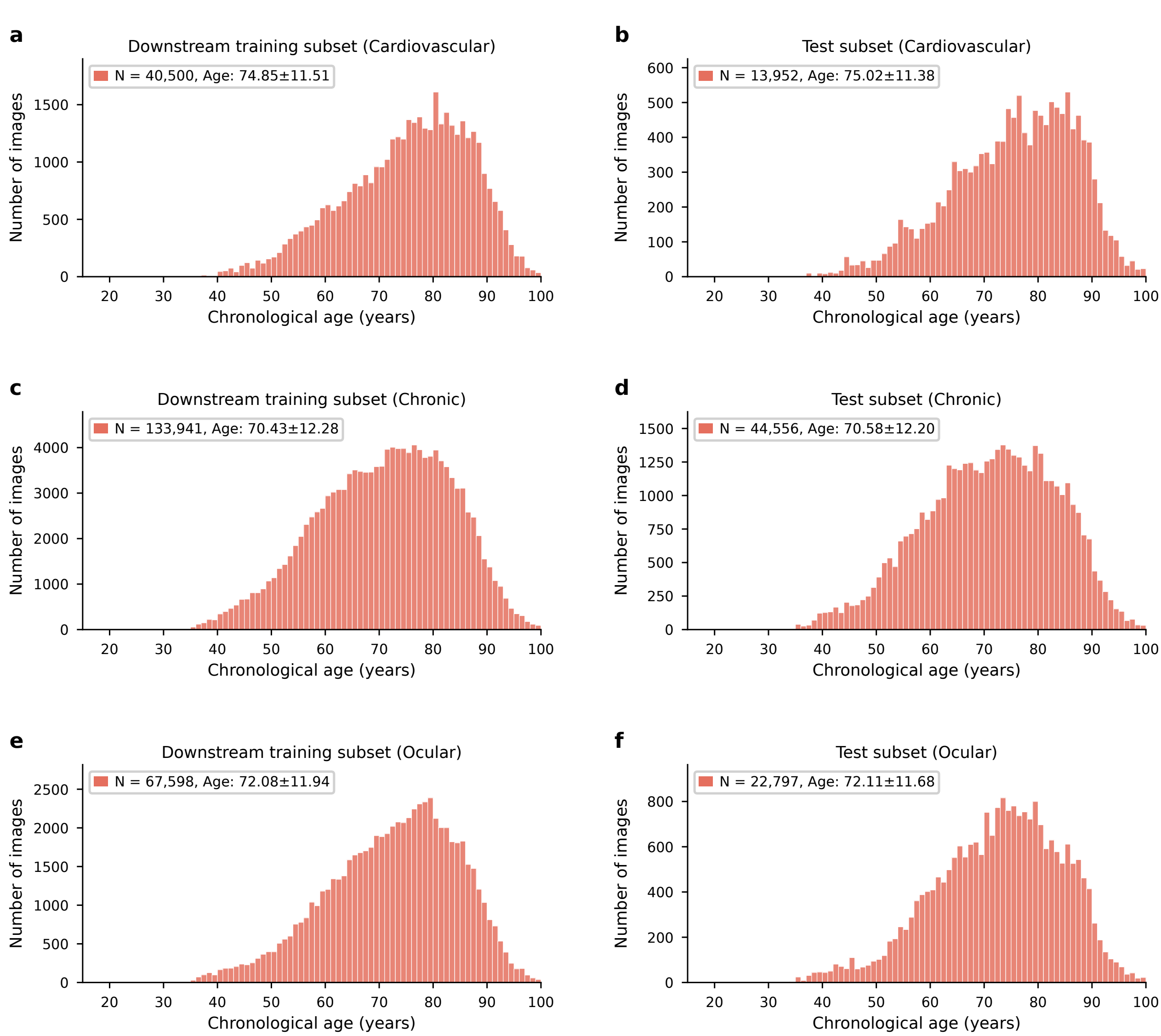

**Supplementary Fig. 8. Chronological age distributions by disease category for AlzEye.** Histograms in 1-year bins of age at imaging for images with a cardiovascular (**a, b**), chronic (**c, d**) or ocular (**e, f**) disease record, the categories shown in Fig. 2g, in the downstream training subset (**a, c, e**) and the test subset (**b, d, f**). An image can belong to more than one category. N and mean±SD in years are annotated.

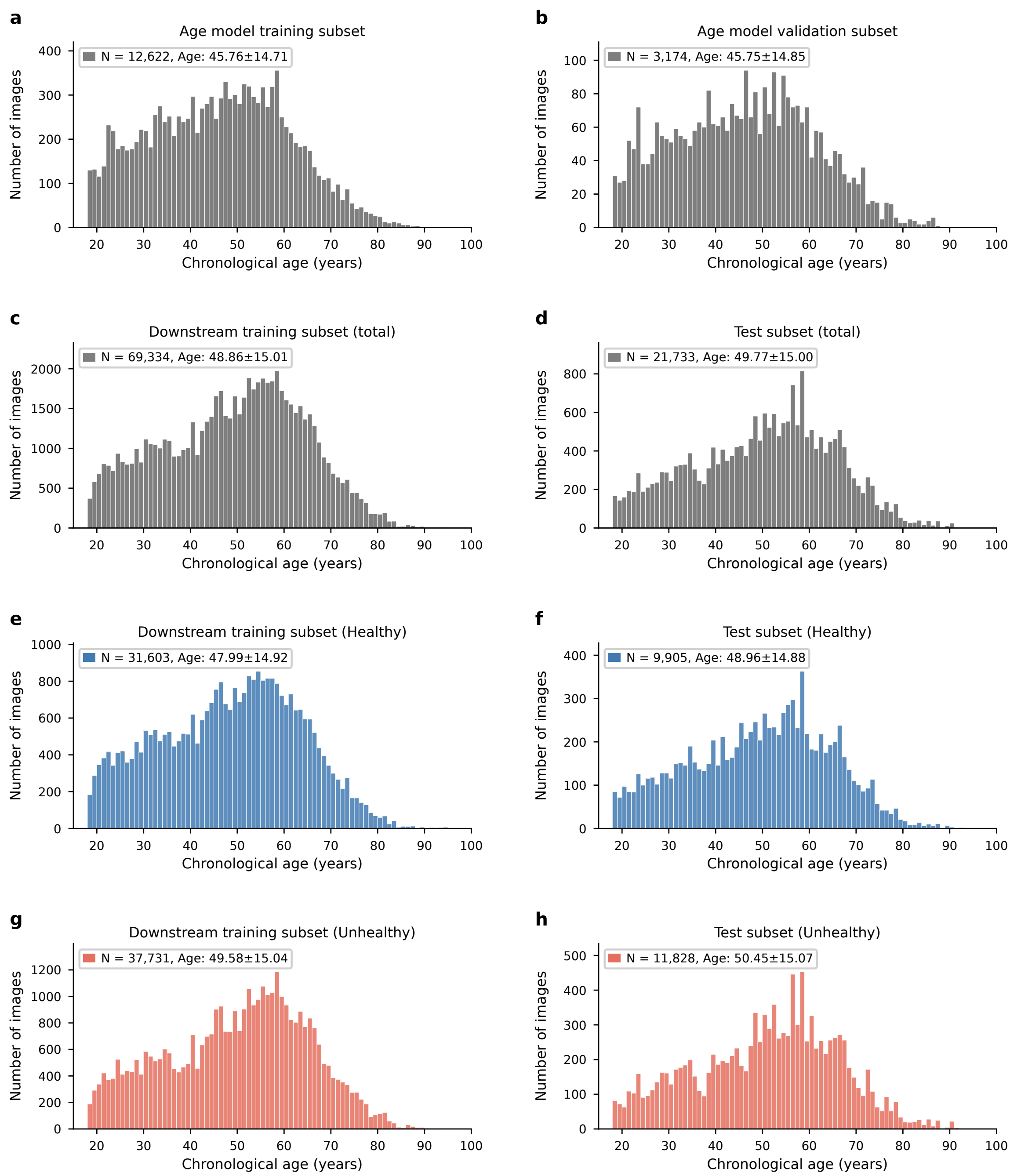

**Supplementary Fig. 9. Chronological age distributions for the ChestX-ray14 subsets.** Histograms in 1-year bins for the age model training subset (**a**) and validation subset (**b**), and for the downstream training subset and the test subset in total (**c, d**) and separately for the healthy (**e, f**) and unhealthy (**g, h**) groups. N and mean±SD in years are annotated.

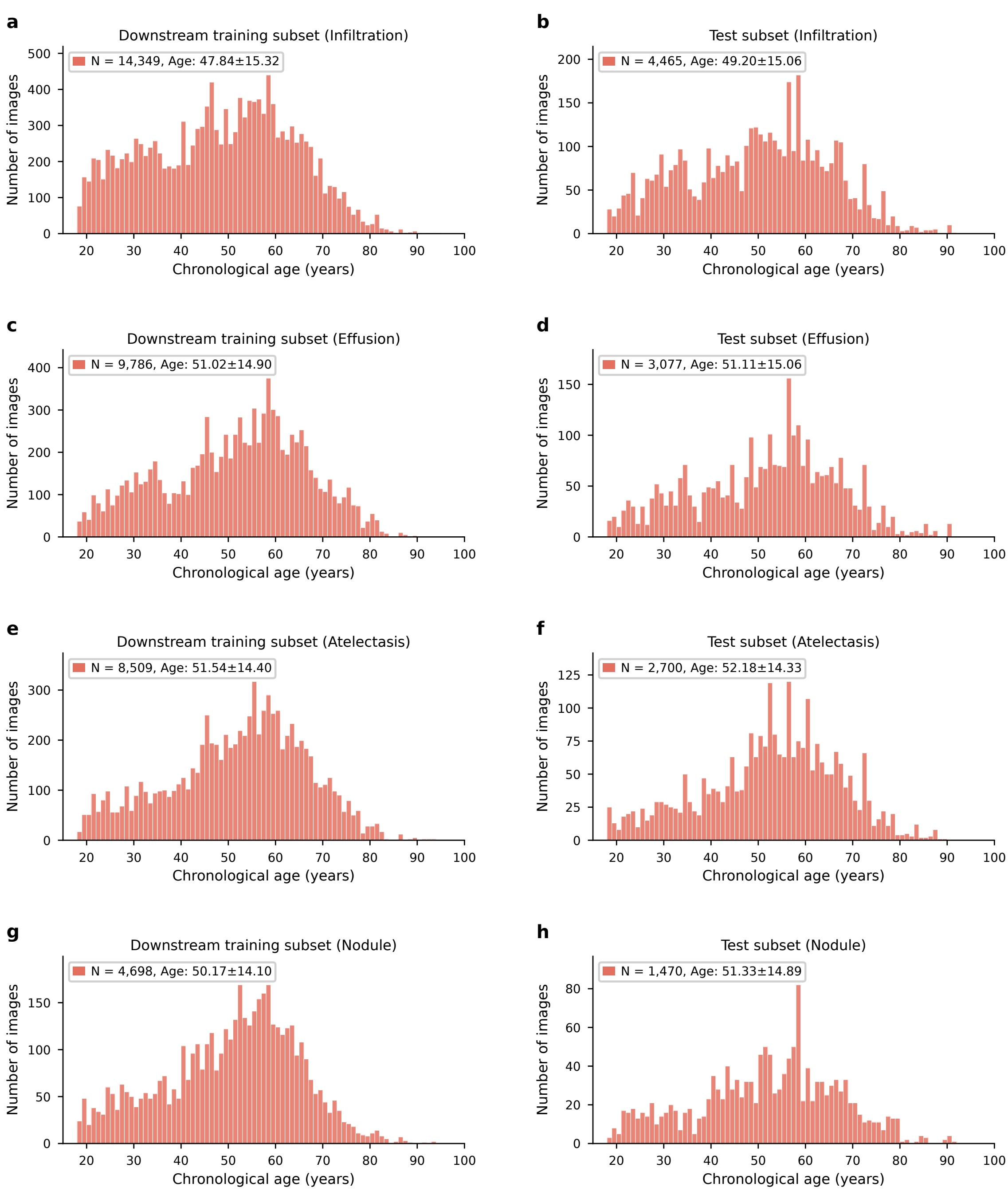

**Supplementary Fig. 10. Chronological age distributions by finding for ChestX-ray14.** Histograms in 1-year bins for images with infiltration (**a, b**), effusion (**c, d**), atelectasis (**e, f**) or a nodule (**g, h**), the four findings shown in Fig. 2h, in the downstream training subset (**a, c, e, g**) and the test subset (**b, d, f, h**). An image can have more than one finding. N and mean±SD in years are annotated.

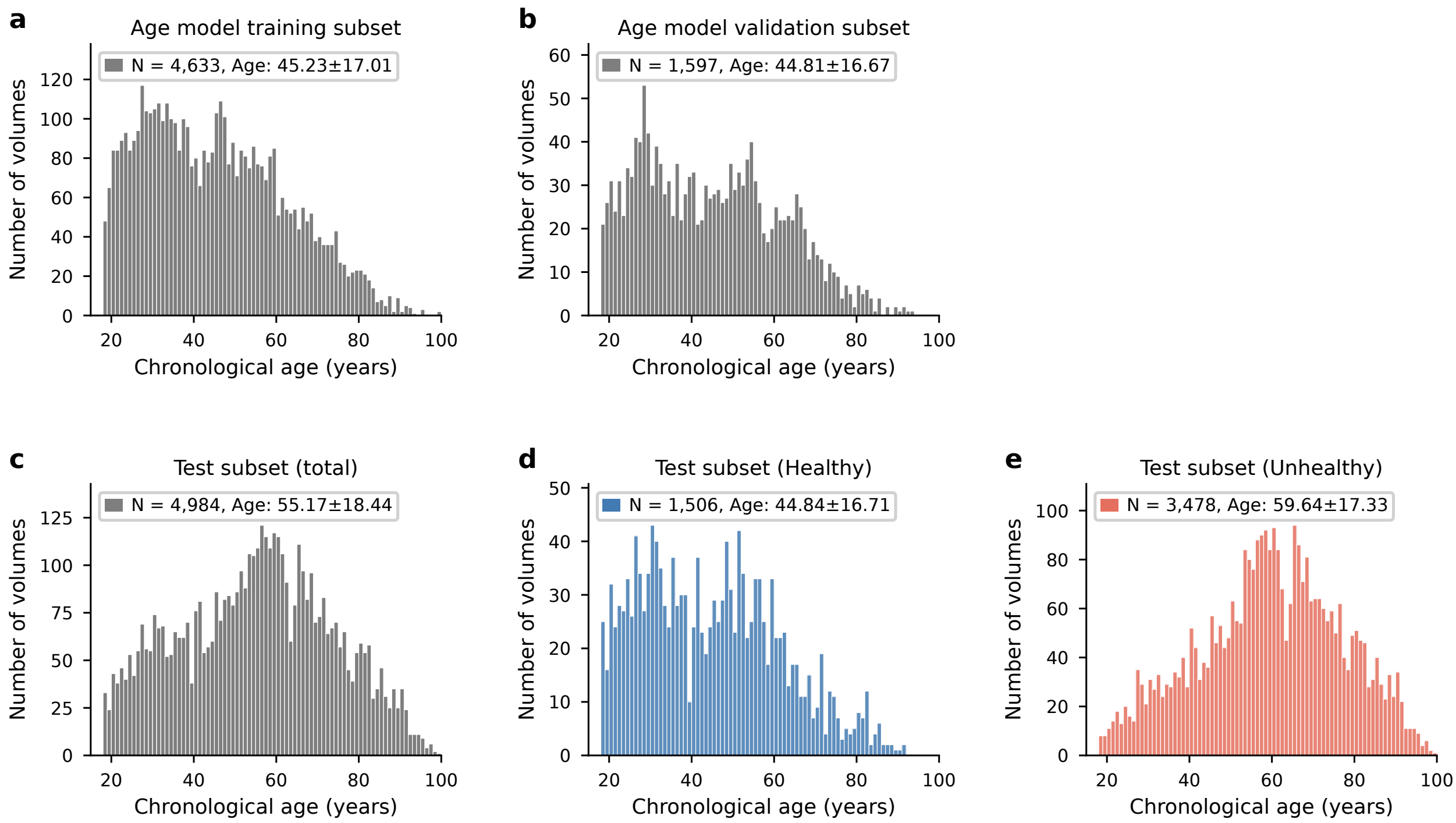

**Supplementary Fig. 11. Chronological age distributions for the Merlin subsets.** Histograms in 1-year bins for the age model training subset (**a**) and validation subset (**b**), and for the test subset in total (**c**) and for the healthy (**d**) and unhealthy (**e**) groups. N and mean±SD in years are annotated.

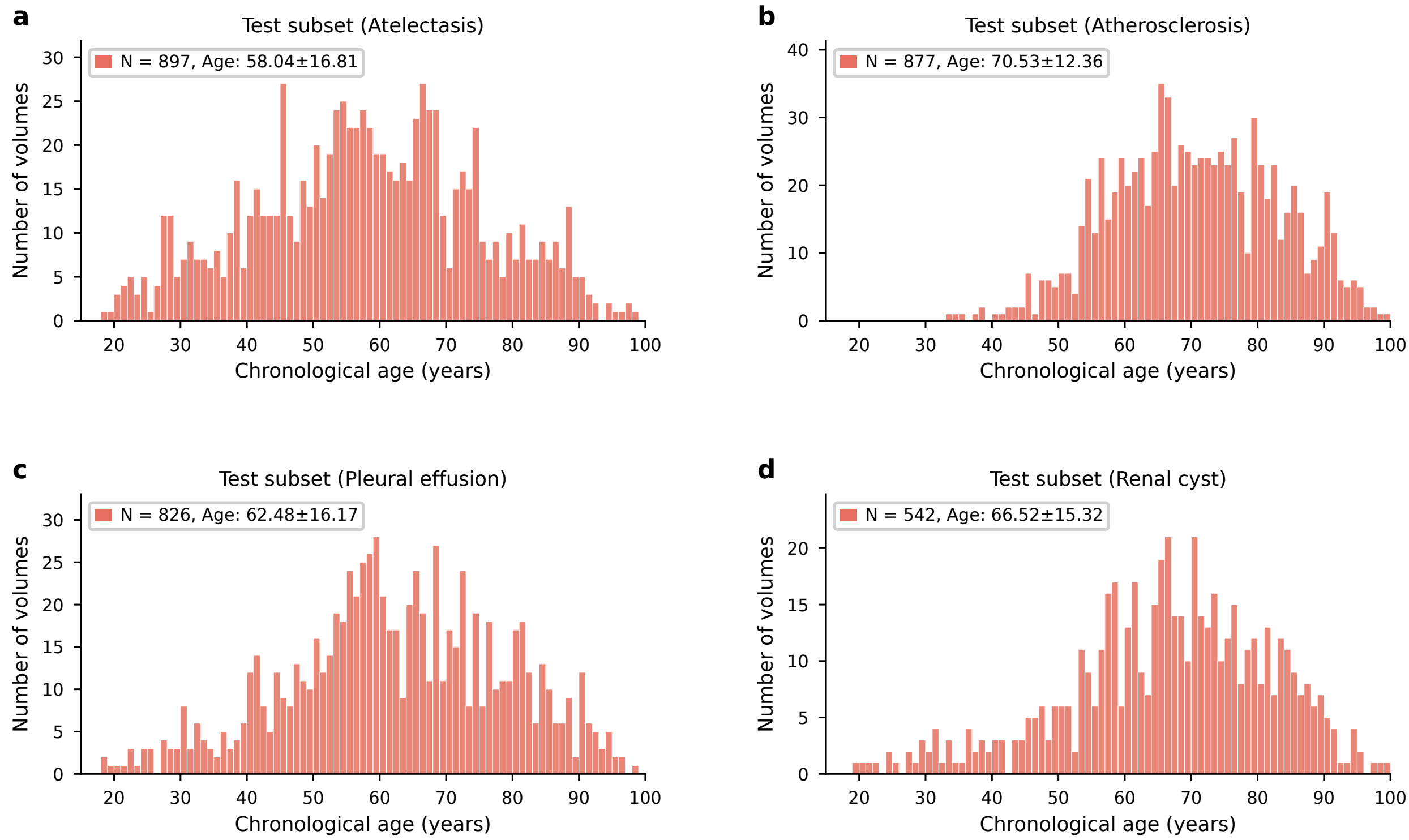

**Supplementary Fig. 12. Chronological age distributions by finding for Merlin.** Histograms in 1-year bins for test subset volumes with atelectasis (**a**), atherosclerosis (**b**), pleural effusion (**c**) or a renal cyst (**d**), the four findings shown in Fig. 2i. A volume can have more than one finding. N and mean±SD in years are annotated.

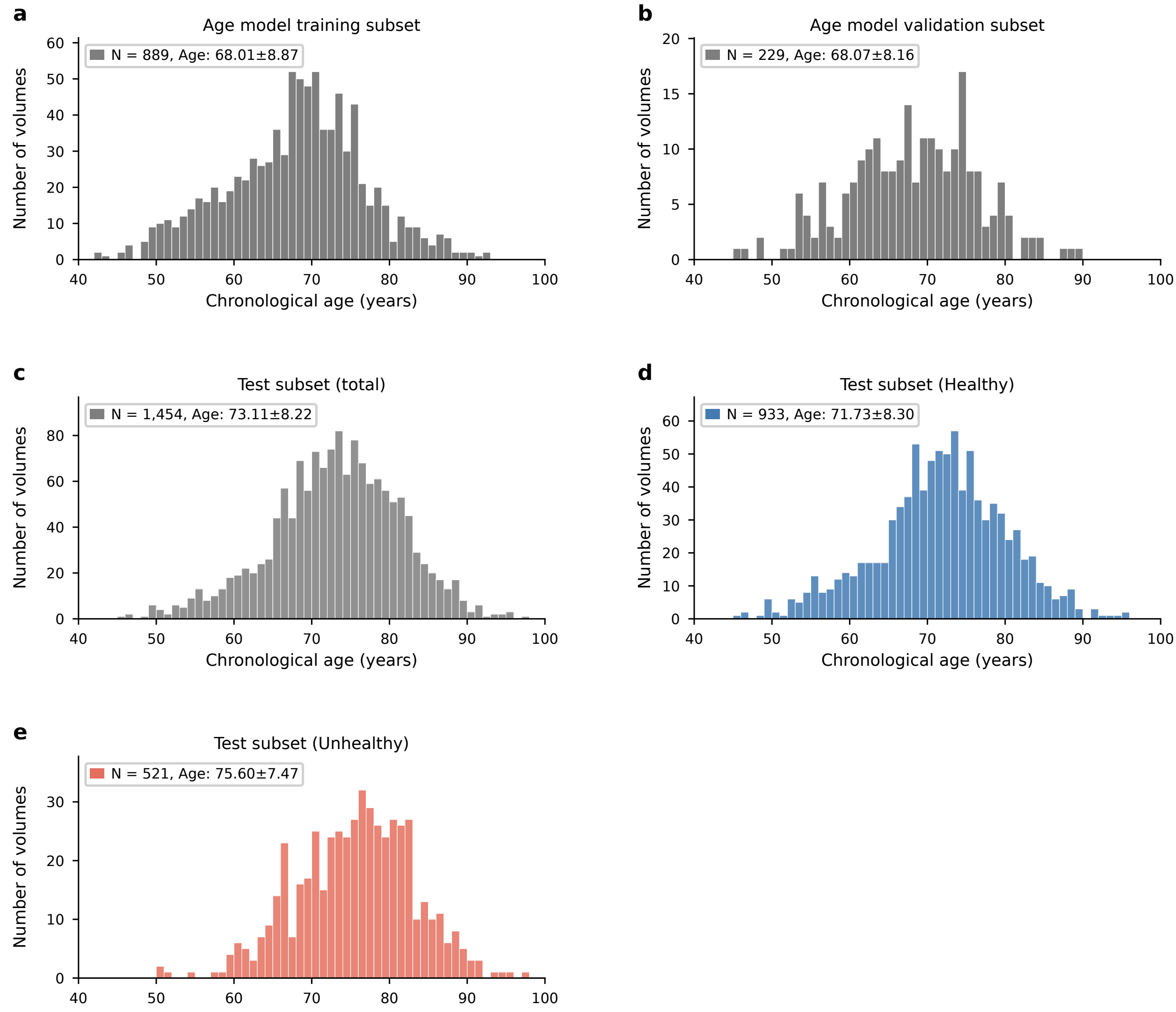

**Supplementary Fig. 13. Chronological age distributions for the OASIS-3 subsets.** Histograms in 1-year bins of age at imaging for the age model training subset (**a**) and validation subset (**b**), and for the test subset in total (**c**) and for the healthy (**d**) and unhealthy (**e**) groups. N and mean±SD in years are annotated.

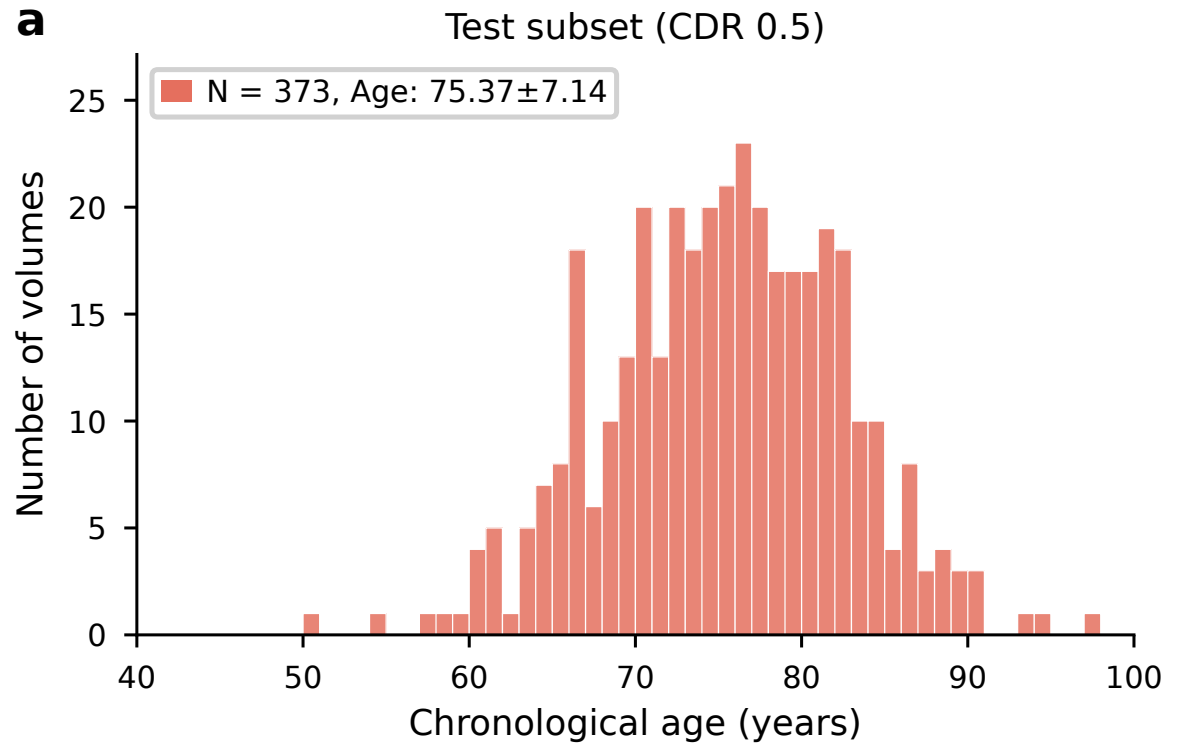

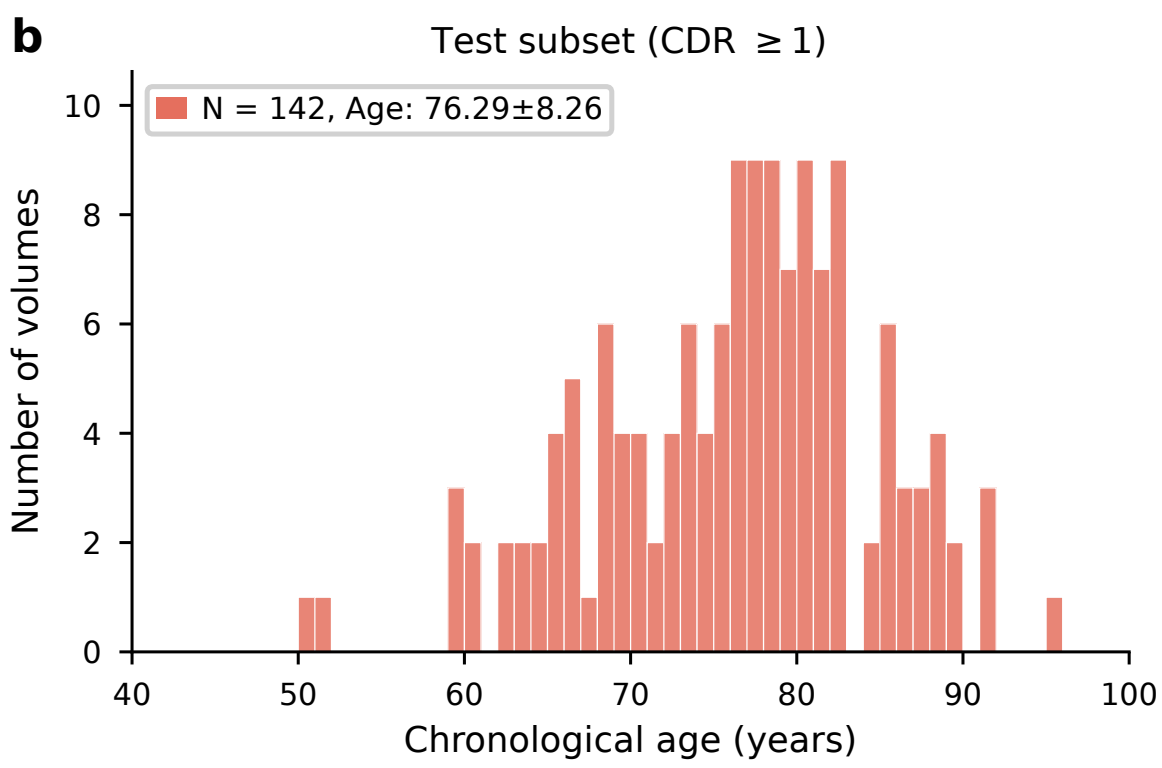

**Supplementary Fig. 14. Chronological age distributions by Clinical Dementia Rating for OASIS-3.** Histograms in 1-year bins for test subset volumes with a Clinical Dementia Rating of 0.5 (**a**) or of at least 1 (**b**), the strata shown in Fig. 2j. N and mean±SD in years are annotated.

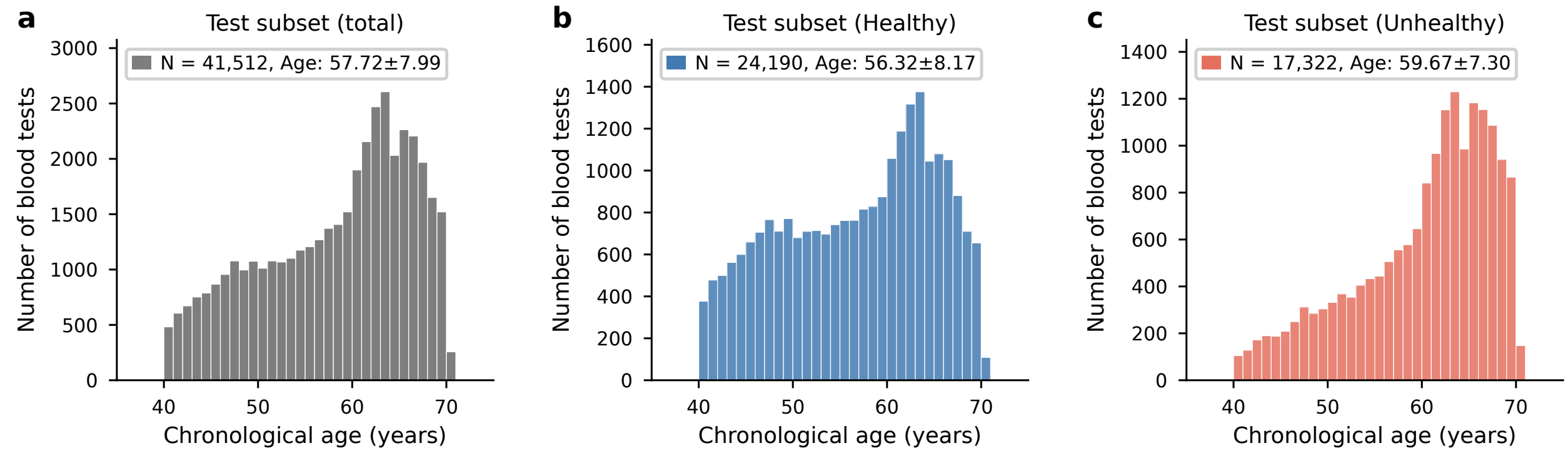

**Supplementary Fig. 15. Chronological age distributions for UK Biobank.** Histograms in 1-year bins for the test subset in total (**a**) and for the healthy (**b**) and unhealthy (**c**) groups. N and mean±SD in years are annotated.

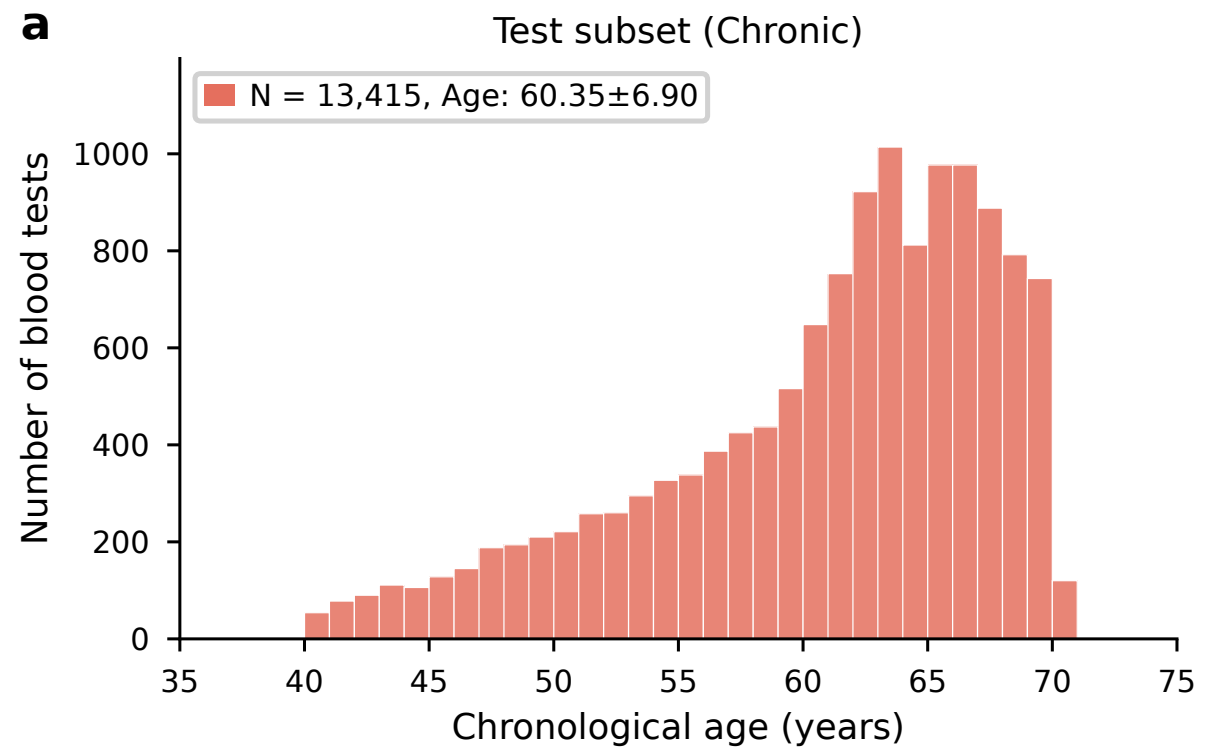

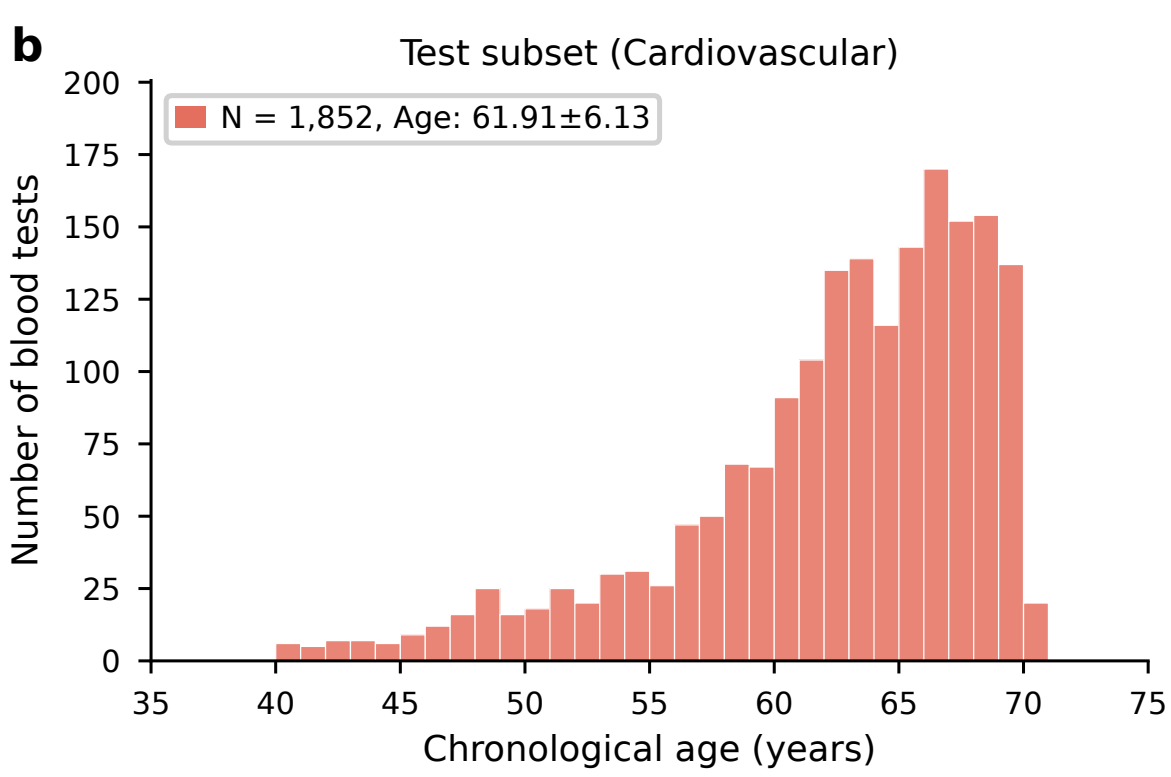

**Supplementary Fig. 16. Chronological age distributions by disease category for UK Biobank.** Histograms in 1-year bins for test subset patients with a chronic (**a**) or cardiovascular (**b**) disease record, the categories shown in Fig. 2k. A patient can belong to both categories. N and mean±SD in years are annotated.

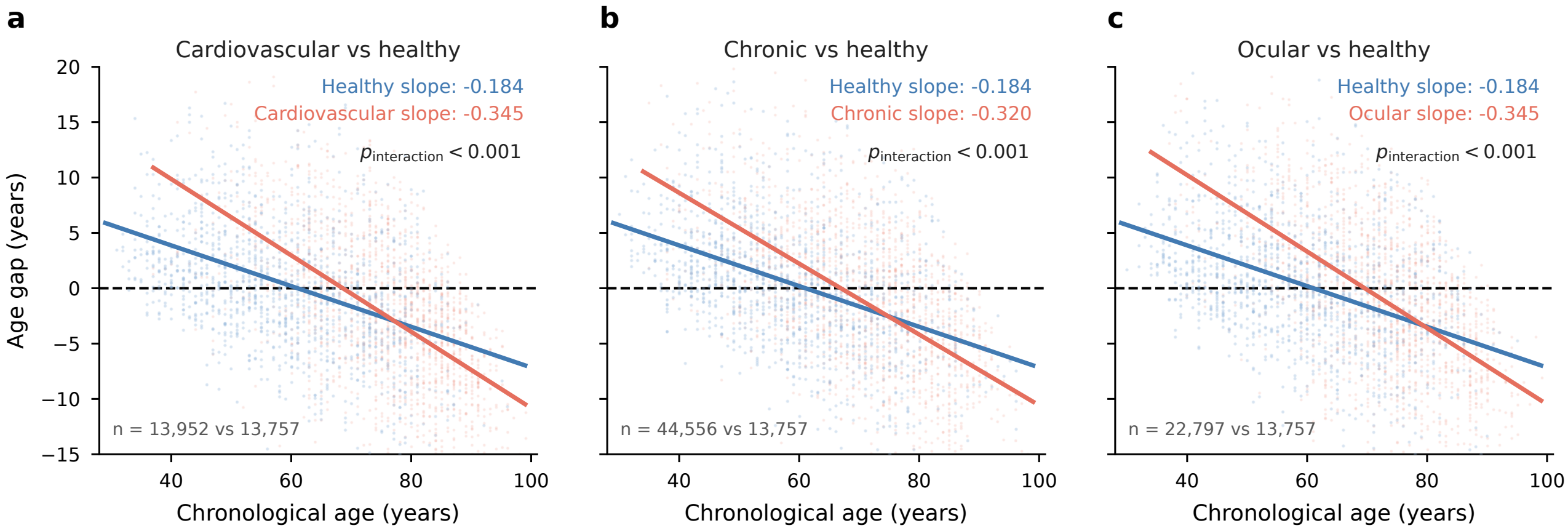

**Supplementary Fig. 17. Differential RTM by disease category for Retinal Age on the AlzEye test subset.** Age gap against chronological age for images of healthy patients and of patients with a cardiovascular (**a**), chronic (**b**) or ocular (**c**) disease record, with the lines fitted by a linear model of the age gap on age, group and their interaction. The slope of each group and the interaction p-value are annotated, and n gives the size of the disease group and of the healthy group. Full results are reported in Supplementary Table 10.

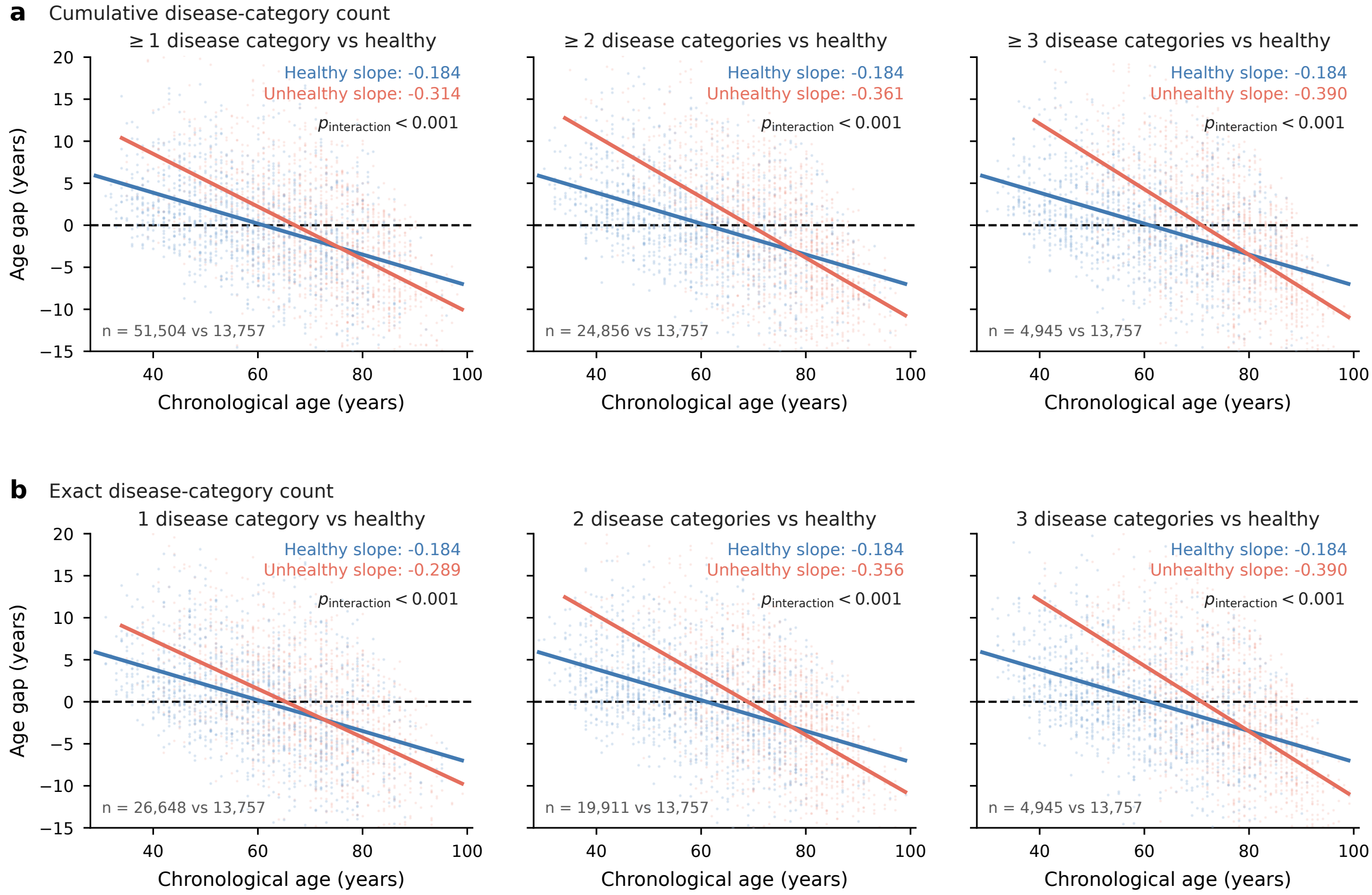

**Supplementary Fig. 18. Differential RTM by number of disease categories for Retinal Age on the AlzEye test subset.** Age gap against chronological age for images of healthy patients and of patients with a record in at least one, two or three (**a**) or in exactly one, two or three (**b**) of the cardiovascular, chronic and ocular categories, with the lines fitted by a linear model of the age gap on age, group and their interaction. Slopes, the interaction p-value and group sizes are annotated as in Supplementary Fig. 17. Full results are reported in Supplementary Table 10.

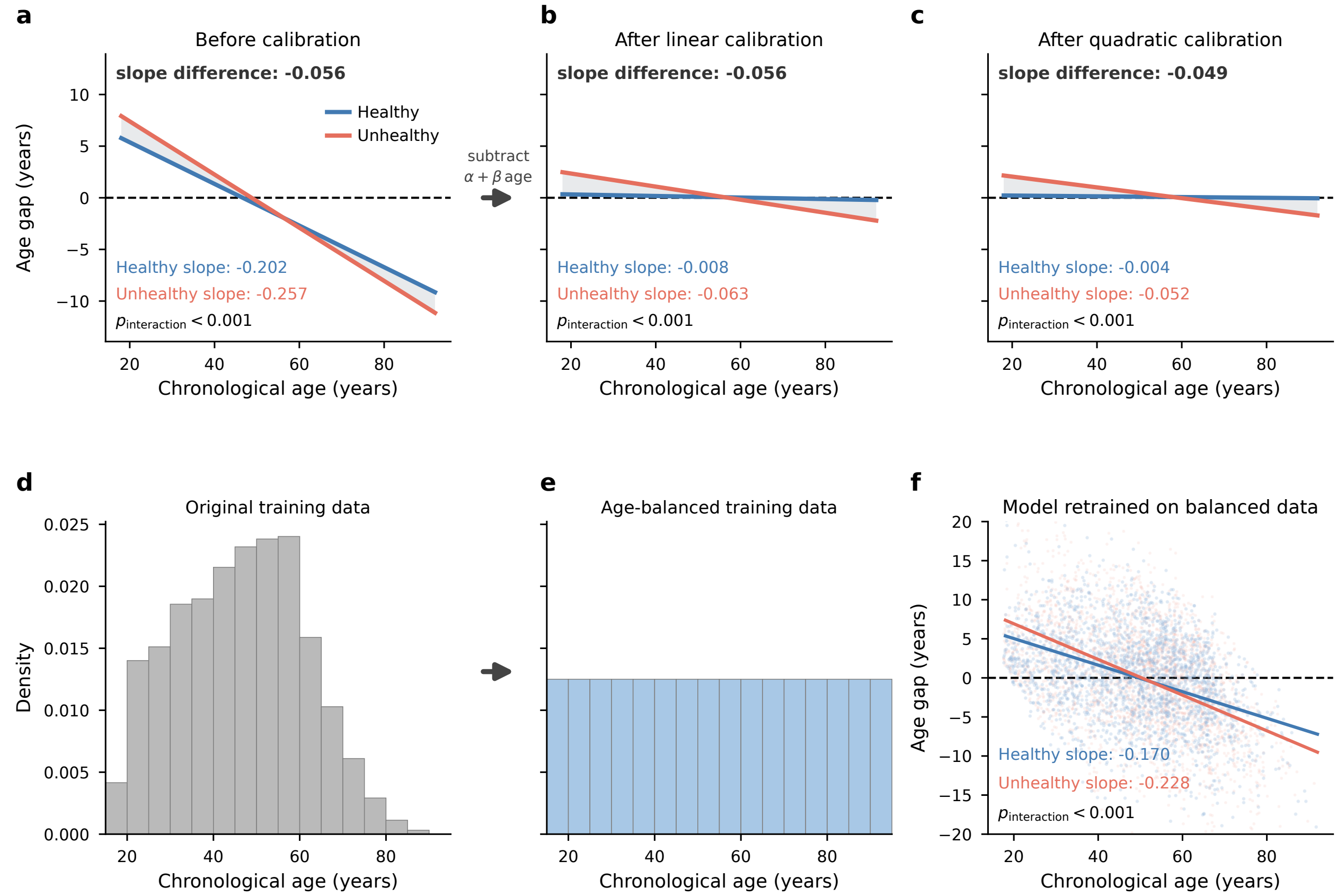

**Supplementary Fig. 19. Differential RTM persists after calibration and balanced training, shown for Chest Age on the ChestX-ray14 test subset. a-c.** Group-specific ordinary least squares fits of the age gap on chronological age before calibration (**a**), after linear calibration (**b**) and after quadratic calibration (**c**). The shaded wedge is the difference between the two groups at each chronological age. The slope of each group (years of age gap per year of chronological age), the difference between the two slopes (unhealthy minus healthy) and the p-value for this difference (the Age × Group interaction) are annotated. A linear calibrator subtracts one function of chronological age from both groups, so both slopes shift by the same amount. The calibrators were fitted on healthy images of the downstream training subset. **d, e.** Balanced training. Age distribution of the age model training subset for the original model (**d**) and after inverse-frequency sampling in 5-year bins (**e**). **f.** Age gap versus chronological age after retraining the Chest Age model with age-balanced sampling, evaluated on the same test subset. Annotated are the two group slopes and the interaction p-value, from a Wald test with cluster-robust standard errors clustered on patient. Full results are reported in Supplementary Tables 7 and 8.

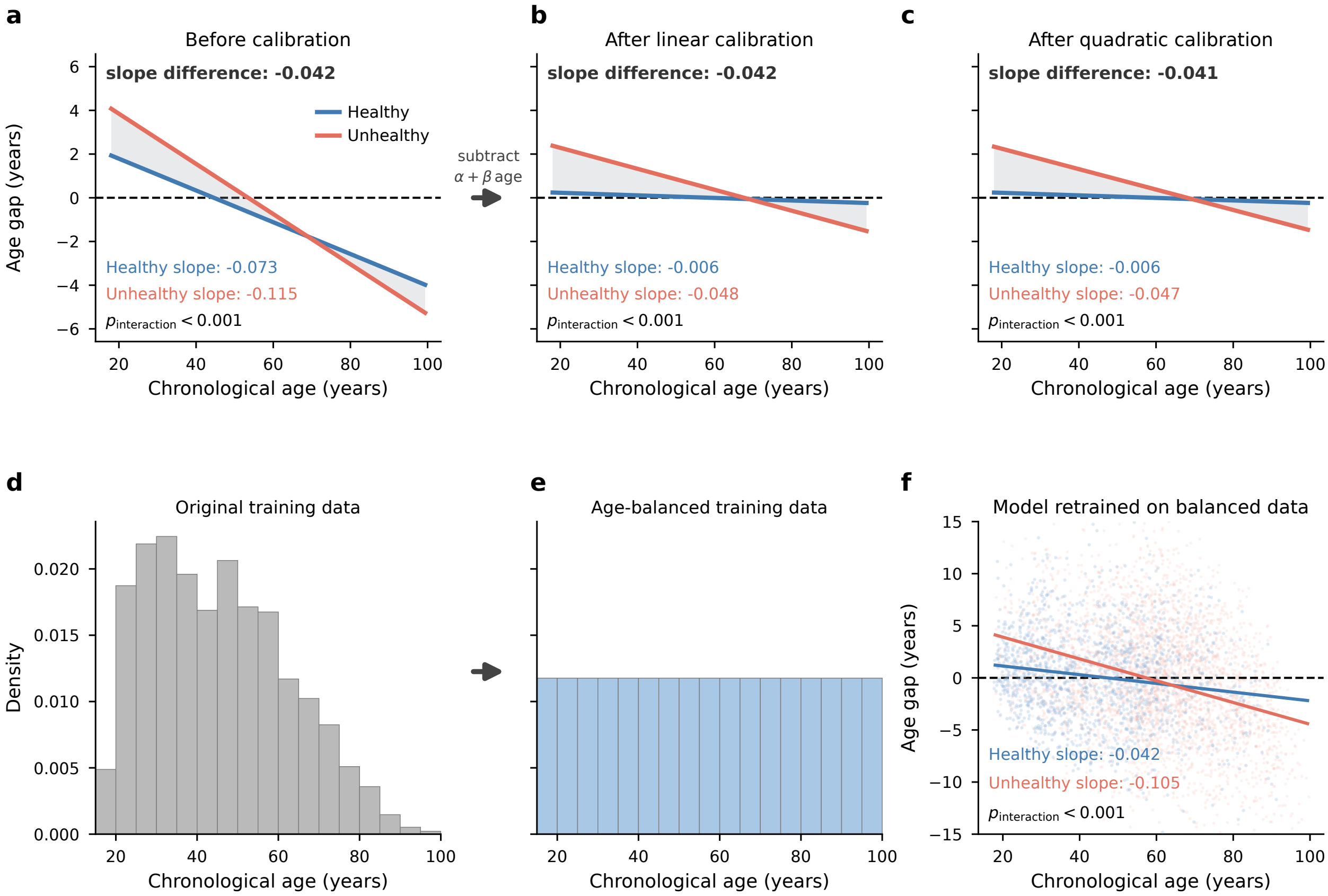

**Supplementary Fig. 20. Differential RTM persists after calibration and balanced training, shown for Abdominal Age on the Merlin test subset. a-c.** Group-specific ordinary least squares fits of the age gap on chronological age before calibration (**a**), after linear calibration (**b**) and after quadratic calibration (**c**). The shaded wedge is the difference between the two groups at each chronological age. The slope of each group (years of age gap per year of chronological age), the difference between the two slopes (unhealthy minus healthy) and the p-value for this difference (the Age × Group interaction) are annotated. A linear calibrator subtracts one function of chronological age from both groups, so both slopes shift by the same amount. The calibrators were fitted on healthy volumes of the age model validation subset. **d, e.** Balanced training. Age distribution of the age model training subset for the original model (**d**) and after inverse-frequency sampling in 5-year bins (**e**). **f.** Age gap versus chronological age after retraining the Abdominal Age model with age-balanced sampling, evaluated on the same test subset. Annotated are the two group slopes and the interaction p-value, from a Wald test with cluster-robust standard errors clustered on volume, because Merlin releases no patient identifiers. Full results are reported in Supplementary Tables 7 and 8.

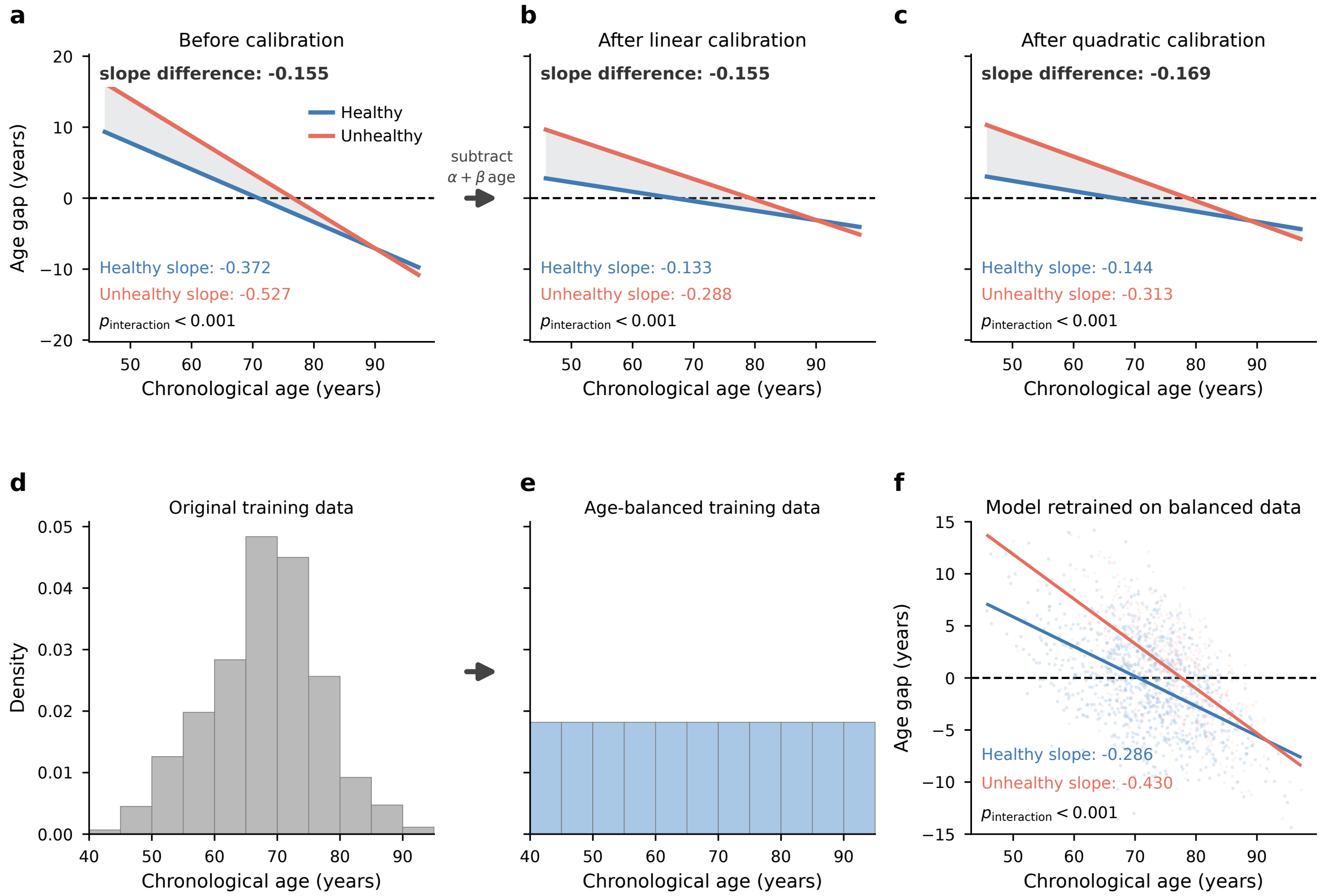

**Supplementary Fig. 21. Differential RTM persists after calibration and balanced training, shown for Brain Age on the OASIS-3 test subset. a-c.** Group-specific ordinary least squares fits of the age gap on chronological age before calibration (**a**), after linear calibration (**b**) and after quadratic calibration (**c**). The shaded wedge is the difference between the two groups at each chronological age. The slope of each group (years of age gap per year of chronological age), the difference between the two slopes (unhealthy minus healthy) and the p-value for this difference (the Age × Group interaction) are annotated. A linear calibrator subtracts one function of chronological age from both groups, so both slopes shift by the same amount. The calibrators were fitted on healthy volumes of the age model validation subset. **d, e.** Balanced training. Age distribution of the age model training subset for the original model (**d**) and after inverse-frequency sampling in 5-year bins (**e**). **f.** Age gap versus chronological age after retraining the Brain Age model with age-balanced sampling, evaluated on the same test subset. Annotated are the two group slopes and the interaction p-value, from a Wald test with cluster-robust standard errors clustered on patient. Full results are reported in Supplementary Tables 7 and 8.

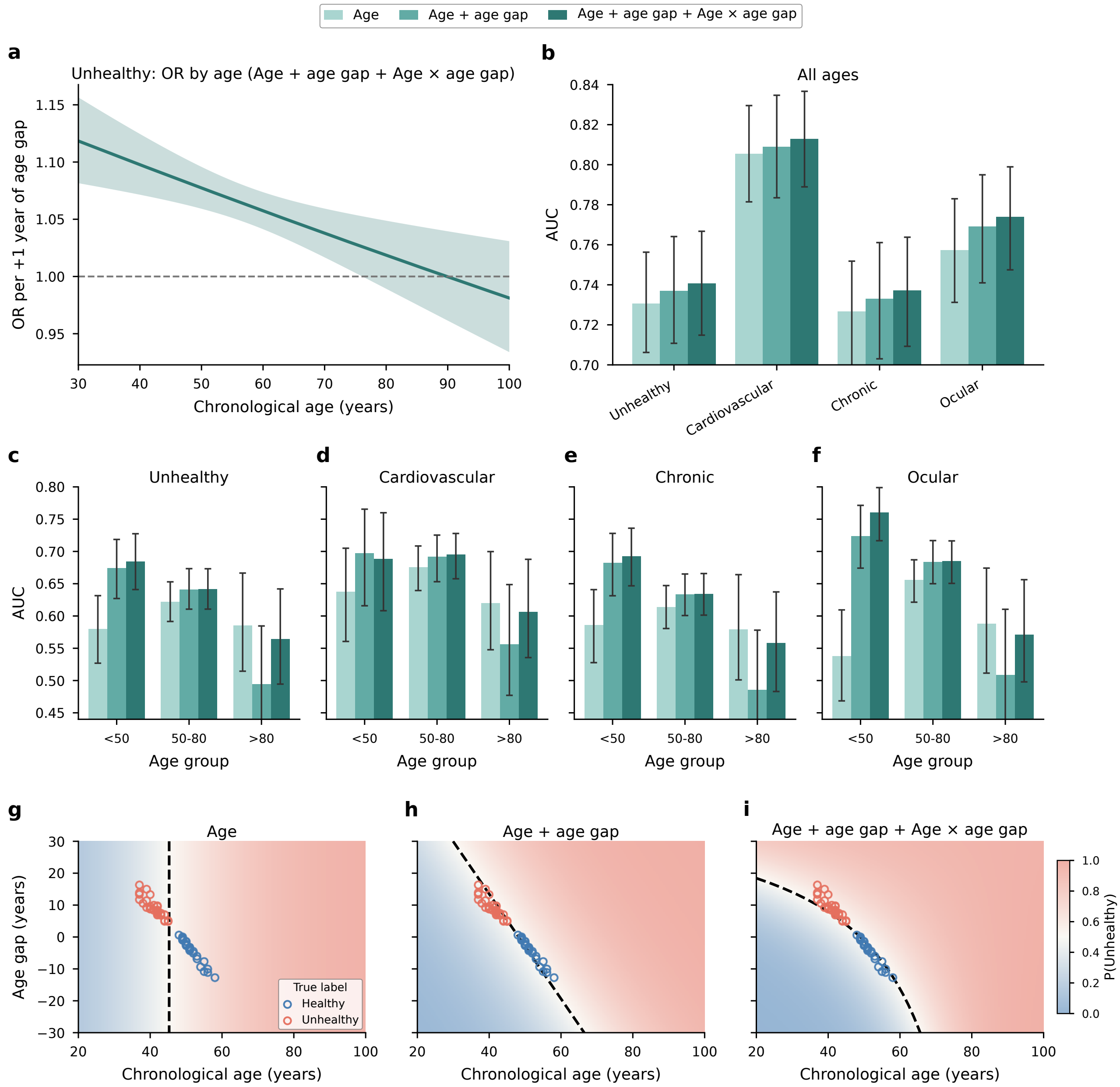

**Supplementary Fig. 22. Including an interaction term between chronological age and the age gap to account for differential RTM, shown for Retinal Age. a.** Odds ratio (OR) for unhealthy status per 1-year increase in the age gap as a function of chronological age, from a logistic regression of unhealthy status on chronological age, the age gap and their interaction, fitted on the AlzEye test subset. The shaded band denotes the 95% confidence interval, cluster-robust at the patient level; the dashed line marks OR = 1. **b.** Overall AUC for unhealthy status and for cardiovascular, chronic and ocular disease prediction on the AlzEye test subset, using three logistic regression models, Age, Age + age gap and Age + age gap + Age × age gap, trained on the AlzEye downstream training subset. Error bars denote 95% confidence intervals. **c-f.** AUC within age groups for unhealthy status (**c**) and for cardiovascular (**d**), chronic (**e**) and ocular (**f**) disease, for the same models as in **b**, showing that the added value of the age gap differs across chronological age. **g-i.** Probability plots in the plane of chronological age and age gap for unhealthy status under the Age (**g**), Age + age gap (**h**) and Age + age gap + Age × age gap (**i**) models, showing the predicted probability of unhealthy status and the corresponding decision boundary. Highlighted points denote test cases that are misclassified by the Age and Age + age gap models but correctly classified by the Age + age gap + Age × age gap model. Full results are reported in Supplementary Tables 11, 12 and 13.

# Supplementary Tables

**Supplementary Table 1. Imaging devices used to acquire the colour fundus images in the original AlzEye dataset.** Shares are percentages of all images in the dataset.

| Device | Images (n) | Share (%) |
|---|---|---|
| Topcon 3D OCT-2000SA | 1,716,282 | 66.44 |
| Topcon Triton plus | 496,344 | 19.22 |
| Topcon FD-OCT | 209,132 | 8.10 |
| Topcon Triton | 76,481 | 2.96 |
| Carl Zeiss Meditec CIRRUS photo 600 | 29,158 | 1.13 |
| Topcon 3D OCT-1000MK2 | 28,988 | 1.12 |
| Topcon 3D OCT-2000 | 26,618 | 1.03 |
| Unknown | 9 | 0.00 |
| Total | 2,583,012 | 100.00 |

**Supplementary Table 2. Age estimation performance of the four organ-derived ageing markers on their test subsets.** Mean absolute error (MAE), root mean squared error (RMSE) and mean age gap (estimated minus chronological age) are in years; $r^2$ is the squared Pearson correlation between estimated and chronological age. Point estimates are those annotated in Fig. 2a-d. The 95% confidence intervals are from a cluster bootstrap with 1,000 resamples of patients (volumes for Merlin, which releases no patient identifiers).

| Ageing marker | Dataset | Test subset | Clusters | MAE [95% CI] | RMSE [95% CI] | $r^2$ [95% CI] | Mean age gap [95% CI] |
|---|---|---|---|---|---|---|---|
| Retinal Age | AlzEye | 65,360 images | 12,141 patients | 5.31 [5.23, 5.40] | 6.80 [6.68, 6.91] | 0.745 [0.732, 0.758] | −0.86 [−1.04, −0.69] |
| Chest Age | ChestX-ray14 | 21,733 images | 4,623 patients | 5.15 [4.95, 5.42] | 6.80 [6.45, 7.26] | 0.796 [0.769, 0.819] | −0.45 [−0.83, −0.08] |
| Abdominal Age | Merlin | 4,984 volumes | 4,984 volumes | 4.04 [3.94, 4.14] | 5.32 [5.17, 5.46] | 0.918 [0.912, 0.923] | −0.50 [−0.64, −0.35] |
| Brain Age | OASIS-3 | 1,454 volumes | 808 patients | 3.81 [3.59, 4.03] | 4.92 [4.65, 5.20] | 0.643 [0.604, 0.680] | −0.01 [−0.35, +0.34] |

**Supplementary Table 3. Association between the age gap and unhealthy status or individual disease categories (Fig. 2f-k).** Odds ratios (OR) per 1-year increase in the age gap are from logistic regressions of each outcome on the age gap, adjusted for chronological age, with the healthy patients of the same test subset as the comparison group. The 95% confidence intervals and Wald p-values use a cluster-robust (CR1) covariance clustered on patient (volume for Merlin, which releases no patient identifiers), as drawn in Fig. 2f-k. Disease categories are those recorded in each cohort (Supplementary Table 15 for AlzEye and UK Biobank) and are not comparable across ageing markers. For PhenoAge the age gap is PhenoAgeAccel.

| Ageing marker (dataset) | Outcome | Cases (n) | Healthy (n) | OR [95% CI] | p |
|---|---|---|---|---|---|
| Retinal Age (AlzEye) | Unhealthy | 51,603 | 13,757 | 1.049 [1.032, 1.067] | $<0.001$ |
| | Chronic | 44,556 | 13,757 | 1.048 [1.031, 1.066] | $<0.001$ |
| | Ocular | 22,797 | 13,757 | 1.065 [1.045, 1.086] | $<0.001$ |
| | Cardiovascular | 13,952 | 13,757 | 1.041 [1.018, 1.065] | $<0.001$ |
| Chest Age (ChestX-ray14) | Unhealthy | 11,828 | 9,905 | 1.011 [1.003, 1.019] | 0.005 |
| | Infiltration | 4,465 | 9,905 | 1.005 [0.994, 1.015] | 0.393 |
| | Effusion | 3,077 | 9,905 | 1.014 [1.000, 1.028] | 0.053 |
| | Atelectasis | 2,700 | 9,905 | 1.019 [1.008, 1.030] | $<0.001$ |
| | Nodule | 1,470 | 9,905 | 1.020 [1.005, 1.035] | 0.008 |
| Abdominal Age (Merlin) | Unhealthy | 3,478 | 1,506 | 1.032 [1.019, 1.045] | $<0.001$ |
| | Atelectasis | 897 | 1,506 | 1.031 [1.011, 1.051] | 0.002 |
| | Atherosclerosis | 877 | 1,506 | 1.062 [1.039, 1.087] | $<0.001$ |
| | Pleural effusion | 826 | 1,506 | 1.012 [0.992, 1.032] | 0.245 |
| | Renal cyst | 542 | 1,506 | 1.057 [1.030, 1.085] | $<0.001$ |
| Brain Age (OASIS-3) | Unhealthy | 521 | 933 | 1.238 [1.182, 1.297] | $<0.001$ |
| | CDR 0.5 | 373 | 933 | 1.196 [1.140, 1.255] | $<0.001$ |
| | CDR $\geq$1 | 142 | 933 | 1.426 [1.313, 1.550] | $<0.001$ |
| PhenoAge (UK Biobank) | Unhealthy | 17,322 | 24,190 | 1.076 [1.072, 1.080] | $<0.001$ |
| | Chronic | 13,415 | 24,190 | 1.088 [1.083, 1.093] | $<0.001$ |
| | Cardiovascular | 1,852 | 24,190 | 1.139 [1.129, 1.149] | $<0.001$ |

**Supplementary Table 4. Mean age gap and association with unhealthy status within age subgroups for the four organ-derived ageing markers (Fig. 3a-d, i-l).** Mean age gap (years) of each health group with a normal-approximation 95% confidence interval (mean±1.96 standard errors). Odds ratios (OR) per 1-year increase in the age gap are from a logistic regression of unhealthy status on the age gap with chronological age as a covariate, fitted within each subgroup; 95% confidence intervals and Wald p-values use a cluster-robust (CR1) covariance clustered on patient (volume for Merlin, which releases no patient identifiers). Age subgroups are as in Fig. 3; n counts images or volumes.

| Ageing marker | Age subgroup | Healthy (n) | Unhealthy (n) | Mean age gap, healthy | Mean age gap, unhealthy | OR [95% CI] | p |
|---|---|---|---|---|---|---|---|
| Retinal Age | $<50$ | 3,796 | 2,572 | +3.36 [+3.21, +3.52] | +6.42 [+6.18, +6.66] | 1.114 [1.085, 1.145] | $<0.001$ |
| | 50-60 | 3,288 | 7,082 | +1.14 [+0.97, +1.32] | +3.62 [+3.49, +3.75] | 1.103 [1.077, 1.129] | $<0.001$ |
| | 60-70 | 3,045 | 12,807 | −0.63 [−0.83, −0.43] | +0.76 [+0.66, +0.86] | 1.048 [1.020, 1.077] | $<0.001$ |
| | 70-80 | 2,606 | 15,469 | −2.32 [−2.55, −2.08] | −1.96 [−2.06, −1.87] | 1.012 [0.982, 1.044] | 0.433 |
| | $\geq 80$ | 1,022 | 13,673 | −4.15 [−4.55, −3.75] | −5.99 [−6.08, −5.89] | 0.955 [0.890, 1.025] | 0.200 |
| Chest Age | $<30$ | 1,240 | 1,345 | +3.85 [+3.55, +4.16] | +5.50 [+5.15, +5.85] | 1.051 [1.024, 1.079] | $<0.001$ |
| | 30-40 | 1,530 | 1,593 | +2.16 [+1.85, +2.46] | +3.26 [+2.93, +3.60] | 1.026 [1.004, 1.047] | 0.018 |
| | 40-50 | 1,994 | 2,193 | +0.67 [+0.43, +0.91] | +1.10 [+0.86, +1.35] | 1.014 [1.000, 1.028] | 0.056 |
| | 50-60 | 2,615 | 3,238 | −0.91 [−1.11, −0.71] | −0.77 [−0.97, −0.57] | 1.005 [0.990, 1.020] | 0.551 |
| | 60-70 | 1,798 | 2,402 | −3.59 [−3.86, −3.33] | −3.72 [−3.96, −3.48] | 0.995 [0.980, 1.010] | 0.490 |
| | $\geq 70$ | 728 | 1,057 | −7.13 [−7.55, −6.71] | −8.35 [−8.74, −7.97] | 0.975 [0.953, 0.998] | 0.035 |
| Abdominal Age | $<40$ | 652 | 513 | +0.99 [+0.71, +1.27] | +1.93 [+1.55, +2.31] | 1.080 [1.048, 1.112] | $<0.001$ |
| | 40-50 | 281 | 449 | −0.11 [−0.58, +0.37] | +0.70 [+0.27, +1.14] | 1.041 [1.007, 1.077] | 0.017 |
| | 50-60 | 287 | 766 | −0.47 [−1.00, +0.06] | −0.13 [−0.51, +0.24] | 1.015 [0.989, 1.041] | 0.255 |
| | 60-70 | 158 | 749 | −1.22 [−1.93, −0.51] | −0.43 [−0.83, −0.02] | 1.032 [1.003, 1.061] | 0.028 |
| | 70-80 | 79 | 538 | −1.44 [−2.56, −0.31] | −1.99 [−2.46, −1.52] | 0.986 [0.948, 1.026] | 0.498 |
| | $\geq 80$ | 49 | 463 | −4.12 [−5.36, −2.88] | −4.89 [−5.43, −4.35] | 1.005 [0.963, 1.049] | 0.804 |
| Brain Age | $<65$ | 169 | 40 | +4.12 [+3.36, +4.87] | +7.49 [+5.64, +9.34] | 1.190 [1.076, 1.316] | $<0.001$ |
| | 65-70 | 193 | 77 | +0.94 [+0.40, +1.47] | +4.48 [+3.60, +5.37] | 1.316 [1.168, 1.483] | $<0.001$ |
| | 70-75 | 245 | 113 | −0.22 [−0.62, +0.18] | +2.45 [+1.86, +3.04] | 1.300 [1.193, 1.416] | $<0.001$ |
| | 75-80 | 184 | 138 | −1.83 [−2.27, −1.38] | −0.07 [−0.58, +0.45] | 1.227 [1.115, 1.351] | $<0.001$ |
| | $\geq 80$ | 142 | 153 | −5.42 [−6.02, −4.83] | −4.22 [−4.80, −3.63] | 1.181 [1.078, 1.294] | $<0.001$ |

**Supplementary Table 5. Group slopes of the age gap on chronological age and age dependence of the association with unhealthy status (Fig. 3 and Fig. 7c, f).** Slopes are from a linear regression of the age gap on chronological age, health group and their interaction; the slope difference is the Age × Group coefficient and $p_{\text{interaction}}$ its Wald p-value. The OR-age trend is the Age × age gap coefficient of a logistic regression of unhealthy status on chronological age, the age gap and their interaction over all ages, that is, the change in the log odds ratio per 1-year increase in the age gap for each additional year of age; $p_{\text{trend}}$ is its 1 degree-of-freedom Wald test. $p_{\text{heterogeneity}}$ is the $k-1$ degree-of-freedom joint Wald test that the $k$ subgroup odds ratios (Supplementary Table 4; Supplementary Table 9 for PhenoAge) are equal. All intervals and p-values use a cluster-robust (CR1) covariance clustered on patient (volume for Merlin, which releases no patient identifiers); n is given with the number of clusters. Slopes are in years of age gap per year of age. For PhenoAge the age gap is PhenoAgeAccel.

| Ageing marker (dataset) | n (clusters) | Healthy slope | Unhealthy slope | Slope difference | $p_{\text{interaction}}$ | OR-age trend | $p_{\text{trend}}$ | $p_{\text{heterogeneity}}$ |
|---|---|---|---|---|---|---|---|---|
| Retinal Age (AlzEye) | 65,360 (12,141) | −0.184 [−0.217, −0.151] | −0.314 [−0.326, −0.303] | −0.130 [−0.165, −0.096] | <0.001 | −0.0019 [−0.0030, −0.0008] | <0.001 | <0.001 |
| Chest Age (ChestX-ray14) | 21,733 (4,623) | −0.202 [−0.223, −0.180] | −0.257 [−0.283, −0.231] | −0.056 [−0.073, −0.039] | <0.001 | −0.0011 [−0.0015, −0.0007] | <0.001 | <0.001 |
| Abdominal Age (Merlin) | 4,984 (4,984) | −0.073 [−0.085, −0.060] | −0.115 [−0.125, −0.105] | −0.042 [−0.058, −0.026] | <0.001 | −0.0013 [−0.0021, −0.0005] | <0.001 | 0.004 |
| Brain Age (OASIS-3) | 1,454 (808) | −0.372 [−0.416, −0.328] | −0.527 [−0.583, −0.471] | −0.155 [−0.222, −0.087] | <0.001 | −0.0008 [−0.0032, +0.0016] | 0.504 | 0.439 |
| PhenoAge (UK Biobank) | 41,512 (41,512) | −0.032 [−0.039, −0.025] | −0.015 [−0.027, −0.003] | +0.017 [+0.003, +0.031] | 0.018 | +0.0003 [−0.0002, +0.0008] | 0.291 | 0.036 |

**Supplementary Table 6. Difference in mean age gap, variance of the age gap and odds ratio within age subgroups for the four organ-derived ageing markers (Fig. 4).** $\Delta$ is the difference in mean age gap between the unhealthy and healthy groups, adjusted for chronological age; $\sigma^2$ is the variance of the residuals of a linear regression of the age gap on chronological age fitted in each group, with the residuals of the two groups combined; $\exp(\Delta/\sigma^2)$ is the odds ratio implied by the two. $\Delta$, $\sigma^2$ and $\exp(\Delta/\sigma^2)$ have 95% confidence intervals from a patient-level cluster bootstrap with 1,000 resamples. The odds ratio (OR) per 1-year increase in the age gap and its cluster-robust Wald interval are as in Supplementary Table 4.

| Ageing marker | Age subgroup | Healthy (n) | Unhealthy (n) | $\Delta$ (years) | $\sigma^2$ (years$^2$) | OR [95% CI] | $\exp(\Delta/\sigma^2)$ [95% CI] |
|---|---|---|---|---|---|---|---|
| Retinal Age | <50 | 3,796 | 2,572 | +3.32 [+2.50, +4.18] | 29.3 [26.6, 31.9] | 1.114 [1.085, 1.145] | 1.120 [1.091, 1.154] |
| | 50-60 | 3,288 | 7,082 | +2.70 [+2.07, +3.30] | 28.8 [26.8, 30.9] | 1.103 [1.077, 1.129] | 1.098 [1.074, 1.123] |
| | 60-70 | 3,045 | 12,807 | +1.50 [+0.64, +2.25] | 32.6 [30.7, 34.2] | 1.048 [1.020, 1.077] | 1.047 [1.020, 1.072] |
| | 70-80 | 2,606 | 15,469 | +0.46 [−0.75, +1.54] | 36.6 [34.5, 38.4] | 1.012 [0.982, 1.044] | 1.013 [0.980, 1.043] |
| | ≥80 | 1,022 | 13,673 | −1.28 [−2.94, +0.66] | 29.2 [27.1, 31.2] | 0.955 [0.890, 1.025] | 0.957 [0.903, 1.024] |
| Chest Age | <30 | 1,240 | 1,345 | +1.72 [+1.11, +2.32] | 35.5 [23.1, 51.1] | 1.051 [1.024, 1.079] | 1.050 [1.030, 1.083] |
| | 30-40 | 1,530 | 1,593 | +1.05 [+0.19, +1.93] | 41.1 [35.1, 48.0] | 1.026 [1.004, 1.047] | 1.026 [1.005, 1.047] |
| | 40-50 | 1,994 | 2,193 | +0.44 [+0.01, +0.89] | 32.1 [29.1, 35.0] | 1.014 [1.000, 1.028] | 1.014 [1.000, 1.029] |
| | 50-60 | 2,615 | 3,238 | +0.13 [−0.28, +0.55] | 28.8 [26.1, 31.4] | 1.005 [0.990, 1.020] | 1.005 [0.990, 1.019] |
| | 60-70 | 1,798 | 2,402 | −0.17 [−0.69, +0.31] | 32.9 [24.4, 44.4] | 0.995 [0.980, 1.010] | 0.995 [0.980, 1.010] |
| | ≥70 | 728 | 1,057 | −0.80 [−1.58, −0.01] | 32.5 [27.1, 38.3] | 0.975 [0.953, 0.998] | 0.976 [0.955, 1.000] |
| Abdominal Age | <40 | 652 | 513 | +1.20 [+0.73, +1.70] | 15.4 [13.8, 16.9] | 1.080 [1.048, 1.112] | 1.081 [1.050, 1.115] |
| | 40-50 | 281 | 449 | +0.79 [+0.16, +1.42] | 19.5 [17.3, 21.7] | 1.041 [1.007, 1.077] | 1.041 [1.009, 1.075] |
| | 50-60 | 287 | 766 | +0.38 [−0.27, +1.09] | 26.1 [23.4, 28.7] | 1.015 [0.989, 1.041] | 1.015 [0.990, 1.043] |
| | 60-70 | 158 | 749 | +0.94 [+0.10, +1.76] | 29.9 [26.3, 33.6] | 1.032 [1.003, 1.061] | 1.032 [1.003, 1.062] |
| | 70-80 | 79 | 538 | −0.43 [−1.68, +0.72] | 30.2 [26.4, 33.7] | 0.986 [0.948, 1.026] | 0.986 [0.947, 1.025] |
| | ≥80 | 49 | 463 | +0.26 [−1.12, +1.63] | 30.1 [23.9, 37.1] | 1.005 [0.963, 1.049] | 1.009 [0.965, 1.060] |
| Brain Age | <65 | 169 | 40 | +4.25 [+2.22, +6.66] | 23.6 [17.8, 28.2] | 1.190 [1.076, 1.316] | 1.198 [1.101, 1.382] |
| | 65-70 | 193 | 77 | +3.49 [+2.22, +4.69] | 14.6 [10.8, 18.3] | 1.316 [1.168, 1.483] | 1.271 [1.172, 1.426] |
| | 70-75 | 245 | 113 | +2.69 [+1.90, +3.49] | 10.1 [8.5, 11.5] | 1.300 [1.193, 1.416] | 1.307 [1.199, 1.451] |
| | 75-80 | 184 | 138 | +1.83 [+1.14, +2.55] | 9.0 [7.3, 10.4] | 1.227 [1.115, 1.351] | 1.226 [1.130, 1.366] |
| | ≥80 | 142 | 153 | +1.25 [+0.61, +1.95] | 7.5 [6.2, 8.6] | 1.181 [1.078, 1.294] | 1.181 [1.085, 1.310] |

**Supplementary Table 7. Group slopes of the age gap on chronological age before and after age calibration and after balanced training for the four organ-derived ageing markers (Fig. 5 and Supplementary Fig. 19-21).** Slopes are from a linear regression of the age gap on chronological age, health group and their interaction, on the test subset; the slope difference is the Age × Group coefficient. Calibration subtracts the calibrator of Supplementary Table 8 from the age gap. Balanced training retrains the age model with inverse-frequency sampling in 5-year age bins and evaluates it on the same test subset. The 95% confidence intervals and Wald p-values use a cluster-robust (CR1) covariance clustered on patient (volume for Merlin, which releases no patient identifiers).

| Ageing marker | Condition | Healthy slope [95% CI] | Unhealthy slope [95% CI] | Slope difference [95% CI] | p |
|---|---|---|---|---|---|
| Retinal Age | Before calibration | −0.184 [−0.217, −0.151] | −0.314 [−0.326, −0.303] | −0.130 [−0.165, −0.096] | <0.001 |
| | Linear calibration | +0.017 [−0.015, +0.050] | −0.113 [−0.124, −0.101] | −0.130 [−0.165, −0.096] | <0.001 |
| | Quadratic calibration | +0.018 [−0.015, +0.051] | −0.101 [−0.113, −0.090] | −0.119 [−0.154, −0.084] | <0.001 |
| | Balanced training | −0.190 [−0.217, −0.162] | −0.344 [−0.355, −0.333] | −0.154 [−0.184, −0.124] | <0.001 |
| Chest Age | Before calibration | −0.202 [−0.223, −0.180] | −0.257 [−0.283, −0.231] | −0.056 [−0.073, −0.039] | <0.001 |
| | Linear calibration | −0.008 [−0.029, +0.014] | −0.063 [−0.089, −0.037] | −0.056 [−0.073, −0.039] | <0.001 |
| | Quadratic calibration | −0.004 [−0.025, +0.018] | −0.052 [−0.078, −0.026] | −0.049 [−0.065, −0.032] | <0.001 |
| | Balanced training | −0.170 [−0.191, −0.149] | −0.228 [−0.255, −0.202] | −0.058 [−0.075, −0.041] | <0.001 |
| Abdominal Age | Before calibration | −0.073 [−0.085, −0.060] | −0.115 [−0.125, −0.105] | −0.042 [−0.058, −0.026] | <0.001 |
| | Linear calibration | −0.006 [−0.018, +0.007] | −0.048 [−0.058, −0.038] | −0.042 [−0.058, −0.026] | <0.001 |
| | Quadratic calibration | −0.006 [−0.018, +0.007] | −0.047 [−0.056, −0.037] | −0.041 [−0.057, −0.025] | <0.001 |
| | Balanced training | −0.042 [−0.055, −0.029] | −0.105 [−0.115, −0.094] | −0.063 [−0.079, −0.046] | <0.001 |
| Brain Age | Before calibration | −0.372 [−0.416, −0.328] | −0.527 [−0.583, −0.471] | −0.155 [−0.222, −0.087] | <0.001 |
| | Linear calibration | −0.133 [−0.177, −0.089] | −0.288 [−0.344, −0.232] | −0.155 [−0.222, −0.087] | <0.001 |
| | Quadratic calibration | −0.144 [−0.188, −0.100] | −0.313 [−0.370, −0.256] | −0.169 [−0.237, −0.101] | <0.001 |
| | Balanced training | −0.286 [−0.330, −0.241] | −0.430 [−0.488, −0.371] | −0.144 [−0.214, −0.074] | <0.001 |

**Supplementary Table 8. Age calibrators.** Each calibrator is a regression of the age gap on chronological age in healthy data outside the test subset, written on the uncentred age scale as Calibrator(Age) $= \alpha + \beta_1 \cdot \text{Age} + \beta_2 \cdot \text{Age}^2$, with $\beta_2 = 0$ for the linear calibrator ($\alpha_l$ and $\beta_l$ in Methods). The calibrated age gap is the age gap minus the calibrator (Supplementary Table 7).

| Ageing marker | Calibrator | $\alpha$ | $\beta_1$ | $\beta_2$ | Fitted on |
|---|---|---|---|---|---|
| Retinal Age | Linear | 12.159 | $-0.2013$ | - | Healthy downstream training images (n=39,867) |
| | Quadratic | 9.899 | $-0.1223$ | $-6.58 \times 10^{-4}$ | Healthy downstream training images (n=39,867) |
| Chest Age | Linear | 8.928 | $-0.1941$ | - | Healthy downstream training images (n=31,603, 9,646 patients) |
| | Quadratic | 2.876 | 0.0899 | $-3.00 \times 10^{-3}$ | Healthy downstream training images (n=31,603, 9,646 patients) |
| Abdominal Age | Linear | 2.898 | $-0.0668$ | - | Healthy volumes of the age model validation subset |
| | Quadratic | 2.775 | $-0.0610$ | $-6.14 \times 10^{-5}$ | Healthy volumes of the age model validation subset |
| Brain Age | Linear | 17.470 | $-0.2385$ | - | Healthy volumes of the age model validation subset (n=229, 103 patients) |
| | Quadratic | 25.176 | $-0.4702$ | $1.72 \times 10^{-3}$ | Healthy volumes of the age model validation subset (n=229, 103 patients) |

**Supplementary Table 9. PhenoAgeAccel within age subgroups in UK Biobank (Fig. 7b, d-f).** Mean PhenoAgeAccel (years) of each health group with a normal-approximation 95% confidence interval. $\Delta$, $\sigma^2$ and $\exp(\Delta/\sigma^2)$ are defined as in Supplementary Table 6, with 95% confidence intervals from a patient-level bootstrap with 1,000 resamples. Odds ratios (OR) per 1-year increase in PhenoAgeAccel are from a logistic regression of unhealthy status on PhenoAgeAccel with chronological age as a covariate within each subgroup, with cluster-robust Wald 95% confidence intervals. Slopes and the tests of age dependence are in Supplementary Table 5.

| Age subgroup | Healthy (n) | Unhealthy (n) | Mean, healthy | Mean, unhealthy | $\Delta$ (years) | $\sigma^2$ (years$^2$) | OR [95% CI] | $\exp(\Delta/\sigma^2)$ [95% CI] |
|---|---|---|---|---|---|---|---|---|
| <50 | 6,131 | 2,143 | −0.25 [−0.37, −0.13] | +1.56 [+1.31, +1.81] | +1.83 [+1.54, +2.10] | 25.7 [24.6, 26.9] | 1.069 [1.059, 1.079] | 1.074 [1.062, 1.085] |
| 50-55 | 3,543 | 1,892 | −0.75 [−0.91, −0.59] | +1.09 [+0.82, +1.35] | +1.84 [+1.55, +2.15] | 26.8 [25.3, 28.2] | 1.069 [1.058, 1.081] | 1.071 [1.060, 1.083] |
| 55-60 | 4,045 | 2,730 | −1.23 [−1.38, −1.09] | +1.10 [+0.87, +1.33] | +2.34 [+2.07, +2.64] | 28.4 [27.0, 29.9] | 1.085 [1.075, 1.096] | 1.086 [1.076, 1.097] |
| 60-65 | 5,984 | 5,177 | −1.16 [−1.27, −1.05] | +0.94 [+0.77, +1.10] | +2.09 [+1.88, +2.28] | 26.8 [25.7, 27.9] | 1.083 [1.075, 1.091] | 1.081 [1.073, 1.089] |
| ≥65 | 4,487 | 5,380 | −0.73 [−0.87, −0.60] | +1.15 [+0.99, +1.31] | +1.87 [+1.67, +2.09] | 28.4 [27.3, 29.7] | 1.071 [1.062, 1.079] | 1.068 [1.061, 1.076] |

**Supplementary Table 10. Regression of the age gap on chronological age by disease category and by number of disease categories for Retinal Age on the AlzEye test subset (Supplementary Fig. 17 and 18).** Each group is compared with the healthy patients in the model age gap = $\beta_0 + \beta_1 \cdot \text{Age} + \beta_2 \cdot \text{Group} + \beta_3 \cdot (\text{Age} \times \text{Group})$, with age centred at the test subset mean of 68.4 years, so that the healthy slope is $\beta_1$ and the group slope $\beta_1 + \beta_3$. The number of disease categories counts the cardiovascular, chronic and ocular categories recorded for the patient. The healthy group has 13,757 images in every model; n is the number of images in the group. The 95% confidence intervals and the Wald p-value of $\beta_3$ use a cluster-robust (CR1) covariance clustered on patient.

| Group | n | $\beta_0$ | $\beta_1$ (healthy slope) | $\beta_2$ | $\beta_3$ (slope difference) | p ($\beta_3$) | Group slope |
|---|---|---|---|---|---|---|---|
| Unhealthy | 51,603 | −1.346 [−1.967, −0.724] | −0.184 [−0.217, −0.151] | +0.932 [+0.292, +1.573] | −0.130 [−0.165, −0.096] | <0.001 | −0.314 [−0.326, −0.303] |
| Cardiovascular | 13,952 | −1.346 [−1.967, −0.724] | −0.184 [−0.217, −0.151] | +1.423 [+0.724, +2.122] | −0.161 [−0.200, −0.122] | <0.001 | −0.345 [−0.365, −0.324] |
| Chronic | 44,556 | −1.346 [−1.967, −0.724] | −0.184 [−0.217, −0.151] | +0.884 [+0.242, +1.526] | −0.136 [−0.171, −0.101] | <0.001 | −0.320 [−0.332, −0.308] |
| Ocular | 22,797 | −1.346 [−1.967, −0.724] | −0.184 [−0.217, −0.151] | +1.768 [+1.099, +2.437] | −0.161 [−0.199, −0.124] | <0.001 | −0.345 [−0.364, −0.327] |
| Exactly 1 disease category | 26,648 | −1.346 [−1.967, −0.724] | −0.184 [−0.217, −0.151] | +0.457 [−0.195, +1.109] | −0.105 [−0.142, −0.069] | <0.001 | −0.289 [−0.305, −0.273] |
| Exactly 2 disease categories | 19,911 | −1.346 [−1.967, −0.724] | −0.184 [−0.217, −0.151] | +1.532 [+0.860, +2.205] | −0.173 [−0.210, −0.135] | <0.001 | −0.356 [−0.374, −0.339] |
| Exactly 3 disease categories | 4,945 | −1.346 [−1.967, −0.724] | −0.184 [−0.217, −0.151] | +2.359 [+1.535, +3.183] | −0.206 [−0.255, −0.158] | <0.001 | −0.390 [−0.426, −0.354] |
| At least 1 disease category | 51,504 | −1.346 [−1.967, −0.724] | −0.184 [−0.217, −0.151] | +0.937 [+0.297, +1.578] | −0.130 [−0.165, −0.095] | <0.001 | −0.314 [−0.325, −0.302] |
| At least 2 disease categories | 24,856 | −1.346 [−1.967, −0.724] | −0.184 [−0.217, −0.151] | +1.675 [+1.009, +2.340] | −0.177 [−0.214, −0.141] | <0.001 | −0.361 [−0.378, −0.345] |
| At least 3 disease categories | 4,945 | −1.346 [−1.967, −0.724] | −0.184 [−0.217, −0.151] | +2.359 [+1.535, +3.183] | −0.206 [−0.255, −0.158] | <0.001 | −0.390 [−0.426, −0.354] |

**Supplementary Table 11. Odds ratio per 1-year increase in the age gap at selected chronological ages for Retinal Age, by disease category (Supplementary Fig. 22a for unhealthy status).** For each category a logistic regression of the category against healthy patients on chronological age, the age gap and Age × age gap was fitted on the AlzEye test subset, giving $\mathrm{OR(Age)} = \exp(\beta_{\text{age gap}} + \beta_{\text{Age}\times\text{age gap}} \cdot \text{Age})$. Values are shown at 10-year intervals from 30 to 100 years, with 95% confidence intervals and Wald p-values from a cluster-robust (CR1) covariance clustered on patient.

| Age (years) | Unhealthy | | Cardiovascular | | Chronic | | Ocular | |
|---|---|---|---|---|---|---|---|---|
| | OR [95% CI] | p | OR [95% CI] | p | OR [95% CI] | p | OR [95% CI] | p |
| 30 | 1.118 [1.082, 1.157] | <0.001 | 1.148 [1.096, 1.203] | <0.001 | 1.119 [1.081, 1.158] | <0.001 | 1.165 [1.120, 1.211] | <0.001 |
| 40 | 1.098 [1.071, 1.125] | <0.001 | 1.119 [1.081, 1.159] | <0.001 | 1.098 [1.071, 1.125] | <0.001 | 1.136 [1.104, 1.169] | <0.001 |
| 50 | 1.077 [1.059, 1.096] | <0.001 | 1.091 [1.064, 1.119] | <0.001 | 1.077 [1.058, 1.096] | <0.001 | 1.108 [1.086, 1.130] | <0.001 |
| 60 | 1.057 [1.041, 1.074] | <0.001 | 1.064 [1.043, 1.085] | <0.001 | 1.057 [1.041, 1.073] | <0.001 | 1.080 [1.062, 1.099] | <0.001 |
| 70 | 1.038 [1.017, 1.059] | <0.001 | 1.037 [1.014, 1.061] | 0.001 | 1.037 [1.016, 1.059] | <0.001 | 1.054 [1.031, 1.077] | <0.001 |
| 80 | 1.019 [0.989, 1.049] | 0.217 | 1.012 [0.980, 1.044] | 0.476 | 1.018 [0.988, 1.048] | 0.247 | 1.027 [0.996, 1.060] | 0.084 |
| 90 | 1.000 [0.961, 1.040] | 0.988 | 0.986 [0.945, 1.029] | 0.527 | 0.998 [0.960, 1.039] | 0.939 | 1.002 [0.961, 1.044] | 0.923 |
| 100 | 0.981 [0.934, 1.031] | 0.451 | 0.962 [0.910, 1.016] | 0.165 | 0.980 [0.932, 1.030] | 0.421 | 0.977 [0.927, 1.030] | 0.391 |

**Supplementary Table 12. Area under the receiver operating characteristic curve (AUC) of the disease classification models overall and within age groups for Retinal Age (Supplementary Fig. 22b-f).** For unhealthy status and for cardiovascular, chronic and ocular disease, three logistic regression models (Age; Age + age gap; Age + age gap + Age × age gap) were fitted on the AlzEye downstream training subset and evaluated on the AlzEye test subset. AUC is reported for all ages and within three age groups, each with a 95% confidence interval from a patient-level cluster bootstrap with 1,000 resamples of patients.

| Outcome | Model | All ages | <50 years | 50-80 years | >80 years |
|---|---|---|---|---|---|
| Unhealthy | Age | 0.731 [0.706, 0.756] | 0.580 [0.527, 0.632] | 0.622 [0.592, 0.653] | 0.585 [0.514, 0.667] |
| | Age + age gap | 0.737 [0.711, 0.764] | 0.674 [0.627, 0.719] | 0.641 [0.611, 0.673] | 0.494 [0.410, 0.585] |
| | Age + age gap + Age × age gap | 0.741 [0.715, 0.767] | 0.685 [0.641, 0.727] | 0.641 [0.611, 0.673] | 0.564 [0.495, 0.642] |
| Cardiovascular | Age | 0.806 [0.781, 0.830] | 0.638 [0.561, 0.705] | 0.676 [0.639, 0.708] | 0.620 [0.548, 0.700] |
| | Age + age gap | 0.809 [0.783, 0.835] | 0.697 [0.616, 0.765] | 0.692 [0.653, 0.725] | 0.556 [0.477, 0.649] |
| | Age + age gap + Age × age gap | 0.813 [0.789, 0.837] | 0.689 [0.608, 0.760] | 0.695 [0.658, 0.728] | 0.606 [0.536, 0.688] |
| Chronic | Age | 0.727 [0.698, 0.752] | 0.586 [0.528, 0.641] | 0.614 [0.581, 0.647] | 0.579 [0.501, 0.664] |
| | Age + age gap | 0.733 [0.703, 0.761] | 0.682 [0.631, 0.728] | 0.633 [0.601, 0.665] | 0.485 [0.404, 0.578] |
| | Age + age gap + Age × age gap | 0.737 [0.709, 0.764] | 0.692 [0.647, 0.736] | 0.634 [0.601, 0.666] | 0.558 [0.483, 0.637] |
| Ocular | Age | 0.758 [0.731, 0.783] | 0.537 [0.468, 0.609] | 0.656 [0.621, 0.687] | 0.588 [0.511, 0.674] |
| | Age + age gap | 0.769 [0.741, 0.795] | 0.723 [0.674, 0.771] | 0.684 [0.650, 0.717] | 0.508 [0.420, 0.610] |
| | Age + age gap + Age × age gap | 0.774 [0.747, 0.799] | 0.760 [0.717, 0.799] | 0.685 [0.651, 0.716] | 0.571 [0.498, 0.656] |

**Supplementary Table 13. Unhealthy classification performance for Retinal Age on the AlzEye test subset.** Two logistic regression models (Age; Age + age gap + Age × age gap) were fitted on the AlzEye downstream training subset. Sensitivity, specificity and F1 score are evaluated at the threshold selected by the Youden criterion on the downstream training subset. The 95% confidence intervals are from a patient-level cluster bootstrap with 1,000 resamples of patients. The p-values compare the two models in AUC, F1 score, sensitivity and specificity, each using the bootstrap standard error of the difference.

| Model | AUC [95% CI] | p (ΔAUC) | F1 [95% CI] | p (ΔF1) | Sensitivity [95% CI] | p (Δsensitivity) | Specificity [95% CI] | p (Δspecificity) |
|---|---|---|---|---|---|---|---|---|
| Age | 0.731 [0.706, 0.756] | <0.001 | 0.777 [0.765, 0.789] | <0.001 | 0.698 [0.684, 0.712] | <0.001 | 0.632 [0.593, 0.674] | 0.353 |
| Age + age gap + Age × age gap | 0.741 [0.715, 0.767] | | 0.789 [0.778, 0.800] | | 0.715 [0.702, 0.728] | | 0.638 [0.599, 0.678] | |

**Supplementary Table 14. Pairwise ordering of healthy and unhealthy patients within age subgroups for the four organ-derived ageing markers (Fig. 6a-d).** For every healthy-unhealthy pair within a subgroup, the percentage of pairs in which the unhealthy patient has the larger age gap and the percentage in which the healthy patient does. Age subgroups are as in Fig. 3.

| Ageing marker | Age subgroup | Healthy (n) | Unhealthy (n) | Unhealthy larger (%) | Healthy larger (%) |
|---|---|---|---|---|---|
| Retinal Age | <50 | 3,796 | 2,572 | 65.1 | 34.9 |
| | 50-60 | 3,288 | 7,082 | 63.1 | 36.9 |
| | 60-70 | 3,045 | 12,807 | 57.8 | 42.2 |
| | 70-80 | 2,606 | 15,469 | 52.2 | 47.8 |
| | ≥80 | 1,022 | 13,673 | 41.1 | 58.9 |
| Chest Age | <30 | 1,240 | 1,345 | 57.5 | 42.5 |
| | 30-40 | 1,530 | 1,593 | 54.8 | 45.2 |
| | 40-50 | 1,994 | 2,193 | 52.1 | 47.9 |
| | 50-60 | 2,615 | 3,238 | 50.4 | 49.6 |
| | 60-70 | 1,798 | 2,402 | 49.5 | 50.5 |
| | ≥70 | 728 | 1,057 | 44.8 | 55.2 |
| Abdominal Age | <40 | 652 | 513 | 55.4 | 44.6 |
| | 40-50 | 281 | 449 | 54.8 | 45.2 |
| | 50-60 | 287 | 766 | 52.0 | 48.0 |
| | 60-70 | 158 | 749 | 54.3 | 45.7 |
| | 70-80 | 79 | 538 | 47.3 | 52.7 |
| | ≥80 | 49 | 463 | 47.5 | 52.5 |
| Brain Age | <65 | 169 | 40 | 66.5 | 33.5 |
| | 65-70 | 193 | 77 | 74.5 | 25.5 |
| | 70-75 | 245 | 113 | 71.9 | 28.1 |
| | 75-80 | 184 | 138 | 67.1 | 32.9 |
| | ≥80 | 142 | 153 | 59.2 | 40.8 |

**Supplementary Table 15. Disease definitions for AlzEye and UK Biobank.** For both cohorts, disease labels were derived from linked hospital records using International Classification of Diseases, 10th Revision (ICD-10) codes, supplemented in UK Biobank by self-reported codes (Data-Field 20002).

| Category | Disease | AlzEye | UK Biobank | |
|---|---|---|---|---|
| | | ICD-10 | ICD-10 | Self-report |
| Cardiovascular | Atrial fibrillation | I48 | I48 | 1471 |
| | Heart failure | I110, I130, I132, I255, I420, I425, I428, I429, I500, I501, I509 | I110, I130, I132, I255, I420, I425, I428, I429, I500, I501, I509 | 1076 |
| | Myocardial infarction | I21, I22, I23, I241, I252 | I21, I22, I23, I241, I252 | 1075 |
| | Intracerebral haemorrhage | I61 | I61 | 1491 |
| | Ischaemic stroke | I63, I64 | I63, I64 | 1583 |
| | Aortic stenosis | I350 | I350 | 1490 |
| | Aortic regurgitation | I351 | I351 | 1587 |
| | Mitral stenosis | I342, I050 | I342, I050 | 1489 |
| | Mitral regurgitation | I340 | I340 | 1585 |
| Chronic | Diabetes mellitus | E08, E09, E10, E11, E13 | E08, E09, E10, E11, E13 | 1220, 1222, 1223 |
| | Hypertension | I10, I11, I12, I13, I15 | I10, I12, I15 | 1065, 1072 |
| Ocular | Glaucoma | H40, H41, H42 | H40 | 1277 |
| | Diabetic retinopathy | E103, E113, H360 | H360 | 1276 |
| | Age-related macular degeneration | H353 | — | — |
| | Cataract | — | H25, H26 | 1278 |
| Neurodegenerative | Dementia | F01, F02, F03 | F01, F02, F03 | 1263 |
| | Parkinson's disease | G20, G21 | G20, G21 | 1262 |

**Supplementary Table 16. The nine blood biomarkers used to calculate PhenoAge in UK Biobank.** Values are median [interquartile range] from the first blood test of the 41,512 patients analysed, in total and in the healthy (24,190) and unhealthy (17,322) groups, and the 0.5th-99.5th percentile range in all patients. C-reactive protein was divided by 10 to give mg/dL, the unit used in the PhenoAge equation.

| Biomarker | UK Biobank field | Unit | All patients | Healthy | Unhealthy | 0.5th-99.5th percentile |
|---|---|---|---|---|---|---|
| Albumin | 30600 | g/L | 45.5 [43.8, 47.3] | 45.6 [43.9, 47.3] | 45.4 [43.6, 47.2] | 38.3-52.7 |
| Creatinine | 30700 | $\mu$mol/L | 71.4 [62.1, 82.2] | 70.4 [61.6, 80.7] | 72.9 [63.2, 84.3] | 43.9-131.5 |
| Glucose | 30740 | mmol/L | 5.02 [4.72, 5.37] | 4.97 [4.70, 5.28] | 5.10 [4.77, 5.53] | 3.67-12.59 |
| C-reactive protein (CRP) | 30710 | mg/L | 1.36 [0.68, 2.84] | 1.20 [0.60, 2.48] | 1.63 [0.81, 3.35] | 0.14-28.84 |
| Lymphocyte percentage | 30180 | % | 28.9 [24.2, 34.1] | 29.3 [24.7, 34.4] | 28.4 [23.5, 33.6] | 10.7-53.3 |
| Mean cell volume (MCV) | 30040 | fL | 92.2 [89.6, 94.8] | 92.2 [89.7, 94.8] | 92.2 [89.5, 94.9] | 73.5-104.4 |
| Red cell distribution width (RDW) | 30070 | % | 13.4 [13.0, 14.0] | 13.4 [13.0, 13.9] | 13.5 [13.0, 14.1] | 12.0-18.4 |
| Alkaline phosphatase (ALP) | 30610 | U/L | 81.6 [68.2, 97.2] | 80.0 [66.8, 95.2] | 83.8 [70.4, 100.0] | 38.8-179.2 |
| White blood cell count (WBC) | 30000 | $10^9$ cells/L | 6.87 [5.81, 8.10] | 6.69 [5.69, 7.87] | 7.12 [6.01, 8.41] | 3.54-13.55 |